\documentclass[a4paper,fleqn,doubleblind]{cas-dc}
\usepackage[authoryear,longnamesfirst]{natbib}

\usepackage{siunitx}
\usepackage{tikz}
\usepackage{pgfplots}
\usepackage{subcaption}
\usepackage{multirow}
\usepackage{tabularx}
\usepackage{longtable}
\usepackage{balance}
\pgfplotsset{compat=newest}
\usepgfplotslibrary{groupplots}
\usepgfplotslibrary{dateplot}
\usetikzlibrary{automata,positioning}
\usetikzlibrary{shapes.geometric, arrows}
\tikzstyle{box} = [rectangle, rounded corners, minimum width=2cm, minimum height=1cm,text centered, draw=black, ]
\usepackage{enumitem}

\def\tsc#1{\csdef{#1}{\textsc{\lowercase{#1}}\xspace}}
\tsc{WGM}
\tsc{QE}

\usepackage{xcolor}
\newcommand{\changed}[1]{{\color{black}{#1}}}
\usepackage{soul}

\begin{document}
\let\WriteBookmarks\relax
\def\floatpagepagefraction{1}
\def\textpagefraction{.001}

\shorttitle{Evaluating Privacy, Security, and Trust Perceptions in Conversational AI}    

\shortauthors{Leschanowsky et al.}

\title [mode = title]{Evaluating Privacy, Security, and Trust Perceptions in Conversational AI: A Systematic Review}  


\affiliation[1]{organization={Fraunhofer IIS},
            addressline={Am Wolfsmantel 33}, 
            city={Erlangen},
            postcode={91058}, 
            country={Germany}}
\affiliation[2]{organization={Aalto University},
            addressline={Konemiehentie 1}, 
           city={Espoo},
            postcode={02150}, 
            country={Finland}}





\author[1]{Anna Leschanowsky}
\ead[url]{anna.leschanowsky@iis.fraunhofer.de}
\cormark[1]
\author[2]{Silas Rech}

\author[1]{Birgit Popp}

\author[2]{Tom Bäckström}
\cortext[cor]{Corresponding author}

\begin{abstract}
\changed{\changed{Conversational AI (CAI) systems which encompass voice- and text-based assistants are on the rise and have been largely integrated into people's everyday lives. Despite their widespread adoption, users voice concerns regarding privacy, security and trust in these systems. However, the composition of these perceptions, their impact on technology adoption and usage and the relationship between privacy, security and trust perceptions in the CAI context remain open research challenges. This study contributes to the field by conducting a Systematic Literature Review and offers insights into the current state of research on privacy, security and trust perceptions in the context of CAI systems. The review covers application fields and user groups and sheds light on empirical methods and tools used for assessment. Moreover, it provides insights into the reliability and validity of privacy, security and trust scales, as well as extensively investigating the subconstructs of each item as well as additional concepts which are concurrently collected. We point out that the perceptions of trust, privacy and security overlap based on the subconstructs we identified. While the majority of studies investigate one of these concepts, only a few studies were found exploring privacy, security and trust perceptions jointly. Our research aims to inform on directions to develop and use reliable scales for users' privacy, security and trust perceptions and contribute to the development of trustworthy CAI systems.}}
\end{abstract}


\maketitle

\section{Introduction}
Conversational AI (CAI) systems use AI to enable natural conversations with users via voice or text. Voice-enabled assistants such as Amazon's Alexa or Apple's Siri have gained widespread adoption and can be integrated into devices such as smart speakers or smartphones. At the same time, text-based CAI systems have been on the rise with the increasing usage of chatbots and the emergence of Large Language Models (LLMs)~\citep{VixenLabs2023}. The integration of CAI systems into Internet of Things (IoT) devices and assisted living technologies further empowers users with seamless control of their devices~\citep{ammariMusicSearchIoT2019a}.  
Despite their remarkable capabilities, CAI systems have been criticised for raising significant privacy concerns and associated threats. Criticism ranges from privacy breaches to the constant listening capabilities and intransparent data handling practices~\citep{eduSmartHomePersonal2021, boltonSecurityPrivacyChallenges2021}. These concerns, in turn, can reduce trustworthiness in the manufacturer and its devices leading to undesired user experiences or their complete abandonment~\citep{dzindolet2003roleoftrustinautomationreliance}. 

\begin{figure}[tb]
    \centering    
    \includegraphics[width=0.5\textwidth]{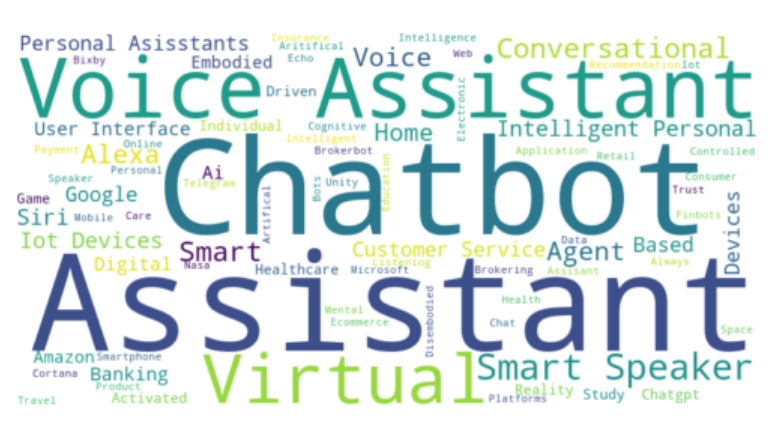}
    \caption{Wordcloud as an illustration of the used terms for researched devices in privacy, security and trust perception papers}
    \label{fig:wordcloudprivacy}
\end{figure}

To enable user-centric development and useful experiences, understanding users' perceptions of privacy, security and trust concerning CAI \changed{is crucial~\citep{panjaitan2023influence}}. As users increasingly engage in interactions with CAI systems, the exchange of personal data has become ubiquitous, necessitating a robust understanding of the factors influencing users' trust. Previous work has emphasized the need for joint investigations of privacy and trust perceptions including domains such as online trading systems~\citep{carlosrocaImportancePerceivedTrust2009}, electronic commerce~\citep{liuConcernPrivacytrustbehavioralIntention2005, Metzger2004ecommerce}, online social networks~\citep{Zlatolas2019AMO}, mobile applications~\citep{kitkowskaShareProtectUnderstanding2023} and smart home automation~\citep{schomakersUsersPreferencesSmart2021}. Yet, the current state of research regarding joint investigations of these constructs in the context of CAI systems remains largely unexplored. 
\changed{Here, we present a systematic review of privacy, security and trust perceptions in CAI.} 

\changed{We argue that trust} in these contexts is not only a function of the security measures protecting privacy but also encompasses the perceived intentions, social components and ethical considerations of the entities collecting and managing personal information. \changed{This can be seen by the influence that trust can have on technology acceptance based on reputation by institutions~\citep{misiolek2002trust}}. Therefore, trust and privacy perceptions have been explored along with other factors in context such as technology adoption and usage by combining Technology Acceptance Models (TAM) and trust perception models~\citep{marangunic2015technologyTAM, ivarsson2023suspicious} or privacy and security factors~\citep{kowalczukConsumerAcceptanceSmart2018a, buteauHeyAlexaWhy2021}. Yet, due to diverse conceptualizations of privacy, security and trust and their contextual dependencies, generalization and evaluation of their impact on other factors present ongoing challenges.

The privacy and security field has long relied on quantitative as well as qualitative research to assess users' perceptions~\citep{distlerSystematicLiteratureReview2021, pattnaikSurveyUserPerspectives2023c}. These methods have also been applied to investigate trust such as interpersonal trust~\citep{rotter1967new, mayer1995integrative} and trust in automation~\citep{lee2004trust, hoff2015trust, jianFoundationsEmpiricallyDetermined2000a}. While qualitative research is usually characterized by collecting non-numerical data such as interview transcripts, quantitative research focuses on the collection of numerical data for example by using questionnaires. When assessing peoples' perceptions through the use of quantitative surveys, the reliability and validity of scales are crucial. While valid and reliable scales can offer valuable insights into users' perceptions and can confidently inform businesses and policymakers, unreliable ones lack predictive power and generalizability~\citep{colnagoItConcernPreferencea}. This could have serious consequences for the credibility of privacy and trust research, investment into privacy-preserving measures, enhancement of trustworthiness in design and influence on policy-making. As privacy and trust have been long and extensively studied, well-established scales play a crucial role in quantifying and understanding the nuances of privacy and trust within technology-mediated relationships. At the same time, reuse and adaption of privacy and trust scales have been critically discussed in the literature~\citep{preibuschGuideMeasuringPrivacy2013, colnagoItConcernPreferencea, Gross2021ValidityAR, chita2021can}, and it remains an open question whether privacy, security and trust scales originally designed for online environments prove reliable and valid in the context of CAI systems. The emergence of CAI-specific usability scales such as the Voice Usability Scale (VUS) or BOT Usability Scale~\citep{zwakmanUsabilityEvaluationArtificial2021c, borsciChatbotUsabilityScale2022b}, underscore the necessity to explore reliability and validity of currently used privacy, security and trust scales and to assess the need of CAI-specific metrics. 

In this work, we conduct a Systematic Literature Review (SLR) to better understand the breadth of research concerning privacy, security and trust perception in the context of CAI systems. In particular, we 1) aim to understand the current state of research concerning the interplay of privacy, security and trust perceptions in the context of CAI and their influence on factors investigated alongside. Moreover, we want to 2) shed light on currently used conceptualizations, measurement methods and tools and their benefits and limitations \changed{, in particular how breaches of privacy can erode trust and how users' perceptions of privacy and trust play an essential role in ensuring the adoption and continuous use of conversational AI systems}. Finally, we hope to 3) inform future work on conducting reliable and valid privacy and trust research in the context of CAI, and provide future research directions \changed{and encourage researchers to critically assess and discuss investigations in this domain}.  

\section{Related Work and Contributions}

\changed{Despite previous research on the interplay of privacy, security and trust perceptions, to the best of our knowledge, systematic reviews of research methodologies and constructs related to privacy, security and trust perceptions have not been conducted.}.

\changed{Trust and privacy have long been explored within the context of online environments, e-commerce, online social networks and Conversational AI~\citep{liuConcernPrivacytrustbehavioralIntention2005, taddei2013privacy, maqableh2021integrating, dekkalFactorsAffectingUser2023a}. These two concepts share similarities, e.g., being essential for human's daily lives, but also differ, for instance, regarding their legislative implementation~\citep{kosa2010vampire}. Numerous attempts have aimed to formalize both privacy and trust and to conceptualize their influence on people's perception and behavior in technological systems. Yet, the interplay between privacy and trust remains an open question. Results vary with trust being treated as an antecedent to privacy, outcome of privacy or as a moderating variable~\citep{smithInformationPrivacyResearch2011}. For instance, \cite{liuConcernPrivacytrustbehavioralIntention2005} proposed a Privacy-trust-behavioral intention model for electronic commerce, showing that privacy significantly influenced behavioral intention mediated by trust. In contrast, in the context of online social networks, ~\cite{taddei2013privacy} compared three different models related to the relationship between control, trust, privacy concerns and online self-disclosure and found that a moderating effect between trust and privacy concerns was most explanatory. As privacy and trust are highly subjective and contextual concepts, it is, however, essential to analyse these findings in the context of their research methodology, measurement scales, and participants sample.} 

\changed{In addition to privacy, the role of security perceptions and their relationship to other concepts in the context of CAI is unclear.} While the concept of security is distinct from privacy~\citep{smithInformationPrivacyResearch2011}, a scoping review on human factors in cybersecurity found that despite their focus on security, privacy was one of the most frequently mentioned words across all texts~\citep{rahmanHumanFactorsCybersecurity2021}. Moreover, studies suggest that people may not clearly distinguish between privacy and security, pooling these concepts in their perception~\citep{bruggemeierPerceptionsReactionsConversational2022b}. \changed{Research on social network sites proposes combined models suggesting that perceived privacy and security influence continuance intention mediated by trust and satisfaction~\citep{maqableh2021integrating}. On the contrary, the Privacy-trust-behavioral intention model conceptualizes security as one dimension of privacy~\citep{liuConcernPrivacytrustbehavioralIntention2005}. Our literature review addresses both privacy and security aspects to cover the full range of privacy and security research and to provide insights into how these concepts have been investigated in the context of CAI.}

\changed{Given the complex interplay of privacy, security and trust, and possible differences concerning technological advances, this systematic literature review aims to shed light on previously explored relationships of trust and privacy perceptions in the context of CAI.
So far, systematic literature reviews in this domain have focused on conversational assistants or IoT devices more generally without a dedicated focus on users' perceptions~\citep{debarcelossilvaIntelligentPersonalAssistants2020a, pattnaikSurveyUserPerspectives2023c}, on privacy and security aspects of CAI systems~\citep{maccarioPrivacySmartSpeakers2023, boltonSecurityPrivacyChallenges2021, eduSmartHomePersonal2021} or empirical methods and measurement scales used in usable privacy and security research~\citep{distlerSystematicLiteratureReview2021, preibuschGuideMeasuringPrivacy2013, rohanSystematicLiteratureReview2023}. Our SLR uniquely contributes to the field by maintaining a specific focus on voice- and text-based CAI systems and the research methodologies employed to assess users' privacy, security and trust perceptions. Consequently, we aim to offer an overview of the varied privacy, security and trust aspects investigated in the context of CAI and their interconnectedness.}
Thereby, we aim to answer the following research questions:

\begin{enumerate}[label={\bfseries RQ\arabic*}]
    \item Which methods are used to research privacy, security and trust perceptions for CAI systems?
    \item What constructs and sub-constructs are researched? 
    \item Which and how often are privacy, security and trust perception scales \changed{used in research on CAI}?
    \item\changed{To what extent have privacy, security and trust perception scales, used in research on CAI, demonstrated reliability and validity?}
    \item Are privacy, security and trust perceptions researched in parallel in the context of CAI? 
    \item Which topics are researched alongside the evaluation of privacy, security and trust perceptions in the context of CAI? 
\end{enumerate}

\section{Method}

We followed the Preferred Reporting Items for Systematic Reviews and Meta-Analysis (PRISMA) method to conduct our systematic literature review on privacy, security and trust perceptions in CAI~\citep{moherPreferredReportingItems2009}. In total, we conducted two individual SLRs with one on the topic of privacy \& security perceptions and one on trust perceptions in CAI systems.
The procedures are outlined in this section and shown in Figure~\ref{fig:prismaprivacy}. We first identified relevant research papers before selecting them based on exclusion and inclusion criteria. Afterwards, we assessed their eligibility and selected appropriate records for the final study.

\subsection{Search Strategy}

\begin{table}
  \caption{Keywords Used in the Search Queries of the SLR on Privacy Perceptions}
  \label{tab:keywordsPrivacy}
  
  \scriptsize
  \begin{tabular}{p{0.05\columnwidth}p{0.85\columnwidth}}
    \toprule
    & (Privacy Perception* OR Perception* of Privacy OR Perceived Privacy \\
    & OR Security Perception* OR Perception* of Security OR Perceived Security) \\
    \midrule
    AND & (Conversational AI OR CAI OR IoT OR Internet of Things OR Voice Assistant* OR Alexa OR Google Assistant \\
    & OR Siri OR Virtual Assistant* OR Smart Speaker* OR Chatbot*) \\
    \bottomrule
\end{tabular}
\end{table} 

\begin{table}
  \caption{Keywords Used in the Search Queries of the SLR on Trust Perceptions}
  \label{tab:keywordsTrust}
  \scriptsize
  
  \begin{tabular}{p{0.05\columnwidth}p{0.85\columnwidth}}
    \toprule
    & (Trust Perception* OR Perception* of Trust OR Perceived Trust OR Trust) \\
    \midrule
    AND & (Conversational AI OR CAI OR IoT OR Internet of Things OR Voice Assistant* OR Alexa OR Google Assistant \\
    & OR Siri OR Virtual Assistant* OR Smart Speaker* OR Chatbot*) \\
    \bottomrule
\end{tabular}
\end{table} 

We included research papers from five databases, i.e. ACM Digital Library, IEEE Xplore, Scopus, Web of Science and SpringerLink.
Table~\ref{tab:keywordsPrivacy} and~\ref{tab:keywordsTrust}
show the keywords used for searching for records on privacy \& security perceptions and trust. While the first subset of keywords captures the aspect of privacy \& security or trust, the second subset of keywords covers a variety of words used to describe text-based as well as voice-based Conversational AI systems. As CAI systems are often integrated into Internet of Things (IoT) devices, e.g. smart speakers, smart fridges or smart vacuum cleaners, we included keywords related to IoT in our search query. 
Similarly to~\cite{pattnaikSurveyUserPerspectives2023c}, we included ACM Digital Library and IEEE Xplore as search databases as they are important publishers in the field of privacy and security and smart appliances \changed{as well as human-computer interaction}. Moreover, we included Elsevier's Scopus and Web of Science as they were shown to be comprehensive yet limited in their overlap~\citep{bar-ilanTaleThreeDatabases2018}. Finally, we included SpringerLink as a journal platform~\citep{gusenbauerWhichAcademicSearch2020} \changed{for their coverage of psychological journals}. We restricted our keyword search to abstracts and included only articles and conference papers. 

\subsection{Study Selection}
\label{sec:exclusion}

By following the PRISMA method, we established exclusion and inclusion criteria to select relevant papers. 
Exclusion criteria were as follows: 

\begin{itemize}
    \item Non-English papers
    \item Workshop descriptions or book chapters
    \item Review papers
    \item Papers that do not cover CAI, e.g. connected cars, smartphone applications, smart meters, blockchain, cryptography, supply chain
    \item Papers that do not include a user study related to privacy, security or trust perceptions
\end{itemize}

After excluding papers based on the criteria described above, we included papers assessing privacy, security and trust perceptions in the context of Conversational AI systems. Finally, we conducted backwards and forward snowballing to identify additional relevant papers using ResearchRabbit~\footnote{https://www.researchrabbit.ai/}. We applied the same inclusion and exclusion criteria and stopped snowballing after one iteration.  

\subsection{Analysis Procedure}

We developed a data chart to extract relevant variables from the research papers. In addition to more general information such as the date of publication, author's affiliation and publication venue, we focused on more specific attributes related to our research questions. 
To gain insight into the current landscape of research methods used in the context of privacy and trust perceptions in CAI, we focused on methodological data, e.g. research method, research platform, number, type and location of research subjects. Moreover, we were interested in the response variables that were assessed by the researchers, the quality measures used to ensure reliability and validity and the techniques applied for analyzing the data. We were particularly interested in the variety of response variables that were evaluated within one study as well as their origins. 

\section{Application Fields and Devices, Research Methods }

\begin{figure*}[ht]
    \centering
    \begin{tikzpicture}
    \node (boxScopus) [box, xshift=-0.5cm] {\shortstack{Scopus \\ $n_{Privacy}=133$ \\
    $n_{Trust}=1880$}};
    \node (boxWoS) [box, right=of boxScopus, xshift=-0.5cm] {\shortstack{Web of Science \\ $n_{Privacy}=62$ \\
    $n_{Trust}=438$} };
    \node (boxSpringer) [box, right=of boxWoS, xshift=-0.5cm] {\shortstack{SpringerLink \\ $n_{Privacy}=121$ \\
    $n_{Trust}=921$} };
    \node (boxIEEE) [box, right=of boxSpringer, xshift=-0.5cm] {\shortstack{IEEEXplore \\ $n_{Privacy}=510$\\
    $n_{Trust}=13$} };
    \node (boxACM) [box, right=of boxIEEE, xshift=-0.5cm] {\shortstack{ACM Digital Library \\ $n_{Privacy}=14$ \\
    $n_{Trust}=133$} };
    \node (boxMerged) [box, below=of boxSpringer] {\shortstack{All search results \\ $n_{Privacy}=840$ \\
    $n_{Trust}=3385$} };
    \node (boxDuplicated) [box, below=of boxMerged, yshift=0.5cm] {\shortstack{Records after removing duplicates \\ $n_{Privacy}=767$ \\
    $n_{Trust}=2911$} };
    \node (boxScreeningOne) [box, below=of boxDuplicated, yshift=0.5cm] {\shortstack{Records after screening titles and abstracts \\ $n_{Privacy}=317$ \\
    $n_{Trust}=168$} };
    \node (boxScreeningTwo) [box, below=of boxScreeningOne, yshift=0.5cm] {\shortstack{Records after full-text screening \\ $n_{Privacy}=\changed{75}$ \\
    $n_{Trust}=51$} };
    \node (boxSnowball) [box, below=of boxScreeningTwo, yshift=0.5cm, xshift=-2cm] {\shortstack{+ Backward Snowballing \\ $n_{Privacy}=15$ \\
    $n_{Trust}=4$} };
    \node (boxForwardSnowball) [box, below=of boxScreeningTwo, yshift=0.5cm, xshift=2cm] {\shortstack{+ Forward Snowballing \\ $n_{Privacy}=10$ \\
    $n_{Trust}=3$} };
    \node (boxFinal) [box, below=of boxScreeningTwo, yshift=-1.5cm] {\shortstack{Final Selection \\ $n_{Privacy}=\changed{100}$ \\
    $n_{Trust}=58$} };

    \draw [-stealth] (boxScopus.south) -- (boxMerged.north);
    \draw [-stealth] (boxWoS.south) -- (boxMerged.north);
    \draw [-stealth] (boxSpringer.south) -- (boxMerged.north);
    \draw [-stealth] (boxIEEE.south) -- (boxMerged.north);
    \draw [-stealth] (boxACM.south) -- (boxMerged.north);
    \draw [-stealth] (boxMerged.south) -- (boxDuplicated.north);
    \draw [-stealth] (boxDuplicated.south) -- (boxScreeningOne.north);
    \draw [-stealth] (boxScreeningOne.south) -- (boxScreeningTwo.north);
    \draw [-stealth] (boxScreeningTwo.south) -- (boxSnowball.north);
    \draw [-stealth] (boxScreeningTwo.south) -- (boxForwardSnowball.north);
    \draw [-stealth] (boxSnowball.south) -- (boxFinal.north);
    \draw [-stealth] (boxForwardSnowball.south) -- (boxFinal.north);
    
    \end{tikzpicture}
    \caption{PRISMA Procedure used for the Systematic Literature Review on privacy, security and trust perceptions.}
    \label{fig:prismaprivacy}
 \end{figure*}
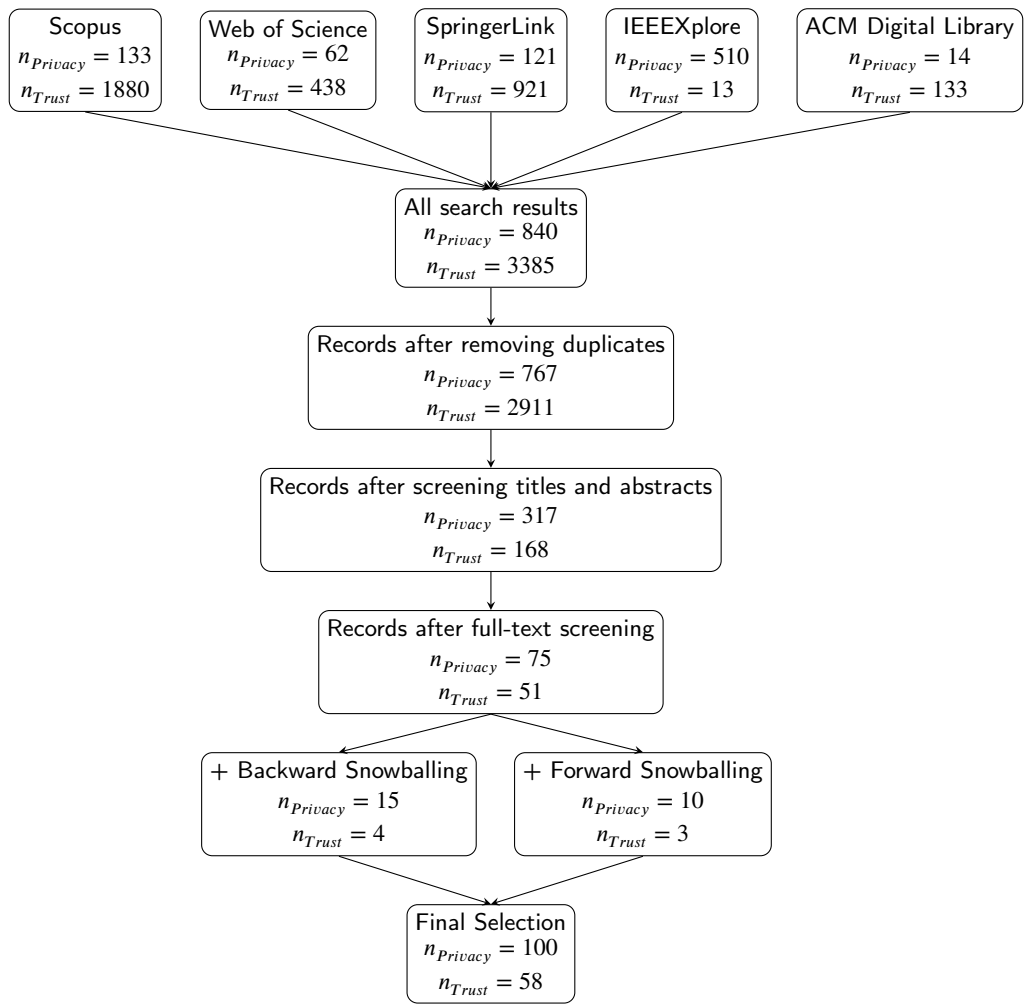
 
\subsection{Selected Papers} 

Our paper selection process is shown in Figure~\ref{fig:prismaprivacy}. Our final selection resulted in \changed{100} papers related to privacy and security perceptions and 58 papers on trust perceptions. While IoT was found a useful keyword in combination with security and privacy perceptions, it led to the exclusion of a majority of papers in combination with trust perceptions. Here, trust in the context of IoT is mostly concerned with zero-trust policies between devices rather than the assessment of trust perceptions with a human counterpart in the interaction with a CAI system. Despite previous research in other domains on the interplay of privacy and trust, surprisingly, we identified only \changed{three} papers that were found by both literature searches \citep{seymourIgnoranceBlissEffect, lappemanTrustDigitalPrivacy2023a, pal2022perceptions}. We will discuss this finding in more detail in upcoming sections.

\subsection{Time Trend}

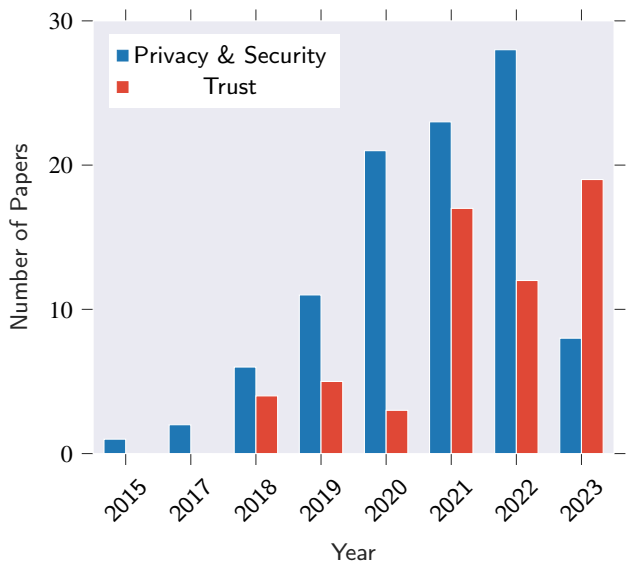
\begin{figure}
    \centering    
\begin{tikzpicture}

\definecolor{darkslategray38}{RGB}{38,38,38}
\definecolor{lavender234234242}{RGB}{234,234,242}
\definecolor{steelblue31119180}{RGB}{31,119,180}
\definecolor{red}{RGB}{224, 72, 54 }

\begin{axis}[
axis background/.style={fill=lavender234234242},
axis line style={white},
tick align=outside,
x grid style={white},
xlabel=\textcolor{darkslategray38}{Year},
xmajorticks=true,
xmin=-0.5, xmax=11.5,
xtick style={color=darkslategray38},
xtick={0.25,1.75,3.25,4.75,6.25,7.75,9.25, 10.75},
xticklabel style={rotate=45.0},
xticklabels={2015,2017,2018,2019,2020,2021,2022,2023},
y grid style={white},
ylabel=\textcolor{darkslategray38}{Number of Papers},
ymajorticks=true,
ymin=0, ymax=30,
ytick style={color=darkslategray38},
legend entries={
    Privacy \& Security,
    Trust
},
legend pos=north west,
legend style={draw=none}
]

\addlegendimage{only marks, mark=square*,color=steelblue31119180}
\addlegendimage{only marks, mark=square*,color=red}

\draw[draw=white,fill=steelblue31119180] (axis cs:-0.25,0) rectangle (axis cs:0.25,1);
\draw[draw=white,fill=steelblue31119180] (axis cs:1.25,0) rectangle (axis cs:1.75,2);
\draw[draw=white,fill=steelblue31119180] (axis cs:2.75,0) rectangle (axis cs:3.25,6);
\draw[draw=white,fill=steelblue31119180] (axis cs:4.25,0) rectangle (axis cs:4.75,11);
\draw[draw=white,fill=steelblue31119180] (axis cs:5.75,0) rectangle (axis cs:6.25,21);
\draw[draw=white,fill=steelblue31119180] (axis cs:7.25,0) rectangle (axis cs:7.75,23);
\draw[draw=white,fill=steelblue31119180] (axis cs:8.75,0) rectangle (axis cs:9.25,28);
\draw[draw=white,fill=steelblue31119180] (axis cs:10.25,0) rectangle (axis cs:10.75,8);

\draw[draw=white,fill=red] (axis cs:0.25,0) rectangle (axis cs:0.75,0);
\draw[draw=white,fill=red] (axis cs:1.75,0) rectangle (axis cs:2.25,0);
\draw[draw=white,fill=red] (axis cs:3.25,0) rectangle (axis cs:3.75,4);
\draw[draw=white,fill=red] (axis cs:4.75,0) rectangle (axis cs:5.25,5);
\draw[draw=white,fill=red] (axis cs:6.25,0) rectangle (axis cs:6.75,3);
\draw[draw=white,fill=red] (axis cs:7.75,0) rectangle (axis cs:8.25,17);
\draw[draw=white,fill=red] (axis cs:9.25,0) rectangle (axis cs:9.75,12);
\draw[draw=white,fill=red] (axis cs:10.75,0) rectangle (axis cs:11.25,19);

\end{axis}

\end{tikzpicture}
    \caption{Number of papers over time with the most recent paper from June 2023}
    \label{fig:papernumber}
\end{figure}

Figure~\ref{fig:papernumber} shows the yearly trend of selected papers for privacy, security and trust. Although we did not restrict our search to a specific timeline, the first papers in our final selection were found in the year \changed{2015} for literature on privacy and security and in the year 2018 for trust. The appearance, as well as the time trend, is in line with previous literature research on privacy in smart speakers~\citep{maccarioPrivacySmartSpeakers2023} and on home networking environments~\citep{pattnaikSurveyUserPerspectives2023c}. Although Apple's Siri was already released in 2011 followed by Amazon's Alexa in 2014, media coverage on privacy concerns around these devices experienced a strong increase with Amazon handing over recordings to investigators~\footnote{\url{https://www.bbc.com/news/technology-39191056}} and increasing awareness about workers listening in on customer's conversations~\footnote{\url{https://www.bloomberg.com/news/articles/2019-04-10/is-anyone-listening-to-you-on-alexa-a-global-team-reviews-audio}}. 
We chose June 2023 as the cut-off for our systematic search. The actual number of papers published on both topics is expected to be higher for all of 2023 which supports the increasing trend of research in this field. The most recent paper was added by snowballing in November 2023.

\subsection{Geographical Distribution}

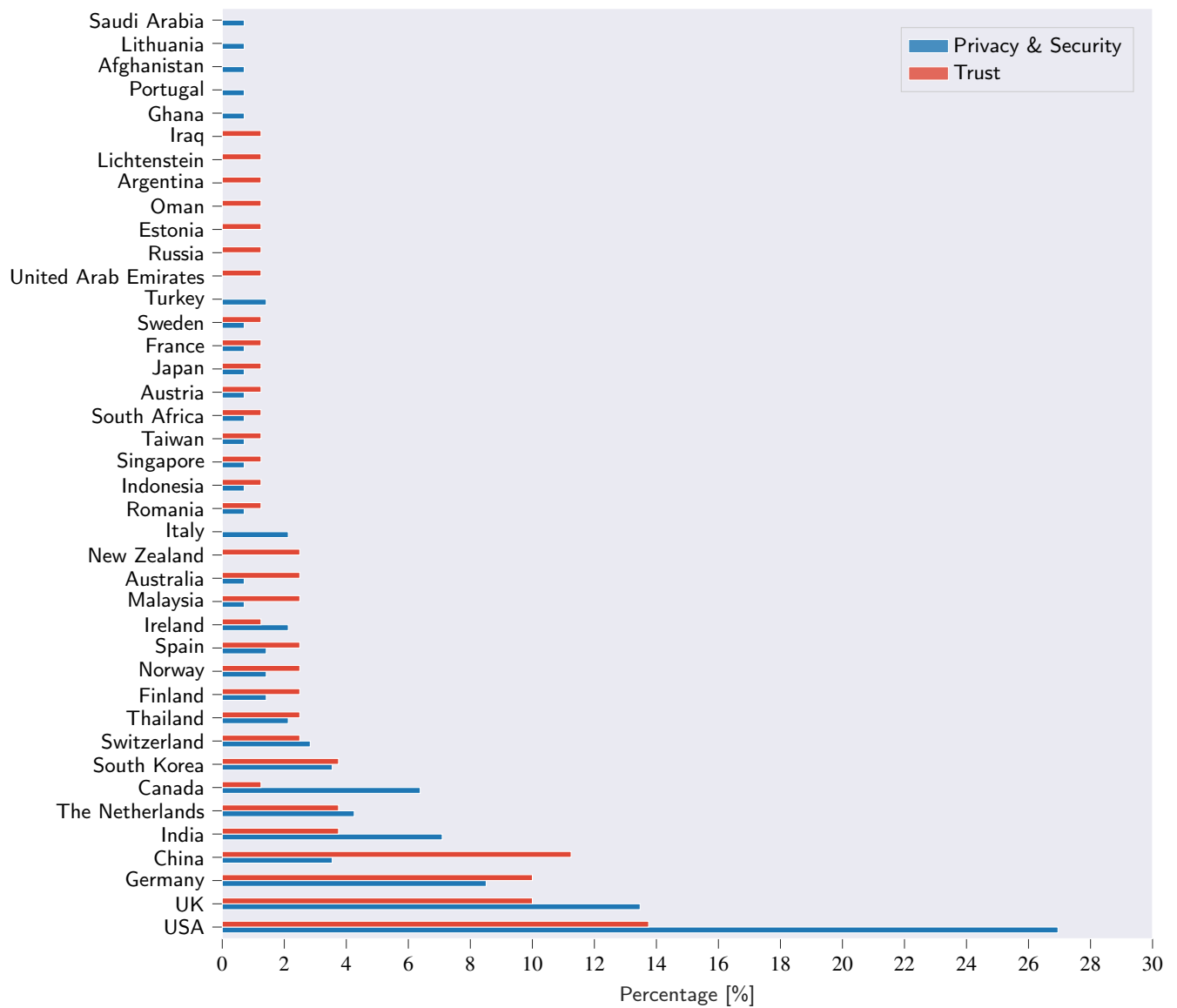
\begin{figure*}[ht]
\centering 
\begin{tikzpicture}

\definecolor{red}{RGB}{255,127,14}
\definecolor{darkslategray38}{RGB}{38,38,38}
\definecolor{lavender234234242}{RGB}{234,234,242}
\definecolor{lightgray204}{RGB}{204,204,204}
\definecolor{steelblue31119180}{RGB}{31,119,180}
\definecolor{red}{RGB}{224, 72, 54 }

\begin{axis}[
height=0.9\textwidth,
width=0.9\textwidth,
axis background/.style={fill=lavender234234242},
axis line style={white},
legend cell align={left},
legend style={
  fill opacity=0.8,
  draw opacity=1,
  text opacity=1,
  draw=lightgray204,
  fill=lavender234234242
},
tick align=outside,
ytick pos=left,
xtick pos=bottom,
x grid style={white},
xlabel=\textcolor{darkslategray38}{Percentage [\%]},
xmajorticks=true,
xmin=0, xmax=30,
xtick style={color=darkslategray38},
y grid style={white},
ymajorticks=true,
ymin=-0.5, ymax=39.5,
ytick style={color=darkslategray38},
ytick={0,1,2,3,4,5,6,7,8,9,10,11,12,13,14,15,16,17,18,19,20,21,22,23,24,25,26,27,28,29,30,31,32,33,34,35,36,37,38,39},
yticklabels={
  USA,
  UK,
  Germany,
  China,
  India,
  The Netherlands,
  Canada,
  South Korea,
  Switzerland,
  Thailand,
  Finland,
  Norway,
  Spain,
  Ireland,
  Malaysia,
  Australia,
  New Zealand,
  Italy,
  Romania,
  Indonesia,
  Singapore,
  Taiwan,
  South Africa,
  Austria,
  Japan,
  France,
  Sweden,
  Turkey,
  United Arab Emirates,
  Russia,
  Estonia,
  Oman,
  Argentina,
  Lichtenstein,
  Iraq,
  Ghana,
  Portugal,
  Afghanistan,
  Lithuania,
  Saudi Arabia
}
]
\draw[draw=white,fill=steelblue31119180] (axis cs:0,-0.25) rectangle (axis cs:26.9503546099291,0);
\addlegendimage{ybar,area legend,draw=white,fill=steelblue31119180}
\addlegendentry{Privacy \& Security}

\draw[draw=white,fill=steelblue31119180] (axis cs:0,0.75) rectangle (axis cs:13.4751773049645,1);
\draw[draw=white,fill=steelblue31119180] (axis cs:0,1.75) rectangle (axis cs:8.51063829787234,2);
\draw[draw=white,fill=steelblue31119180] (axis cs:0,2.75) rectangle (axis cs:3.54609929078014,3);
\draw[draw=white,fill=steelblue31119180] (axis cs:0,3.75) rectangle (axis cs:7.09219858156028,4);
\draw[draw=white,fill=steelblue31119180] (axis cs:0,4.75) rectangle (axis cs:4.25531914893617,5);
\draw[draw=white,fill=steelblue31119180] (axis cs:0,5.75) rectangle (axis cs:6.38297872340426,6);
\draw[draw=white,fill=steelblue31119180] (axis cs:0,6.75) rectangle (axis cs:3.54609929078014,7);
\draw[draw=white,fill=steelblue31119180] (axis cs:0,7.75) rectangle (axis cs:2.83687943262411,8);
\draw[draw=white,fill=steelblue31119180] (axis cs:0,8.75) rectangle (axis cs:2.12765957446809,9);
\draw[draw=white,fill=steelblue31119180] (axis cs:0,9.75) rectangle (axis cs:1.41843971631206,10);
\draw[draw=white,fill=steelblue31119180] (axis cs:0,10.75) rectangle (axis cs:1.41843971631206,11);
\draw[draw=white,fill=steelblue31119180] (axis cs:0,11.75) rectangle (axis cs:1.41843971631206,12);
\draw[draw=white,fill=steelblue31119180] (axis cs:0,12.75) rectangle (axis cs:2.12765957446809,13);
\draw[draw=white,fill=steelblue31119180] (axis cs:0,13.75) rectangle (axis cs:0.709219858156028,14);
\draw[draw=white,fill=steelblue31119180] (axis cs:0,14.75) rectangle (axis cs:0.709219858156028,15);
\draw[draw=white,fill=steelblue31119180] (axis cs:0,15.75) rectangle (axis cs:0,16);
\draw[draw=white,fill=steelblue31119180] (axis cs:0,16.75) rectangle (axis cs:2.12765957446809,17);
\draw[draw=white,fill=steelblue31119180] (axis cs:0,17.75) rectangle (axis cs:0.709219858156028,18);
\draw[draw=white,fill=steelblue31119180] (axis cs:0,18.75) rectangle (axis cs:0.709219858156028,19);
\draw[draw=white,fill=steelblue31119180] (axis cs:0,19.75) rectangle (axis cs:0.709219858156028,20);
\draw[draw=white,fill=steelblue31119180] (axis cs:0,20.75) rectangle (axis cs:0.709219858156028,21);
\draw[draw=white,fill=steelblue31119180] (axis cs:0,21.75) rectangle (axis cs:0.709219858156028,22);
\draw[draw=white,fill=steelblue31119180] (axis cs:0,22.75) rectangle (axis cs:0.709219858156028,23);
\draw[draw=white,fill=steelblue31119180] (axis cs:0,23.75) rectangle (axis cs:0.709219858156028,24);
\draw[draw=white,fill=steelblue31119180] (axis cs:0,24.75) rectangle (axis cs:0.709219858156028,25);
\draw[draw=white,fill=steelblue31119180] (axis cs:0,25.75) rectangle (axis cs:0.709219858156028,26);
\draw[draw=white,fill=steelblue31119180] (axis cs:0,26.75) rectangle (axis cs:1.41843971631206,27);
\draw[draw=white,fill=steelblue31119180] (axis cs:0,27.75) rectangle (axis cs:0,28);
\draw[draw=white,fill=steelblue31119180] (axis cs:0,28.75) rectangle (axis cs:0,29);
\draw[draw=white,fill=steelblue31119180] (axis cs:0,29.75) rectangle (axis cs:0,30);
\draw[draw=white,fill=steelblue31119180] (axis cs:0,30.75) rectangle (axis cs:0,31);
\draw[draw=white,fill=steelblue31119180] (axis cs:0,31.75) rectangle (axis cs:0,32);
\draw[draw=white,fill=steelblue31119180] (axis cs:0,32.75) rectangle (axis cs:0,33);
\draw[draw=white,fill=steelblue31119180] (axis cs:0,33.75) rectangle (axis cs:0,34);
\draw[draw=white,fill=steelblue31119180] (axis cs:0,34.75) rectangle (axis cs:0.709219858156028,35);
\draw[draw=white,fill=steelblue31119180] (axis cs:0,35.75) rectangle (axis cs:0.709219858156028,36);
\draw[draw=white,fill=steelblue31119180] (axis cs:0,36.75) rectangle (axis cs:0.709219858156028,37);
\draw[draw=white,fill=steelblue31119180] (axis cs:0,37.75) rectangle (axis cs:0.709219858156028,38);
\draw[draw=white,fill=steelblue31119180] (axis cs:0,38.75) rectangle (axis cs:0.709219858156028,39);
\draw[draw=white,fill=red] (axis cs:0,0) rectangle (axis cs:13.75,0.25);
\addlegendimage{ybar,area legend,draw=white,fill=red}
\addlegendentry{Trust}

\draw[draw=white,fill=red] (axis cs:0,1) rectangle (axis cs:10,1.25);
\draw[draw=white,fill=red] (axis cs:0,2) rectangle (axis cs:10,2.25);
\draw[draw=white,fill=red] (axis cs:0,3) rectangle (axis cs:11.25,3.25);
\draw[draw=white,fill=red] (axis cs:0,4) rectangle (axis cs:3.75,4.25);
\draw[draw=white,fill=red] (axis cs:0,5) rectangle (axis cs:3.75,5.25);
\draw[draw=white,fill=red] (axis cs:0,6) rectangle (axis cs:1.25,6.25);
\draw[draw=white,fill=red] (axis cs:0,7) rectangle (axis cs:3.75,7.25);
\draw[draw=white,fill=red] (axis cs:0,8) rectangle (axis cs:2.5,8.25);
\draw[draw=white,fill=red] (axis cs:0,9) rectangle (axis cs:2.5,9.25);
\draw[draw=white,fill=red] (axis cs:0,10) rectangle (axis cs:2.5,10.25);
\draw[draw=white,fill=red] (axis cs:0,11) rectangle (axis cs:2.5,11.25);
\draw[draw=white,fill=red] (axis cs:0,12) rectangle (axis cs:2.5,12.25);
\draw[draw=white,fill=red] (axis cs:0,13) rectangle (axis cs:1.25,13.25);
\draw[draw=white,fill=red] (axis cs:0,14) rectangle (axis cs:2.5,14.25);
\draw[draw=white,fill=red] (axis cs:0,15) rectangle (axis cs:2.5,15.25);
\draw[draw=white,fill=red] (axis cs:0,16) rectangle (axis cs:2.5,16.25);
\draw[draw=white,fill=red] (axis cs:0,17) rectangle (axis cs:0,17.25);
\draw[draw=white,fill=red] (axis cs:0,18) rectangle (axis cs:1.25,18.25);
\draw[draw=white,fill=red] (axis cs:0,19) rectangle (axis cs:1.25,19.25);
\draw[draw=white,fill=red] (axis cs:0,20) rectangle (axis cs:1.25,20.25);
\draw[draw=white,fill=red] (axis cs:0,21) rectangle (axis cs:1.25,21.25);
\draw[draw=white,fill=red] (axis cs:0,22) rectangle (axis cs:1.25,22.25);
\draw[draw=white,fill=red] (axis cs:0,23) rectangle (axis cs:1.25,23.25);
\draw[draw=white,fill=red] (axis cs:0,24) rectangle (axis cs:1.25,24.25);
\draw[draw=white,fill=red] (axis cs:0,25) rectangle (axis cs:1.25,25.25);
\draw[draw=white,fill=red] (axis cs:0,26) rectangle (axis cs:1.25,26.25);
\draw[draw=white,fill=red] (axis cs:0,27) rectangle (axis cs:0,27.25);
\draw[draw=white,fill=red] (axis cs:0,28) rectangle (axis cs:1.25,28.25);
\draw[draw=white,fill=red] (axis cs:0,29) rectangle (axis cs:1.25,29.25);
\draw[draw=white,fill=red] (axis cs:0,30) rectangle (axis cs:1.25,30.25);
\draw[draw=white,fill=red] (axis cs:0,31) rectangle (axis cs:1.25,31.25);
\draw[draw=white,fill=red] (axis cs:0,32) rectangle (axis cs:1.25,32.25);
\draw[draw=white,fill=red] (axis cs:0,33) rectangle (axis cs:1.25,33.25);
\draw[draw=white,fill=red] (axis cs:0,34) rectangle (axis cs:1.25,34.25);
\draw[draw=white,fill=red] (axis cs:0,35) rectangle (axis cs:0,35.25);
\draw[draw=white,fill=red] (axis cs:0,36) rectangle (axis cs:0,36.25);
\draw[draw=white,fill=red] (axis cs:0,37) rectangle (axis cs:0,37.25);
\draw[draw=white,fill=red] (axis cs:0,38) rectangle (axis cs:0,38.25);
\draw[draw=white,fill=red] (axis cs:0,39) rectangle (axis cs:0,39.25);
\end{axis}

\end{tikzpicture}
\caption{Author's affiliation by country for privacy, security and trust perception papers \changed{in percentage. Due to the different number of papers for Privacy \& Security and Trust, papers can have multiple authors from varying countries.}}
\label{fig:countryAffiliation}
\end{figure*}

Figure~\ref{fig:countryAffiliation} shows the author's affiliation by country for each of the constructs. It is not surprising that the majority of authors are affiliated with a US institution (27\%) followed by the UK (13\%) and Germany (\changed{9\%}) for privacy \& security. Similar results were shown by \cite{maccarioPrivacySmartSpeakers2023} when surveying privacy research for smart speakers. The results show that the headquarters of large CAI manufacturers, such as Amazon, Google or Apple are not directly linked to the author's affiliation by country.

A similar phenomenon can be observed for trust in CAI where the US leads with 13.75\%, China (11.25\%) and Germany (10\%). We find that there is a significant difference between countries researching privacy and trust in terms of the number of publications and global diversification. In particular, the numbers suggest that the political system and possibly associated status of privacy influence whether research is conducted on privacy or trust perceptions. Another reason could be that trust is a more generic term contrary to privacy and security and its dominance across several research fields influences publication counts per country. Moreover, the concept of trust positively impacts technology adoption~\citep{choung2023trust}, while privacy and security are often seen as inhibitors to adoption and innovation which could be another reason for researchers to focus on one or the other.
It needs to be mentioned that the research communities vary greatly on a per-country basis, thus the number of reported papers is not representative of the overall research output from a particular country. \changed{Additionally, we are aware that reporting the overall research output per country does not consider the research capacity each country has based on the number of researchers in each country. Thus, we can not draw any conclusions on how relevant "privacy \& security" and "trust" research in each country is.} However, it allows a comparison between the research communities of "privacy \& security" and "trust" perceptions.

\subsection{Devices and Modalities}

In our literature review, we focus on Conversational AI systems -- systems that leverage natural language processing capabilities to converse with humans. Thereby, we include text-based as well as voice-based systems. While we aimed to cover the most prevalent terms for CAI systems in our keyword search, a detailed analysis of the devices researched by the identified papers shows the diversity of terms used to describe CAI systems. Figure~\ref{fig:wordcloudprivacy} shows a word cloud of researched devices as described by the authors of the paper in their titles or abstracts. In combination with privacy and security perceptions, the word ``smart speakers" was used most (\changed{15}\%) when describing researched devices, followed by ``voice assistants" (\changed{10}\%) and ``IoT devices" (\changed{6}\%) \changed{and ``Intelligent Personal Assistants" (6\%)}. In particular, we find that all studies on ``IoT devices" include smart speakers as researched devices in addition to other IoT devices~\citep{alghamdiUserStudyEvaluate2022a, al-ameenLookUserPrivacy2021a, ahmadTangiblePrivacyUserCentric2020b}. Thus, increasing the number of studies that research ``smart speakers" to 21\%. In the context of trust perceptions, the word "assistant" is most frequently used with 41\%, either in combination with "virtual"(48\%) or "voice"(52\%). This is followed by "chatbots" as research devices with 22\% and then by "Alexa"as a research device (12\%).

We further categorize the researched devices depending on their modality. We find that a majority of privacy and security-related papers focus on voice as a modality (\changed{76}\%), while \changed{16}\% focus on text as the main modality. Thereby, all of the studies researching privacy perceptions in text-based CAI systems refer to their systems as ``bot" or ``chatbot"~\citep{patilCanWeTrust2022a,vandergootExploringAgeDifferences2020, chengHowAIdrivenChatbots2020}. Yet, some acknowledge that chatbot can be used as a more general term including text as well as voice-based conversations. The remaining papers do not explicitly specify the modality or refer to their researched devices more generally as ``virtual assistants" or ``agents". Similarly, in trust research 50\% of papers conduct research on voice-based systems. 36\% focus on primarily text-based systems, which are exclusively called chatbots just as in privacy research. Additionally, in trust papers, we identified 14\% of papers as virtual reality-related. Since all papers on virtual reality use voice as a mode, the total percentage of voice-based systems increases to 64 \%.
 
\subsection{Application Fields}

Table~\ref{tab:ApplicationFields} shows the application fields targeted by the reviewed studies. To identify application fields, we relied on keywords, titles and abstracts. We refer to ``General" whenever an application field is not explicitly mentioned. For privacy \& security perceptions, \changed{58}\% did not target a specific application field. A similar trend can be observed in the context of trust perceptions. Nevertheless, additional application fields such as virtual reality (VR) or space flight were identified. This finding leaves room for future research to explore privacy, security and trust perceptions of CAI systems targeted towards under-researched applications. In particular, it might be valuable to explore fields that have been researched in the context of trust perceptions but not privacy and security perceptions and vice versa to better understand their interplay and contextual dependencies. Moreover, only a few papers target more than one application field or compare privacy, security and trust perceptions among them. For example,~\cite{bruggemeierPerceptionsReactionsConversational2022b} investigate conversational privacy across three different application areas, i.e. banking, tax advice and music control.

\begin{table}[ht]
  \caption{Application Fields for privacy \& security and trust perception in the context of CAI. \changed{Application fields were identified via keywords, titles and abstracts.}}
  \label{tab:ApplicationFields}
  \scriptsize
  \begin{tabular}{p{0.3\columnwidth}p{0.25\columnwidth}p{0.25\columnwidth}}
    \toprule
    Application Field & Privacy \& Security - Frequency (Percentage) & Trust - Frequency (Percentage)\\
    \midrule 
    General & \noindent\changed{58 (58\%)} & 37 (64\%) \\
    Smart Home & \noindent\changed{15 (15\%)} & 4 (7\%) \\
    Healthcare & \noindent\changed{6 (6\%)} & 2 (3\%) \\
    Retail & \noindent\changed{5 (5\%)} & 5 (9\%)\\
    Finance & \noindent\changed{5 (5\%)} & 3 (5\%) \\
    Customer Service & \noindent\changed{5 (5\%)} & 2 (3\%) \\
    Education & \noindent\changed{3 (3\%)} & 2 (3\%) \\
    Workplace & \noindent\changed{1 (1\%)} & - \\
    Insurance & \noindent\changed{1 (1\%)} & - \\
    \noindent\changed{Hospitality} & \noindent\changed{1 (1\%)} & - \\
    VR & - & 3 (5\%) \\
    Space Flight & - & 2 (3\%) \\
    Self Driving & - & 1 (2\%) \\
    \bottomrule
\end{tabular}
\end{table} 

\subsection{Research Methods}
\label{sec:researchMethods}

\begin{table}[ht]
  \caption{Research Methods in our sample}
  \label{tab:ResearchMethods}
  \scriptsize
  
  \begin{tabular}{p{0.3\columnwidth}p{0.25\columnwidth}p{0.25\columnwidth}}
    \toprule
    Research Method & Privacy \& Security - Frequency (Percentage) & Trust - Frequency (Percentage)\\
    \midrule 
    Quantitative & \noindent\changed{60 (60\%)} & 47 (81\%) \\
    \midrule
    Survey & \noindent\changed{45 (45\%)} & 22 (38\%) \\
    Experiment and Survey & \noindent\changed{12 (12\%)} & 11 (19\%) \\
    Vignette Study & 3 (4\%) & 14 (24\%) \\  
    Trust Game & - & 2 (3\%) \\
    \midrule
    Mixed & \noindent\changed{19 (19\%)} & 6 (10\%) \\
    \midrule 
    Qualitative & \noindent\changed{21 (21\%)} & 5 (9\%) \\
    \bottomrule
\end{tabular}
\end{table} 

One of the goals of our literature review was to identify common research methods used to study privacy, security and trust perceptions in CAI. We show an overview of research methods used by the analysed papers in Table~\ref{tab:ResearchMethods}. A more detailed overview is provided in the Appendix in Table~\ref{tab:ResearchMethodsExtensive}. To classify the study methods, we relied on previous classification schemes used to review empirical methods in usable privacy and security research~\citep{distlerSystematicLiteratureReview2021}. Therefore, we refer to \textit{experiments} whenever the effect of experimental conditions was measured either through oral examination or written questionnaire form. We further distinguish between \textit{survey studies} that use written questionnaires to interrogate people and \textit{interview studies} with an oral interaction between the researcher and participants. 
We coded qualitative methods that included a survey to identify participants, e.g. smart speaker users, as qualitative methods only because the surveys served only the purpose of identification and were usually not analysed further. Finally, we explicitly coded less common study methods and combinations of those, e.g. focus groups, co-design methods or trust games. Trust games originate, among others, from the investment game introduced by~\cite{berg1995trustgame} to measure trust in economic settings. In this paradigm, one participant allocates funds to a different location, where a second participant deliberates on whether and to what extent to reciprocate the amount to the first participant. The underlying objective is to directly gauge levels of trust and the willingness to undertake risks for the prospect of a favourable outcome.

We found that in both cases, i.e., privacy \& security and trust perceptions, a majority of analyzed papers used quantitative methods. \changed{While quantitative methods can be used to test hypotheses, qualitative methods can provide rich, detailed data that helps in understanding the complexity of a particular issue and can be used to form hypotheses~\citep{lazarResearchMethodsHumanComputer2017}. Thus, diversification of research methods is generally desirable as both methods complement each other and help to advance the research field.}

Interestingly, we identified a significant difference in the number of papers using qualitative and mixed research methods to assess privacy and trust perceptions. While \changed{21\% and 19\%} respectively applied qualitative or mixed methods to assess privacy and security perceptions, only 10\% of trust papers were qualitative, and only 9\% were mixed methods. \changed{For research on trust perceptions, questionnaires were identified as the most common form of measurement possibly due to the difficulty of measuring trust indirectly from other sources such as bio-signals or other behavior tracking~\citep{bauer2018measuring}.}
In the context of privacy and security perceptions, mixed study methods including multiple different methods are common. For example, \cite{kowalczukConsumerAcceptanceSmart2018a} analysed customer reviews and Twitter data to identify factors influencing smart speaker usage intentions before evaluating the acceptance model through a user study. In another study,~\cite{malkinPrivacyAttitudesSmart2019} applied experience sampling methodology to expose participants to their past smart speaker interactions followed by privacy perception and preference assessment. In contrast, research on trust perceptions applied less diverse study methods in mixed-method settings and relied predominantly on \textit{mixed surveys} that contained additional open-ended questions in the survey which were investigated qualitatively~\citep{nordheim2019initialtrust}. 

\changed{Our sample showed a predominance of survey methods which is in line with~\cite{distlerSystematicLiteratureReview2021}. While their review focused on usable privacy and security research, they showed that research on privacy and security perceptions, attitudes and behavior was dominated by descriptive studies. These studies were also less likely to include prototypes and relied heavily on self-reports. We agree with~\cite{distlerSystematicLiteratureReview2021} that methods that have not been deployed frequently could not only help advance the usable privacy and security research field but help to gather more diverse insights on privacy, security and trust perceptions. For example, long-term perceptions of privacy and trust can be assessed via diary studies -- a research method that has been rarely used in our reviewed sample. Moreover, triangulation of various methods can deepen insights and help to understand results~\citep{pettersson2018}, for instance, by combining interviews and analysis of log files~\citep{ammariMusicSearchIoT2019a}. Here, research on trust perceptions could rely on mixed-method studies already applied in the field of privacy and security. Finally, the usage of prototypes can also enhance studies on privacy and trust perceptions by making scenario-based situations more concrete~\citep{distlerSystematicLiteratureReview2021}. We will discuss tools in more depth in Section~\ref{sec:Tools}.}

\subsection{Recruitment Methods}

\begin{table}
  \caption{Recruitment Methods for participants used in the analyzed papers. Papers can use multiple recruitment methods.}
  \label{tab:recruitmentParticipants}
  \scriptsize
  \begin{tabular}{p{0.3\columnwidth}p{0.25\columnwidth}p{0.25\columnwidth}}
    \toprule
    Recruitment Method & Privacy \& Security - Frequency (Percentage) & Trust - Frequency (Percentage) \\
    \midrule 
    Crowdsourcing & \noindent\changed{38 (29\%)} & 23 (32\%) \\
    Internet and Social Media & \noindent\changed{27 (21\%)}& 9 (14\%) \\
    University & \noindent\changed{23 (17\%)} & 8 (11\%)\\
    not reported & \noindent\changed{16 (12\%)} & 23 (32\%) \\
    Personal Network & \noindent\changed{14 (11\%)} & 7 (10\%) \\ 
    Special Facilities & \noindent\changed{9 (7\%)} & - \\
    Paper Advertisement & \noindent\changed{4 (3\%)} & 1 (1\%) \\
    \bottomrule
\end{tabular}
\end{table} 

We further analyzed recruitment strategies addressed by our sampled papers to enhance our understanding of target groups as shown in Table~\ref{tab:recruitmentParticipants}. Thereby, we listed all used recruitment methods for each study as it was not uncommon that multiple methods were applied.
As privacy, security and trust perceptions are highly subjective, recruitment strategies are crucial to identify to what extent study results are generalizable. As an easily accessible recruitment method, it is not surprising that crowdsourcing has been applied most frequently using platforms such as Amazon Mechanical Turk, Qualtrics or marketing research agencies. Social media and online platforms, including blogs and social networks, also featured prominently, while traditional methods such as paper advertisements were notably infrequent and often only used in the context of recruitment of older adults. Interestingly, some of the analyzed papers investigating privacy and security perceptions relied on special facilities to recruit specific target groups such as patients or seniors while analysed papers on trust perceptions did not. 

Crowdsourcing predominated in quantitative studies, while qualitative and mixed-method studies leaned towards personal networks, paper advertisements, and university channels. This was also confirmed by the difference in average participant numbers, with quantitative studies totalling an average of 
\changed{427} and
273 participants, in privacy \& security and trust research respectively compared to the \changed{26} to 29 participants in qualitative studies. It is noteworthy, that the lowest quantitative study found in the context of trust perceptions has 13 participants which is lower than the average number of participants in the qualitative studies~\citep{gupta2019ai}. 

\subsection{Participants}

\begin{table}
  \caption{Type of participants researched in the analyzed papers. Papers can use multiple different types. \changed{Participants can be part of more than one category.}}
  \label{tab:TypeParticipants}
  \scriptsize
  \begin{tabular}{p{0.3\columnwidth}p{0.25\columnwidth}p{0.25\columnwidth}}
    \toprule
    Type of Participants & Privacy \& Security - Frequency (Percentage) & Trust - Frequency (Percentage) \\
    \midrule 
    Technology Users & \noindent\changed{48 (32\%)} & 2 (3\%)\\
    Crowd workers & \noindent\changed{37 (26\%)} & 31 (50\%)\\
    Students & \noindent\changed{17 (12\%)} & 18 (29\%) \\
    Special User Group & \noindent\changed{15 (10\%)} & -\\
    not reported & \noindent\changed{11 (7\%)} & 7 (11\%) \\
    Experts & 9 (6\%) & 4 (7\%)  \\
    Non-Users & \noindent\changed{6 (4\%)} & - \\
    Employees & \noindent\changed{5 (3\%)} & -  \\
    \bottomrule
\end{tabular}
\end{table} 

As shown in Table~\ref{tab:TypeParticipants}, we grouped participants into more specific groups. \changed{It needs to be noted that participants can be part of more than one category, e.g., they might have been students or crowd workers as well as technology users.} We find that a majority of papers evaluating privacy and security perceptions specifically recruited technology users for their studies while research on trust perceptions did not. \changed{Frequently, crowd workers were screened for their technology usage before participation in the study.} For trust, a more general type of participants is desired as can be seen by the majority of participants in trust research being crowd workers. Overall, students form another significant research group and are generally a result of convenience sampling. Special user groups encompass patients, older adults or visitors of smart homes while the group of experts consists of developers, medical professionals as well as teachers. Finally, a few studies in the domain of privacy and security perceptions, recruited employees including university staff, forming a convenience sample~\citep{Mols2021, liaoUnderstandingRolePrivacy2019b, hornungAIInvadingWorkplace2022b, fahnTimeGetConversational2021a}.

\section{Measurements of Privacy, Security and Trust Perceptions}

Next to comparing meta-information of the studies, like author affiliations, participants, researched devices and application fields in privacy, security and trust research, we investigate the developed and used privacy, security and trust perception measurements in the reviewed papers.

\subsection{Qualitative Privacy, Security and Trust Assessment}

The assessment of qualitative studies was conducted by grouping their findings. Within the qualitative trust literature, there were no papers that defined response variables, or trust-related questions used in the interviews. Instead, relevant items were identified after the interviews were conducted with audio recordings, transcriptions or field notes among others. Across all papers that used a qualitative approach, we could not identify any overlapping theme or item used across papers but the perception of giving up privacy when using voice assistants~\citep{perez2018buildingQualitative, fakhimi2023effectsQualitative}. \cite{perez2018buildingQualitative} also note a questionnaire with 500 hundred individuals. However, there is neither a statistical analysis of the questionnaire nor discussion of individual items.

Further, the benefits of interactions, in terms of utilitarian, hedonic and social aspects, were included in~\cite{fakhimi2023effectsQualitative} which emphasize the more social influence of trust in interactions. Similarly, the effect of anthropomorphised voice assistants was subject to the interviews of this paper. Next to anthropomorphised analysis stands a design approach that was taken in~\cite{durall2023designing}. The differences between research approaches make the comparison and analysis of qualitative papers challenging.

Due to the large differences in researched items within the qualitative studies, as well as in comparison to the quantitative approach, we did not include the qualitative findings further into the literature review of this study.

\subsection{Quantitative Privacy, Security and Trust Assessment}

\changed{Next, we aim to understand how privacy, security and trust perceptions are quantitatively measured in the context of CAI and which constructs are assessed}. 

\subsubsection{Privacy, Security and Trust Scales and Constructs}

We analyse the privacy, security and trust scales used to assess peoples' perceptions in quantitative and mixed-method studies. Table~\ref{tab:OverviewConstructs} provides an overview of constructs assessed by studies evaluating privacy, security and trust perceptions. 
We extracted response variables from the detailed questionnaires that were published by a majority of papers. It is noteworthy that in general more than 80\% of papers reported on the detailed items and questions assessed in the surveys. 
Nevertheless, with the remaining papers not reporting on used scales, there is still room for improving transparency and reproducibility for the privacy and trust research community. We extracted response variables from the published research models, survey and results sections for this subset. Moreover, we found that detailed scales were often referenced from the source neglecting adaptations made by the authors and hindering reuse and replicability of scales. We can see 56 of all papers referencing trust as a construct in their work which is followed by \changed{40} papers referencing privacy concerns directly. Further, we distinguished between trust and trust \& technology constructs to highlight the difference between scales used in the CAI context. As CAI systems are technical systems, it is concerning that only a minority of papers use trust scales developed for the technological context.

\begin{table}[ht]
  \caption{Overview of privacy, security and trust constructs used in mixed-method and quantitative studies. Studies can assess multiple constructs. A detailed overview is provided in Appendix Table~\ref{tab:TypePrivacyConstructs} and Table~\ref{tab:TypeTrustConstructs}.}
  \label{tab:OverviewConstructs}
  \begin{tabular}{lc}
    \toprule
    Construct & Frequency \\
    \midrule 
    Trust & 55 \\
    Privacy Concerns & \noindent\changed{40} \\
    Affect, Humanness and Anthropomorphism & 29 \\
    Risk Perception & \noindent\changed{29} \\
    Character and Personality & 14 \\
    Security and Security Perception & \noindent\changed{11} \\
    Control & \noindent\changed{8} \\
    Trust \& Technology & 11 \\
    Privacy and Privacy Perception & 4 \\
    Interactivity & 3 \\
    \bottomrule
\end{tabular}
\end{table}

\paragraph{Privacy and Privacy Perception}

We find that the reviewed literature explores diverse privacy and privacy perception constructs. As no generally agreed-upon definition of privacy exists, studies must be as explicit and detailed about their privacy \changed{conceptualizations and} measures as possible to allow traceability and reproducibility. For example, in the study by ~\cite{furiniUsageSmartSpeakers2020b} participants are asked about the importance they attribute to privacy and their willingness to use a service capable of surveillance in particular in the context of violating COVID-19 lockdown measures. \changed{However, neither in their questionnaire nor in the paper, do they provide a definition or explanation of privacy}. Conversely,~\cite{kreyVoiceAssistantsHealthcare2022b} focus on unique characteristics of voice assistants to assess privacy by asking hospital patients about their privacy concerns related to the passive-listening capabilities of the device, the necessity of providing information about privacy risks and concerns around the voice assistants disclosing patient's personal information loudly. \changed{Based on these three questions, privacy is narrowly defined within the study context, yet, comparability to established privacy conceptualizations is limited~\citep{smithInformationPrivacyResearch2011}. Similarly,~\cite{patilCanWeTrust2022a} do not define privacy but rely on a previously used privacy scale including questions on disclosure, the importance of regulations and trustworthiness. In contrast, \cite{bruggemeierPerceptionsReactionsConversational2022b} investigate conversational privacy and provide a conceptualization by drawing from previous literature. Thus, they refer to conversational privacy whenever CAI systems ``express privacy-related information in dialogue form"~\citep{bruggemeierPerceptionsReactionsConversational2022b, lauAlexaAreYou2018c, harkous2016pribots}.}

\paragraph{Control}
\changed{Six studies specifically investigate privacy control, perceived control, controllability and privacy boundary control~\citep{alghamdiUserStudyEvaluate2022a, linTransferabilityPrivacyrelatedBehaviours2020a, pal2020personal, purwanto2020interactive, kang2023communication, ahmadTangiblePrivacySmart2022}. \cite{alghamdiUserStudyEvaluate2022a} and \cite{pal2020personal} conceptualize control as ``users' ability to control their personal information". Importantly, they focus on control over existing as well as future generated data~\citep{pal2020personal}.~\cite{linTransferabilityPrivacyrelatedBehaviours2020a} investigated perceptions of control and compared privacy control usage relying on existing privacy controls such as voice authentication or muting and unplugging of devices. Finally, \cite{kang2023communication} draw from Communication Privacy Management (CPM) theory and conceptualize boundary control as ``users' control over the process of withdrawing and controlling personal information". While the aforementioned studies conceptualize control as control over personal information, \citep{purwanto2020interactive} describes controllability as one facet of interactivity and as a general feature that allows users to manipulate the interaction, e.g. content and timing, with a digital assistant.

As studies keep investigating privacy control, it is necessary to state what is being controlled and which dimensions of control are considered. For instance, \citeauthor{Brandimarte2013} argue that one needs to distinguish between control regarding the release of personal information, access to it as well as further usage. While people may be most likely to focus on control regarding the disclosure of personal information, control over access and usage might be more relevant to their privacy. Despite the limited exploration of privacy as control in the context of CAI systems in our sample, the studies that do exist, assess control indeed across various dimensions. 
While two studies focus on control over the collection of personal information and its influence on disclosure,~\citep{linTransferabilityPrivacyrelatedBehaviours2020a} provides a detailed analysis of privacy controls encompassing control over data access and data retention. 

The majority of studies investigating privacy control in our sample refer to a common conceptualization of privacy, i.e., users' ability to control their information~\cite{smithInformationPrivacyResearch2011}. However, scholars have argued that control should not serve as a proxy for privacy as privacy situations can exist without someone perceiving or exercising control~\citep{Laufer1977}. For instance, choosing privacy-preserving technology in the first place reduces the need to control information during usage. Thus, individuals might not perceive control at the time of usage while still perceiving a high level of privacy. In addition, what matters to privacy is the amount of information others have access to and the appropriateness of the flow of information~\citep{nissenbaum2009privacy}. This can be seen as independent of whether an individual controls the information flow or someone else, e.g., data protection authorities monitor the application and if necessary apply corrective measures. 
The above discussion necessitates a detailed and explicit description of constructs and methodological transparency when assessing privacy control.}

\paragraph{Privacy Concerns}
The predominantly assessed construct in the context of privacy is privacy concerns. While most studies assess privacy concerns more generally,~\cite{lutzPrivacySmartSpeakers2021} evaluate seven distinct types of privacy concerns among smart speaker users. Their findings reveal that institutional actors such as third-party companies lead to stronger privacy concerns compared to social actors such as household members. Along the same lines, \cite{pal2022perceptions} investigate the impact of various types of privacy concerns on trust and usage of voice-activated personal assistants. They find that privacy concerns related to the voice assistants and the household have a significant impact on users' emotional and cognitive trust. However, vendor and third-party privacy concerns only significantly affect emotional trust while government privacy concerns have no significant effect. While most reviewed papers investigate general privacy concerns, the previously discussed studies suggest various dimensions of privacy concerns should be considered. In particular, distinctions between horizontal or social and vertical or institutional privacy concerns could be beneficial to enhance understanding of the influence of privacy concerns on trust and in specific contexts, e.g., public domain or multi-user context~\citep{Masur2019, Ayalon2019}. 

\paragraph{Security and Security Perception}

In addition to exploring privacy perceptions, investigations into security and security perceptions in the context of CAI systems have been conducted. It is crucial to recognize that privacy and security are distinct concepts. While security is considered a necessity for privacy, security does not necessarily hinder subsequent usage of personal information, potentially leading to privacy breaches~\citep{Culnan2009, Ackerman2004}. While various definition of security exists, it commonly refers to the protection of personal information concerning integrity, authentication, and confidentiality~\citep{smithInformationPrivacyResearch2011}. Therefore, it comes as no surprise that security perceptions in our identified studies have been assessed in the context of attack susceptibility of voice assistants~\citep{mccarthyShoutingLetterboxesStudy2020b, kreyVoiceAssistantsHealthcare2022b}, or authentication methods~\citep{renzAuthenticationMethodsVoice2023}, and in contexts with heightened sensitivity such as banking~\citep{fahnTimeGetConversational2021a, bruggemeierPerceptionsReactionsConversational2022b}, \changed{shopping~\citep{aw2022alexa}} or device sharing scenarios~\citep{lee2019adoption}. Furthermore, most studies assessed security perceptions alongside privacy perceptions or privacy concerns while acknowledging the differences between these concepts. \changed{For instance,~\cite{buteauHeyAlexaWhy2021} define privacy concerns according to ~\cite{dinevInternetPrivacyConcerns2005a} as ``perceived vulnerability and ability to control ones' information" while they refer to perceived security as ``perceptions regarding reliability of mechanisms of data transmission and storage" according to~\cite{flavianConsumerTrustPerceived2006}.} 
Their results show that privacy concerns negatively influence attitudes towards using voice assistants while security perceptions contribute positively. They particularly highlight participants' distinct understanding of privacy and security as separate concepts. Another study investigated the effect of conversational privacy on both privacy and security perceptions in three distinct contexts, i.e., banking, tax and music~\citep{bruggemeierPerceptionsReactionsConversational2022b}. While they found that conversational privacy can positively impact privacy and security perceptions, no significant differences between privacy and security perceptions were reported. This shows that the type of privacy construct studied alongside security perceptions influences the understanding of the complex relationship between privacy and security perceptions in CAI systems. 

\paragraph{Risk Perception}

As shown in Table~\ref{tab:OverviewConstructs} risk perceptions are frequently mentioned, in total \changed{29} times. Notably, the majority of papers \changed{assess risk perception in the context of structural models that aim to evaluate factors impacting CAI adoption, usage and more. Many of these studies build upon the Technology Adoption Model (TAM) or related conceptual models and extend them by adding factors like perceived privacy risk, perceived security risk, or safety risk. For instance,~\cite{kowalczukConsumerAcceptanceSmart2018a} evaluated security and privacy risks as a combined construct conceptualizing them as consumers' fears of personal information leakage or hacking. Similarly,~\citep{hanUnderstandingAdoptionIntelligent2018a} jointly assess security and privacy risks but using only a subsection of an established scale. We will discuss extensions of technology adoption models that include risk perception in more detail in Section~\ref{sec:Adoption}.} 

In our sample, only \changed{few} papers independently discussed risks.~\cite{nordheim2019initialtrust} investigated factors including risks that influence trust in customer service chatbots. Based on previous literature, they conceptualize risk as ``Users' perceptions regarding the likelihood of an undesirable outcome". Their findings show that perceived risk significantly influences users' trust in chatbots. In another study, focused on privacy and security labels for smart devices, including smart speakers,~\cite{emami-naeiniWhichPrivacySecurity2021b} assessed participants' risk perception of label attribute-value pairs, e.g. security update - automatic. They found that the risk perception was influenced by the displayed label pairs, effectively conveying information. \changed{\cite{song2022should} conduct a scenario-based experiment investigating the self-recovery approaches of chatbots on consumer satisfaction while considering perceived privacy risks. They identified a moderating role of robot intelligence on perceived privacy risks in chatbot self-recovery.}

The variation in risk perceptions observed in our sample mirrors the fragmented notion of privacy perceptions. \changed{Moreover, none of the identified studies distinguishes dimensions of risks nor risks, consequences and sources~\cite{Lim2003}. Given the various application areas of CAI, future research should investigate more fine-grained risk perceptions.} Additionally, on a conceptual level privacy perceptions, privacy concerns and privacy risks are distinctively different from privacy risks relating to an individual's perception of risks in a specific state. In contrast, privacy concerns should generally be treated as a trait-like characteristic~\citep{smithInformationPrivacyResearch2011}. Therefore, when evaluating systems or designs concerning privacy, it is advisable to assess perceived privacy risks, while privacy concerns are neglectable.

\paragraph{Trust}
Trust was the most assessed construct in the reviewed literature, with a total of 56 papers. However, it needs to be pointed out that using the overall term ``trust" can be misleading since the definition of trust largely differs on a per-paper basis. Some just define trust as how safe a mobile payment is and how little it will hurt privacy~\citep{jameel2022perceived} while others include brand trust and emotional components as in~\cite{lappemanTrustDigitalPrivacy2023a}. Others define trust as ``the belief that an entity will
act cooperatively to fulfil clients’ expectations without exploiting their vulnerabilities"~\citep{seymourIgnoranceBlissEffect}. Further, trust is broken down into sub-constructs, such as competence, which is already built-in in the model of~\cite{mayer1995integrative}. This can lead to confusion when trust is measured as a single item, or as competence like in~\cite{pesonen2021you}.

Trust models consists of items, which directly ask the participants about their trust perception for which an interpersonal scale is often used, as the machine-part is seen as another person. Hence, there is no difference between human-human and human-machine communication~\citep{weidmuller2022human}. \cite{weidmuller2022human} includes human-like and machine-like trustworthiness as items. Further, trust items can be investigated more specifically. \cite{lappemanTrustDigitalPrivacy2023a} include brand trust in their questionnaire design as a relevant factor for overall trust levels in the interaction. On the other hand,~\cite{jiang2023make} include task-specific trust in their work as a component, since there is the assumption that one might show trust to a person well-qualified for a task. However, this does not guarantee that the same person will be given the same trust during a different task. 

\paragraph{Trust (Mayer)}
The Mayer model of trust is highly accepted in trust measurement such that a multitude of papers used items directly from that scale. Therefore, items like integrity, benevolence, competence, technical competence, perceived benefits and reputation are all included in the trust scales. This subgroup emphasises the influence that the trust model of Mayer has on the development of trust scales over the last years. It raises the question of to what degree the Mayer model should be used in trust in technology context -- particularly in the context of CAI -- since the original model has an organisational, interpersonal focus. 

\paragraph{Trust and Technology} 
A larger subgroup of trust can be trust and technology, which all contain items relevant to technology. Following the origins of the constructs used in the studies revealed that there were only two scales used. First, the scale by~\cite{jianFoundationsEmpiricallyDetermined2000a} was developed in 2000 -- a time when neither AI nor Chatbots were as sophisticated as today -- and second, the scale by~\citep{gupta2019ai} marks one of the most recent development steps of a human-computer trust scale within the last years. Yet, for the latter one, it needs to be mentioned that three items (benevolence, competence and perceived risk) stem from the Mayer model. Further, we found the scale developed by~\cite{jianFoundationsEmpiricallyDetermined2000a} being mentioned in both, trust in automation and trust scales. This leads to confusion as the context in which the scale was developed matters. 

\paragraph{Interactivity}
A smaller subgroup, namely three papers use interactivity as one of the constructs in their scales~\cite{purwanto2020interactive, prakash2023determinants, hu2021dual}. This is surprising, especially when considering that interactiveness is essential in the perception of a human-like interaction between a CAI system and a human. The trust research community could benefit from an additional focus on interactivity, especially in the context of CAI systems.

\paragraph{Affect, Humanness, Antropomorphism}
A crucial and often overlooked aspect of trust, next to task-solving capabilities, is the integration of affect reaction in CAI systems. We cluster perception of humanness by assistants as one group which contains items like social presence, which is often used in technology acceptance studies, humanlikeness~\citep{lv2022research}, understanding humanness~\citep{hu2021dual} or antropomorphism~\citep{chen2021anthropomorphism}. Interestingly, perceived humanity, perceived social presence and perceived social interactivity are originally part of a technology acceptance study~\citep{venkatesh2000theoretical}. Items like faith and personal attachment stem from the development of a human-computer trust scale~\citep{madsen2000measuring}, but they were only used in one study~\citep{lee2022testing}. 

\paragraph{Character and Personality}
Next to items directly related to trust, another group of items identified in a variety of papers was items related to character and personality. Already \cite{mayer1995integrative} identified that to some degree depends on the 'propensity to trust' which is highly personality dependent. This indicates that there might be other factors related to character and personality, that could be relevant for trust measures. \cite{muller2019chatbot} integrated a complete personality questionnaire into their study to see what effect personality has on trust measures and development. This was further supported by items like self-esteem or disposition to trust.

\subsubsection{Scale Development and Reuse}

After discussing which constructs have been researched to assess users' privacy, security and trust perceptions, we were further interested in which scales were used or referred to in detail (RQ3). Therefore, we analysed the extent to which studies developed or reused scales to investigate privacy, security and trust perceptions. 

We find that research on trust perceptions largely reuses and adapts scales. Similar observations were made for privacy and security scales with more than \changed{80\%} of papers relying on scales used in prior work. Whenever scales were not explicitly reused or adapted from previous work, researchers mostly used single items or multiple distinct questions. For example, \cite{gauder2023towards} and \cite{tenhundfeld2021myISMYSIRI} use a single item to evaluate confidence and trust. Similarly, ~\cite{joyInvestigatingUsersPrivacy2022b} rely on a single question to measure participants' level of data privacy concerns. Yet, such single-question instruments are rarely developed methodologically and lack reliability and consistency measures~\citep{preibuschGuideMeasuringPrivacy2013}. 

We were further interested in whether specific scales dominated the assessment of privacy, security and trust perception research in the context of CAI. \changed{While~\cite{lankton2015technology} and~\cite{jianFoundationsEmpiricallyDetermined2000a} were frequently cited as a source and origin for trust in technology in Table~\ref{tab:TypeTrustConstructs}(Appendix), the same does not hold for privacy and security perception measures.} This is surprising as, despite the existence of established scales for measuring privacy concerns~\citep{preibuschGuideMeasuringPrivacy2013}, only a few papers report references to these scales. 

Only \changed{three studies} specifically cite the well-known privacy concern scale (CFIP)~\citep{uysalTrojanHorseUseful2022a, vimalkumarOkayGoogleWhat2021c, park2021users, smithInformationPrivacyMeasuring1996b}. However, while the original CFIP scale consists of four sub-scales related to collection, errors, unauthorized secondary use and improper access,~\cite{uysalTrojanHorseUseful2022a} utilized only one subscale, consisting of four items, to evaluate privacy concerns. Similarly,~\cite{choWillDeletingHistory2020e} rely on a scale originally designed with thirteen items to measure privacy concerns but uses only three. In general, altering well-established scales should be done with caution as it may invalidate existing reliability and validity~\citep{preibuschGuideMeasuringPrivacy2013}. 
Nevertheless, using well-established scales in new contexts such as CAI systems, could help in reevaluating the scales' validity. In addition, it could improve and foster cross-study comparison and enhance understanding of contextual influences on privacy, security and trust measures. 

\subsubsection{Reported Scale Reliability and Validity}
\label{sec:ReliabilityValidityScales}

To address the fourth research question and investigate reliability and validity scores that were reported on privacy, security and trust metrics in the context of CAI. Therefore, an evaluation framework developed by~\cite{rohanSystematicLiteratureReview2023} was used to assess information security scales. \changed{Their framework draws from widely accepted research on scale development and evaluation and can be generally applied to assess the reliability and validity of measurement scales. This framework was only applied to studies relying on multi-item scales, \changed{as commonly used reliability measures such as Cronbach's alpha require at least two items}. The total scores on scale validation, i.e., reliability and validity, are calculated as follows: Reliability testing is based on the calculation of Cronbach's alpha and composite reliability each contributing 0.5 to a possible full score of 1. Convergent validity assessment is again based on two measures, namely on factor loadings and average variance extracted (AVE) coefficient. Finally, a full score of 1 is given if discriminant validity has been assessed by the study.} Thereby, the respective values need to be within widely accepted thresholds, e.g. above 0.7 for Cronbach's Alpha. It needs to be mentioned that for trust and privacy \& security scales, none of the papers reported values outside their threshold. If papers initially found reliability or validity issues, measures were taken to resolve them. 

We want to highlight that by applying this framework we do not judge the overall quality of reviewed studies. Instead, our analysis serves the purpose of evaluating and deepening our understanding of privacy, security and trust perception scales used in the context of CAI. While we acknowledge that reliability and validity measures can take various forms, the framework helps to assess whether researchers followed best practices of scale validation. This is crucial as research on privacy, security and trust can have serious consequences in informing businesses and policymaking. 

Table~\ref{tab:ReliabilityConstructs} shows the total scores for the validation of internal consistency, discriminant validity and convergent validity for privacy \& security and trust scales. \changed{The individual assessments of the scales can be found in the Appendix in Table~\ref{tab:TypePrivacyConstructs} and~\ref{tab:TypeTrustConstructs}. The overall percentage was computed by averaging the individual scores per paper based on the framework provided by~\cite{rohanSystematicLiteratureReview2023}. As previously mentioned, we only considered papers that used multi-item scales. Therefore, 100\% would have been achieved if all papers using multi-item scales had calculated internal consistency as required by the framework.}

First, there is a drastic difference between papers extracted by our SLR on security and privacy perceptions and the one on trust. Research on privacy and security perceptions assessed reliability and validity almost twice as often as research on trust perceptions in the context of CAI. Moreover, two different trends are observable. First, for the literature on trust, it was found that 25\% of the papers calculated only Cronbach's alpha resulting in overall values of 0.5 for the measure of internal consistency. In contrast, papers evaluating privacy and security perceptions predominantly evaluated both, Cronbach's alpha and composite reliability, resulting in values of 1 for internal consistency evaluation. Thus, in total, \changed{60\%} of papers sufficiently assessed the scale's reliability. 

Second, for both literature reviews, it was observable that validity was only evaluated in cases in which also internal consistency was assessed. In turn, reliability was assessed the most to validate privacy, security and trust scales. 
Interestingly, for research on privacy perceptions, the assessment of reliability and validity measures is highly correlated with structural equation modelling methods for data analysis. This makes sense as structural equation modelling requires explicit quality assessment of the measurement model~\citep{hairPartialLeastSquares2021}. These findings show that reliability testing seems to be the default while validity testing is not. Therefore, studies on privacy, security and trust perceptions could benefit from the well-established practices in the context of structural equation modelling to ensure the usage of valid and reliable scales.

\begin{table}[ht]
  \caption{\changed{Reliability and validity assessment of papers using multi-item scales based on a framework by~\cite{rohanSystematicLiteratureReview2023}. Individual assessments of the scales are shown in the Appendix in Table~\ref{tab:TypePrivacyConstructs} and~\ref{tab:TypeTrustConstructs}}. The overall percentage was computed by averaging the individual scores per paper.}
  \label{tab:ReliabilityConstructs}
  \begin{tabular}{p{1.7cm}|p{1.7cm}|p{1.7cm}|p{1.7cm}}
    \toprule
    Fulfilled Criteria & Internal Consistency & Discriminant Validity & Convergent Validity  \\
    \midrule
    Privacy \& Security & \noindent\changed{76\%} & \noindent\changed{63\%} &  \noindent\changed{63\%} \\
    \midrule
    Trust & 45\% & 27\% & 27\% \\
    \bottomrule
\end{tabular}
\end{table}

\subsection{Privacy, Security and Trust as an Influence on CAI Adoption}
\label{sec:Adoption}

Roughly one-third of reviewed papers on privacy and security perceptions assessed perceptions in the context of CAI adoption. They predominantly drew upon the Technology Acceptance Model (TAM) or related conceptual frameworks. This does not hold for trust research, which only includes TAM in about 15\% of studies. \changed{Nonetheless, TAM studies considered the constructs of trust and privacy \& security in combination and are thus, investigated more closely on their interplay between trust, privacy and security.} An overview of the studies along with their assessed variables is provided in Table~\ref{tab:TAMStudies} \changed{and~\ref{tab:TAMPlusStudies}}. It is crucial to highlight that all of the studies are based on one or more existing research model and usually provide insightful extensions to these frameworks in the context of CAI. The underlying research models (as shown in Table~\ref{tab:TAMStudies} \changed{and~\ref{tab:TAMPlusStudies}}) have been extracted from the descriptions and references found in the respective papers. \changed{Thereby, we only consider studies that explicitly stated that they built upon or extended an existing model.}

\subsubsection{Technology Acceptance Model (TAM)}
\label{sec:TAMConcept}
The Technology Acceptance Model is one of the major frameworks for understanding factors affecting the potential adoption of technology~\citep{marangunic2015technologyTAM}. At its core, the basic TAM explains the adoption of a system through three key factors: ease of use, perceived usefulness and attitude toward usage. We identified \changed{13} studies that build upon the original TAM to investigate adoption factors of CAI systems including various privacy, security and trust constructs. While the majority treats these constructs as independent variables,~\cite{chaSustainabilityCalculusAdopting2021} hypothesize that perceived privacy risks not only negatively influence the intention to adopt smart speakers but also exert a moderating effect. Their findings reveal cultural differences between South Korean and US samples. In a unique contribution, ~\cite{acikgozRolePrivacyCynicism2022} assesses the relatively novel concept of privacy cynicism and its impact on attitude and trust in the context of CAI systems. Surprisingly, their findings indicate that privacy cynicism negatively influences attitudes towards voice assistant usage but contrary to existing literature positively affects users' trust. \changed{In addition to purely quantitative TAM studies,~\cite{Tennant2022Caregiver} conducted a mixed-method study on caregivers' expectations of voice assistants in home care and rely on TAM to analyze their qualitative findings.}

Finally, a study by~\cite{dekkalFactorsAffectingUser2023a} draws on TAM but concentrates mainly on factors typically overlooked in common technology acceptance models. They explore drivers such as practicity, personalisation and enjoyment as well as inhibitors, such as privacy concerns and creepiness, influencing trust and adoption in insurance chatbots. They find that privacy concerns do not significantly influence trust while creepiness does. Furthermore, both privacy concerns and creepiness have a marginal influence on adoption intention. In resonance with previously discussed studies, trust lacks a significant effect on adoption intention but is influenced by a moderating impact of technological anxiety. 

\changed{Table~\ref{tab:TAMPlusStudies} provides an overview of studies that contrast or incorporate other conceptual models into the TAM framework. \cite{reseChatbotsRetailersCustomer2020a} contrast TAM and the Uses and Gratifications Theory (U\&GT) in the context of chatbot adoption in online retailers and find that both models showed similar predictive power.}~\cite{kwangsawadOvercomingCustomerInnovation2022} integrates TAM and Diffusion of Innovation theory (DOI) to evaluate consumers' intention to utilize chatbot services. Meanwhile,~\cite{patilCanWeTrust2022a} builds on TAM and the Health Belief Model (HBM) to examine the use of health and wellness chatbots considering factors such as severity of potential health problems and benefits and barriers to chatbot adoption.
Another study combines TAM and the Information Systems Success Model (D\&M) to validate the acceptance of cognitive chatbots in a B2B context. They confirm that privacy risks have a moderating effect on aspects of information system quality. Finally,~\cite{pitardi2021alexa} draw from various theories including human-computer interaction theories and para-social relationship theories such as the Extended Privacy Calculus Model (EPCM) and Service Robot Acceptance model (sRAM). 

\begin{table*}[ht]
\scriptsize
\caption{\changed{Overview of studies in our reviewed sample that build upon and extend the Technology Acceptance Model (TAM) to research adoption intention in the context of privacy, security and trust perceptions. Only studies that explicitly stated the reliance or extension of TAM are considered. Detailed explanations are provided in Section~\ref{sec:TAMConcept}.}}
\label{tab:TAMStudies}
\begin{tabular}{p{0.05\textwidth}p{0.3\textwidth}p{0.1\textwidth}p{0.125\textwidth}p{0.125\textwidth}p{0.125\textwidth}}
\toprule
\noindent\changed{Underlying Research Model} & Independent Variable(s) & Moderating Variable     & Mediating Variable & Dependent Variable & Reference \\ 
\midrule
TAM & Technology Optimism, System Diversity, System Quality, Perceived Enjoyment, Perceived Usefulness, Perceived Ease of Use, Risk, \textbf{Surveillance Anxiety}, \textbf{Security/Privacy Risk} & & & Behavioral Intention & \cite{kowalczukConsumerAcceptanceSmart2018a} \\
\midrule
TAM & Perceived Usefulness, Perceived Ease of Use, Attractiveness, \textbf{Trust}, \textbf{Mistrust}, Hedonic Quality, Pragmatic Quality, \textbf{Perceived Security} & & & Intent &  \cite{fahnTimeGetConversational2021a} \\
\midrule
TAM & Perceived Usefulness, Perceived Ease of Use, Attitude, \textbf{Trust Perception} & & & Usage Intention &  \cite{choung2023trust} \\
\midrule
TAM & Perceived Usefulness, Perceived Ease of Use, Perceived Enjoyment, Price Consciousness, \textbf{Perceived Risk}, \textbf{Trust}, Personal Innovativeness & & Attitude & Intention to Use & \cite{kasilingamUnderstandingAttitudeIntention2020a}\\
\midrule
TAM & Perceived Usefulness, Perceived Ease of Use, Personal Norms, Societal Norms, \textbf{Privacy Concerns}, \textbf{Perceived Security} & & Attitude toward using Voice Assistants & Intention to Use & \cite{buteauHeyAlexaWhy2021} \\ 
\midrule
TAM & Perceived Usefulness, Perceived Ease of Use, \textbf{Perceived Privacy Risk}, Innovativeness, Perceived Enjoyment, Social Attraction, Task Technology Fit  &  & Attitude towards smart speakers & Intention to Use & \cite{aiolfiHowShoppingHabits2023} \\ 
\midrule
TAM & Perceived Usefulness, Perceived Ease of Use, \textbf{Privacy Cynicism} & & \textbf{Trust towards using VAs}, Attitude towards VAs & Habit of Using & \cite{acikgozRolePrivacyCynicism2022} \\ 
\midrule
TAM & Perceived Usefulness, Perceived Ease of Use, Perceived Enjoyment, \textbf{Perceived Privacy Risks}  & \textbf{Perceived Privacy Risks} &  & Adoption Intention & \cite{chaSustainabilityCalculusAdopting2021} \\
\midrule
TAM & Perceived Ease of Use, \textbf{Perceived Trust}, Perceived Enjoyment, Perceived Usefulness  &  &  & Behavioral Intention to Use & \cite{jameel2022perceived} \\
\midrule
TAM & Perceived Usefulness, Perceived Ease of Use, Perceived Enjoyment, Information Quality, Service Quality, Interface and Design, \textbf{Perceived Risk}, Structural Assurances, \textbf{Privacy and Security Concerns}, \textbf{Disposition to Trust}, Technology Fear, Ubiquity  &  & \textbf{Chatbot Trust}
  & Attitude, Intention, Customer Satisfaction & \cite{alagarsamy2023exploring} \\
\midrule
\noindent\changed{TAM} & \noindent\changed{Perceived Usefulness, Perceived Ease of Use, \textbf{Perceived Privacy Risk}, Perceived Compatibility, Awareness of Service} & & & \noindent\changed{Behavioral Intention to Use} & \noindent\changed{\cite{alt2021identifying}} \\
\midrule
\noindent\changed{TAM} & \noindent\changed{Perceived Usefulness, Perceived Ease of Use, \textbf{Concerns}, Excitement, Cost, Prior Experience} & & & \noindent\changed{Attitudes towards Use} & \noindent\changed{\cite{Tennant2022Caregiver}} \\
\midrule
TAM & Practicity (Perceived Ease of Use/Perceived Usefulness), Personalization, Enjoyment, \textbf{Privacy Concerns}, Creepiness & Technology Anxiety & \textbf{Trust}   & \textbf{Trust}, Adoption Intention  & \cite{dekkalFactorsAffectingUser2023a}\\ 
\bottomrule
\end{tabular}
\end{table*}

\begin{table*}[ht]
\scriptsize
\caption{\changed{Overview of studies in our reviewed sample that contrast or extend the Technology Acceptance Model (TAM) with other conceptual frameworks to research adoption intention in the context of privacy, security and trust perceptions. Only studies that explicitly stated the reliance or extension of the underlying research model are considered. Detailed explanations are provided in Section~\ref{sec:TAMConcept}.}}
\label{tab:TAMPlusStudies}
\begin{tabular}{p{0.05\textwidth}p{0.3\textwidth}p{0.1\textwidth}p{0.125\textwidth}p{0.125\textwidth}p{0.125\textwidth}}
\toprule
\indent\changed{Underlying Research Model} & Independent Variable(s) & Moderating Variable     & Mediating Variable & Dependent Variable & Reference \\ 
\midrule
\noindent\changed{TAM, UTAUT} & \noindent\changed{Perceived Usefulness, Perceived Ease of Use, Social Influence, Hedonic Motivation, \textbf{Perceived Privacy Concerns, Perceived Trust}} & & & \noindent\changed{Behavioral Intention to Use} & \cite{munoz2023voice} \\
\midrule
\noindent\changed{TAM, U\&GT} & \noindent\changed{Perceived Usefulness, Perceived Ease of Use, Convenience, Authenticity of Conversation, Enjoyment, Pass Time, \textbf{Privacy Concerns}, Immature Technology} & & & \noindent\changed{Behavioral Intention to Use} & \cite{reseChatbotsRetailersCustomer2020a} \\
\midrule
TAM, IDP & Perceived Time Risk, \textbf{Perceived Privacy Risk}, Perceived Usefulness, Perceived Ease of Use, Perceived Convenience, Perceived Information Quality, Technological Anxiety, Openness to Experience & & Attitude toward continued & Intention  & \cite{kwangsawadOvercomingCustomerInnovation2022} \\ 
\midrule
TAM, HBM & \textbf{Propensity to trust technology}, Social Influence, Perceived Usefulness, Perceived ease of use, \textbf{Privacy}, \textbf{Safety Risk}, Facilitating Condition, Perceived Susceptibility, Perceived Severity, Perceived Benefit, Perceived Barrier & & \textbf{Trust Belief in Technology} & Adoption Intention & \cite{patilCanWeTrust2022a} \\ 
\midrule
TAM, D\&M & Perceived Information Quality, Perceived System Quality, Perceived Service Quality, Perceived Ease of Use, Perceived Usefulness, \textbf{Perceived Trust} & \textbf{Perceived Risk} & Customer Experience, Attitude Towards Technology & Adoption Intention & \cite{beheraCognitiveChatbotPersonalised2021a} \\ 
\midrule
TAM, SRAM, EPCM & Perceived Usefulness, Perceived Ease of Use, Enjoyment, Social Presence, Social Cognition, \textbf{Privacy Concern} & & Attitude, \textbf{Trust} & Intention to Use & \cite{pitardi2021alexa} \\
\midrule
TAM, SRAM, EPCM & Behavioral Traits, Humanlike Traits & & Perceived Usefulness, Perceived Ease of Use, Social Presence, \textbf{Trusting Belief}  & Continued Usage Intention & \cite{prakash2023determinants} \\
\midrule
TAM, SRAM & Perceived Usefulness, Perceived Ease of Use, Perceive Humanity, Perceived Social Interactivity, Perceived Social Presence & & \textbf{Trust}  & Acceptance of Artificial Intelligence Virtual Assistant & \cite{zhang2021motivation} \\
\bottomrule
\end{tabular}
\end{table*}

\subsubsection{Unified Theory of Acceptance and Use of Technology (UTAUT)} 
\label{sec:UTAUT}

\changed{As shown in Table~\ref{tab:UTAUTStudies}, six} reviewed papers build on the Unified Theory of Acceptance and Use of Technology (UTAUT) \changed{and UTAUT2} to examine adoption factors in the context of CAI. Developed by~\cite{venkatesh2000theoretical}, UTAUT and its successor UTAUT2, evolved through review, comparison and integration of technology acceptance models. Thereby, the original UTAUT includes constructs on performance expectancy, effort expectancy, social influence and facilitating conditions to influence behavioural intention and thus, actual use of technology. In the context of virtual assistants for cancer patients,~\cite{vanbusselAnalyzingDeterminantsAccept2022b} \changed{deployed UTAUT} by introducing additional factors such as self-efficacy, trust and resistance to change. 

\changed{Four papers build on } UTAUT2 in the context of CAI systems which modifies and extends UTAUT by introducing three new constructs: hedonic motivation, price value and habit~\citep{venkatesh2012consumer}. For instance,~\cite{garciadeblanessebastianApplicationExtensionUTAUT22022} found that trust significantly impacts users' intention to use virtual assistants while perceived privacy risks did not show the same influence. Similarly,~\cite{vimalkumarOkayGoogleWhat2021c} examined the adoption of voice-based digital assistants based on an extension of UTAUT2. Notably, they are among the few to explicitly examine the relationship between perceived trust, perceived privacy risks and perceived privacy concerns. Their results show that perceived privacy risks significantly influence perceived privacy concerns and trust. In line with \cite{garciadeblanessebastianApplicationExtensionUTAUT22022}, they find that perceived trust does positively influence behavioral intention while perceived privacy risks do not. Moreover, they discovered that perceived privacy concerns do not significantly impact behavioral intention. 

In a distinct approach,~\cite{palEffectTrustIts2022a} integrate UTAUT2 with parasocial relationship theory (PSR) and privacy aspects to investigate behavioral intention to use voice-based consumer electronic devices. Their results suggest that perceived privacy risks significantly influence perceived privacy concerns and trust. Yet, in stark contrast to previous research, they found that trust did not significantly influence behavioral intention. Additionally, they observed that privacy concerns did not significantly influence trust. 

Generally, these findings suggest that perceived privacy risk may not have a direct influence on adoption but may be mediated through consumer trust~\citep{vimalkumarOkayGoogleWhat2021c}. Nevertheless, the studies also reveal conflicting results concerning the influence of trust on behavioral intention to use CAI systems. Their research specifically highlights the importance of a comprehensive assessment of privacy, security and trust constructs to enhance understanding of adoption factors within the context of CAI systems considering diverse situations and user groups. 

\begin{table*}[ht]
\scriptsize
\caption{\changed{Overview of studies in our reviewed sample that build on the Unified Theory of Acceptance and Use of Technology (UTAUT) and UTAUT2 framework to research adoption intention in the context of privacy, security and trust perceptions. Detailed explanations are provided in Section~\ref{sec:UTAUT}.}}
\label{tab:UTAUTStudies}
\begin{tabular}{p{0.05\textwidth}p{0.3\textwidth}p{0.1\textwidth}p{0.125\textwidth}p{0.125\textwidth}p{0.125\textwidth}}
\toprule
\noindent\changed{Underlying Research Model} & Independent Variable(s) & Moderating Variable & Mediating Variable & Dependent Variable & Reference \\ 
\midrule
UTAUT & Self-Efficacy, Performance Expectancy, Effort Expectancy, Social Influence, Facilitating Conditions, \textbf{Trust}, Resistance to Change & & Effort Expectancy & Behavioral Intention & \cite{vanbusselAnalyzingDeterminantsAccept2022b} \\
\midrule
\noindent\changed{UTAUT} & \noindent\changed{Effort Expectancy, Effort Expectancy, \textbf{Perceived Security}, Social Influence, Facilitating Condition, Perceived Anthropomorphism, Perceived Animacy, Perceived Intelligence} & & \noindent\changed{Parasocial Interactions, Smart-Shopping Perception, AI-Enabled Customer Experience} & \noindent\changed{Continuance Intention} & \cite{aw2022alexa} \\
\midrule
\noindent\changed{UTAUT2} & \noindent\changed{Performance Expectancy, Effort Expectancy, Social Influence, Facilitating Conditions, Hedonic Motivation, Price Value, Habit, \textbf{Privacy Concerns}} & & & \noindent\changed{Intention to Use} & \cite{farooq2022voice} \\
\midrule
UTAUT2  & Performance Expectancy, Effort Expectancy, Social Influence, Facilitating Conditions, Personal Innovativeness, \textbf{Trust}, \textbf{Perceived Privacy Risk}, Hedonic Motivation, Habit, Price/Value  & & & Behavioral Intention  & \cite{garciadeblanessebastianApplicationExtensionUTAUT22022} \\ 
\midrule
UTAUT2 & Performance Expectancy, Effort Expectancy, Social Influence, Hedonic Motivation, Price/Value, Facilitating Conditions, \textbf{Perceived Privacy Risks} & & \textbf{Perceived Trust}, \textbf{Perceived Privacy Concerns}, Behavioral Intention & Adoption & \cite{vimalkumarOkayGoogleWhat2021c} \\ 
\midrule
UTAUT2, PSR & Performance Expectancy, Effort Expectancy, Hedonic Motivation, \textbf{Perceived Privacy Risk}, Perceived Social Presence, Perceived Humanness, Social Cognition  & & \textbf{Perceived Privacy Concern}, \textbf{Trust}  & Behavioral Intention & \cite{palEffectTrustIts2022a}\\ 
\bottomrule
\end{tabular}
\end{table*}

\begin{table*}[h!]
\scriptsize
\caption{\changed{Overview of studies in our reviewed sample that build on conceptual models beyond TAM, UTAUT and UTAUT2 to research adoption intention in the context of privacy, security and trust perceptions. Detailed explanations are provided in Section~\ref{sec:OtherConceptualModels}.}}
\label{tab:OtherConceptualModels}
\begin{tabular}{p{0.05\textwidth}p{0.3\textwidth}p{0.1\textwidth}p{0.125\textwidth}p{0.125\textwidth}p{0.125\textwidth}}
\toprule
\noindent\changed{Underlying Research Model} & Independent Variable(s) & Moderating Variable & Mediating Variable & Dependent Variable & Reference \\ 
\midrule
PSR & Task Attraction, Social Attraction, Physical Attraction, \textbf{Security/Privacy Risk} & & Parasocial Relationship, Satisfaction & Adoption Intention & \cite{hanUnderstandingAdoptionIntelligent2018a} \\
\midrule
VAM & Perceived Usefulness, Perceived Enjoyment, \textbf{Perceived Security Risk}, Perceived Technicality, Perceived Cost, Social Influence, Usage & Gender & Perceived Value & Adoption Intention & \cite{Shofolahan2018} \\ 
\midrule
\noindent\changed{SVAM} & \noindent\changed{Self-efficacy, Functional Value, Emotional Value, Social Value, \textbf{Privacy Risk}} & & & \noindent\changed{Adoption Intention} & \cite{cao2022adoption} \\
\bottomrule
\end{tabular}
\end{table*}

\subsubsection{Other Conceptual Models}
\label{sec:OtherConceptualModels}

Some of the reviewed papers investigate the adoption of CAI systems drawing insights from various other conceptual models \changed{as shown in Table~\ref{tab:OtherConceptualModels}}.~\cite{hanUnderstandingAdoptionIntelligent2018a} extend the \changed{Parasocial Relationship theory (PSR)} to investigate factors influencing adoption intentions. They find that privacy and security risks negatively affected parasocial relationships.~\cite{Shofolahan2018} build upon the \changed{Value-based Adoption Model (VAM)} taking into account factors such as perceived security risk, perceived technicality, perceived cost as well as perceived usefulness and enjoyment. Surprisingly, only perceived cost was found to negatively influence perceived value while security risk and technicality do not. \changed{Along similar lines,~\cite{cao2022adoption} explores the adoption intention of voice assistants among Airbnb guests by drawing on the Self-Efficacy based Value Adoption Model (SVAM) and identified privacy risks as a significant determinant.}

\subsection{Beyond Adoption: Influence on Privacy, Security and Trust on Usage, Disclosure and More}

In addition to theoretical models that examine the influence of privacy, security and trust perception on the adoption of CAI systems, we identified various studies that \changed{did not build upon TAM, UTAUT or UTAUT2 and investigated variables beyond adoption such as usage or disclosure. Thereby, they do not always build upon theoretically grounded research models but conceptualize and validate individual models.}

\subsubsection{\changed{Usage, Disclosure and Attitude}}
\label{sec:UsageDisclosure}

\begin{table*}[ht]
\caption{Overview of studies in our reviewed sample that research privacy, security and trust perceptions in the context of usage, disclosure and attitude towards Conversational AI systems. Detailed explanations are provided in Section~\ref{sec:UsageDisclosure}.}
\label{tab:UsageDisclosure}
\scriptsize
\begin{tabular}{p{0.08\textwidth}p{0.3\textwidth}p{0.1\textwidth}p{0.125\textwidth}p{0.125\textwidth}p{0.125\textwidth}}
\toprule
\noindent\changed{Underlying Research Model} & Independent Variable(s) & Moderating Variable & Mediating Variable & Dependent Variable & Reference \\ 
\midrule
U\&GT & Utilitarian Benefits, Hedonic Benefits, Symbolic Benefits, Social Presence, Social Attraction & \textbf{Perceived Privacy Risk}  &  & Usage & \cite{mcleanHeyAlexaExamine2019a} \\ 
\midrule
\noindent\changed{U\&GT} & \noindent\changed{Utilitarian Uses, Hedonic Uses, Subjective Norms, Perceived Critical Mass, \textbf{Perceived Privacy Concerns}} & \noindent\changed{\textbf{Perceived Privacy Concerns}} & & \noindent\changed{Continuance Use Intention} & \cite{kefi2021privacy} \\ 
\midrule
\noindent\changed{U\&GT} & Information, Entertainment, Media Appeal, Social Presence, \textbf{Privacy Risk} & & Satisfaction, Customer Loyality & Continued Use Intention & \cite{chengHowAIdrivenChatbots2020} \\
\midrule
\noindent\changed{U\&GT, Signaling and Prospect Theory} & \noindent\changed{Utility Features, Hedonic Features, Social Presence, \textbf{Perceived Privacy Risk}} & \noindent\changed{Gender, Brand Credibility} & \noindent\changed{Overall Perceived Value} & \noindent\changed{Continued Usage Intention} & \cite{jain2022interactive} \\
\midrule
\noindent\changed{U\&GT, Signaling and Prospect Theory} & \noindent\changed{Utility Features, Hedonic Features, Social Presence, \textbf{Perceived Privacy Risk}, Irritation} & \noindent\changed{Gender, Brand Credibility, Irritation} & \noindent\changed{Perceived Value} & \noindent\changed{Continued Usage Intention, Brand's Loyalty} & \cite{maroufkhani2022interactive} \\
\midrule
\noindent\changed{PAMISC} & \noindent\changed{Personal Innovativeness, Technology Anxiety} & & \noindent\changed{Confirmation, Privacy Value, Hedonic Motivation, Compatibility, Perceived Security, Satisfaction} & \noindent\changed{Continuance Intention, Intention to Recommend Others} & \cite{lee2021continuation} \\
\midrule 
n.a. & Hedonic Motivation, Compatability, \textbf{Perceived Security}, Diversity Use & & Satisfaction, Habit & Continuance Use & \cite{leeHeyAlexaMagic2020a} \\
\midrule
n.a. & \textbf{VAPA Privacy Concern}, \textbf{Household Members Privacy Concern}, \textbf{Government Privacy Concern}, \textbf{Vendor and Third Party Privacy Concern}, Perceived Anthropomorphism, Perceived Intelligence & Perceived Intrusiveness & \textbf{Emotional Trust}, \textbf{Cognitive Trust} & Usage of VAPA's & \cite{pal2022perceptions} \\
\midrule
\midrule
SCM, RFT & Regulatory Focus, Interaction Style
 &  & \textbf{Trust} & Willingness to Self-Disclose & \cite{choi2023inducing} \\
\midrule
\cite{saffarizadeh2017conversational} & \textbf{Brand Trust}, \textbf{Digital Privacy Concerns} & & \textbf{Cognitive Trust}, \textbf{Emotional Trust} & User Self-Disclosure & \cite{lappemanTrustDigitalPrivacy2023a} \\
\midrule
n.a. & Human-like Chatbot vs. Machine-like Chatbot vs. Website,  & & \textbf{Privacy Concerns}, Perceived Anthropomorphism & Information Disclosure, Attitudes, Recommendation Adherence & \cite{ischenPrivacyConcernsChatbot2020b} \\
\midrule
\midrule
n.a. & Modality & \textbf{Privacy Concerns} & Perceived Social Presence & Attitude toward the Voice Assistant & \cite{choHeyGoogleCan2019b} \\
\bottomrule
\end{tabular}
\end{table*}

\changed{Table~\ref{tab:UsageDisclosure} provides an overview of reviewed papers investigating usage, disclosure and attitude towards CAI systems. Uses and Gratification Theory (U\&GT) is among the mostly employed theories to understand usage intention in our sample. For instance, ~\cite{mcleanHeyAlexaExamine2019a} explore utilitarian, hedonic and symbolic benefits in addition to social benefits such as social presence and attraction and their impact on voice assistant usage. Their findings suggest positive effects on voice assistant use across all factors with perceived privacy risks exerting a significant negative moderating effect. Another study drawing on U\&GT investigated continuance use intentions of virtual assistants across generations and identified a non-linear relationship between privacy concerns and generations~\citep{kefi2021privacy}.~\cite{leeHeyAlexaMagic2020a} explore the effect of habit and continuance use on the perception of group harmony in situations where voice assistants are shared. In addition, they examine hedonic motivation, compatibility and perceived security as antecedents to satisfaction. Their results indicate that perceived security influences satisfaction only marginally while habit and continuance use significantly impact group harmony in a multi-person context. Similarly, in the context of chatbots,~\cite{chengHowAIdrivenChatbots2020} found that privacy risks negatively influence satisfaction. In turn, satisfaction positively affects both continued use and customer loyalty. Three studies do not rely on U\&GT to investigate usage. Instead,~\cite{lee2021continuation} extends the Post-Acceptance Model of Information System Continuance (PAMISC) to investigate the influence of personal traits on the continuation and recommendation intention of voice assistants. They found that perceived security had no significant influence on satisfaction possibly a result of investigating continuance intention rather than initial adoption intention. Based on their own conceptualization, \cite{pal2022perceptions} show that both cognitive as well as emotional trust significantly influence the usage of voice assistants. Interestingly, as opposed to the hypothesis, perceived anthropomorphism did not significantly affect emotional trust.} 

\changed{Three studies explore the impact of privacy and trust on self-disclosure in the context of CAI.~\cite{lappemanTrustDigitalPrivacy2023a} investigate three dimensions of trust and privacy concerns in financial chatbots suggesting that while privacy concerns negatively influence users' self-disclosure, brand trust was not found to influence self-disclosure in South Africa.} In another study, \cite{ischenPrivacyConcernsChatbot2020b} compared two types of chatbots, i.e. human-like, machine-like chatbots, and a traditional website concerning privacy concerns and information disclosure. Their findings show that the human-like chatbot led to higher perceived anthropomorphism and fewer privacy concerns, consequently resulting in higher information disclosure in comparison to machine-like chatbots. Yet, the results do not hold when compared to a common website. Finally, \citep{choHeyGoogleCan2019b} investigate the effect of modality and device on users' perceived social presence and attitude towards voice assistants including usability aspects. They found that voice positively contributes to social presence and thus, a positive attitude towards the assistant only under certain circumstances, i.e., low information sensitivity and low privacy concerns, suggesting that text-based interactions can similarly induce human-like perceptions. 

Generally, these findings imply a transferability of insights from conceptual models developed for text-enabled CAI systems to investigating adoption, usage and other patterns in voice-based CAI systems. Factors like privacy risks seem to continuously impact satisfaction and thus usage of assistants while the interplay of anthropomorphic factors, modality and perceptions requires further investigation. As voice-based CAI systems continue to evolve, drawing from previous work on text-based CAI and traditional interactive systems can guide researchers and contribute to a more comprehensive understanding of factors shaping perceptions in the CAI context. 

\changed{\subsubsection{Advertising, Brand Loyalty and Privacy Management}}

\changed{Table~\ref{tab:AdvertisingMore} provides an overview of studies employing conceptual models to research privacy, security and trust perceptions in the context of privacy management, advertising, and brand loyalty.}
Only one paper investigated potential advertising acceptance in smart speakers. \cite{guerreiroAdvertisingAcceptanceSmart2022} show that usefulness and hedonic motivations significantly influence advertising acceptance with a moderating effect of privacy risks on smart speaker usefulness. In a case study on Siri and an investigation on brand loyalty, ~\cite{hasanConsumerTrustPerceived2021} show that perceived risks negatively affect brand loyalty. Interestingly, this effect is moderated by employment with a greater negative influence observed for unemployed consumers. Furthermore,~\cite{xuSmartSpeakersRequire2022} investigated the effect of gratification on privacy concerns and medium-as-social actor presence and their effects on privacy setting review and ownership protection. Their findings suggest that personal utility, relaxation, enjoyment and status affected privacy concerns. Moreover, privacy concerns had a positive effect on both privacy setting review and ownership protection. 

As CAI systems continue to be integrated further into everyday life, researchers need to move beyond factors influencing adoption and usage. Therefore, gaining insights into factors shaping advertising acceptance and the acceptance of privacy management options in the context of CAI systems is crucial to advancing technologies and formulating design recommendations. While our review uncovered only a limited number of studies investigating these factors based on conceptual and theoretical models, future work should deepen understanding in these domains. 
\label{sec:AdvertisingMore}

\begin{table*}[h!]
\caption{Overview of studies in our reviewed sample that research privacy, security and trust perceptions in the context of privacy management, advertising and brand loyalty in Conversational AI systems. Detailed explanations are provided in Section~\ref{sec:AdvertisingMore}.}
\label{tab:AdvertisingMore}
\scriptsize
\begin{tabular}{p{0.05\textwidth}p{0.3\textwidth}p{0.1\textwidth}p{0.125\textwidth}p{0.125\textwidth}p{0.125\textwidth}}
\toprule
\noindent\changed{Underlying Research Model} & Independent Variable(s) & Moderating Variable & Mediating Variable & Dependent Variable & Reference \\ 
\midrule
SOR & Perceived Warmth, Communication Delay, Perceived Competence, Communal relationship, Exchange Relationship & & \textbf{Trust in Chatbots}
 & Intention to Switch & \cite{cheng2022human} \\
\midrule
\midrule 
\noindent\changed{U\&GT, Media Equation, CPM} & Personal Utility, Info Seeking, Relaxation, Enjoyment, Status, Music Exploration, Multitasking & & \textbf{Privacy Concerns}, Medium-As-Social-Actor Presence & \textbf{Privacy Setting Review}, \textbf{Ownership Protection} & \cite{xuSmartSpeakersRequire2022} \\
\midrule 
\noindent\changed{SOR} & \noindent\changed{Privacy Concerns} &  & \noindent\changed{Emotional Reactions} & \noindent\changed{Problem-focused Coping, Emotion-focused Coping} & \cite{park2021users} \\
\midrule
\midrule
n.a. & \textbf{Perceived Trust} &  & Controllability, Perceived Performance, Bidirectionality, Synchronicity  & Customer Satisfaction & \cite{purwanto2020interactive} \\
\midrule
n.a. & Smart Speaker Ease of Use, Advertising Functionality, Advertising Relevance, Advertising Format & \textbf{Privacy Risk} & Smart Speaker Usefulness, Hedonic Motivation & Advertising Acceptance & \cite{guerreiroAdvertisingAcceptanceSmart2022} \\ 
\midrule
n.a. & \textbf{Trust}, Interaction, \textbf{Perceived Risk}, Novelty Value & & & Brand Loyalty & \cite{hasanConsumerTrustPerceived2021} \\
\bottomrule
\end{tabular}
\end{table*}

\subsubsection{\changed{Trust as a Dependent Variable}}
\label{sec:DependentTrust}
\changed{Next to gathering trust as a direct item in questionnaires, seven papers used trust as the dependent variable (as shown in Table~\ref{tab:DependentTrust}). Due to the difficulty of defining trust, the objective of these studies was either to find antecedents to trust and their influence on trust or to use proxies instead of trust. For instance, in \citep{gulati2018modelling}, the antecedents of trust and their impact on user trust were investigated. \citep{muller2019chatbot} took a different approach and focused on which character profiles show the highest trust towards chatbots. This was based on the HEXACO personality profile \cite{lee2004psychometricHEXACO} which is an extension to the well-known Big Five personality traits. This stands in line with the studies using trust as a mediating variable, where the independent variable was mostly antecedents of trust with an additional dependent variable. 

It seems beneficial to use proxies, or additional measures of trust in the collection of trust as they provide a context into the study as well as being more straightforward to grasp. For example, predictability, or reputation are concepts which seem more natural to reflect on. 
}
\begin{table*}[h!]
\caption{Overview of studies in our reviewed sample that use trust or trustworthiness as the dependent variable. Detailed explanations are provided in Section~\ref{sec:DependentTrust}.}
\label{tab:DependentTrust}
\scriptsize
\begin{tabular}{p{0.05\textwidth}p{0.3\textwidth}p{0.1\textwidth}p{0.125\textwidth}p{0.125\textwidth}p{0.125\textwidth}}
\toprule
\noindent\changed{Underlying Research Model} & Independent Variable(s) & Moderating Variable & Mediating Variable & Dependent Variable & Reference \\ 
\midrule
SPT & Motivation, Competence, Benevolence, Predictability, Honesty, Reciprocity, Structural Assurance, Willingness & &
 & \textbf{Trust} & \cite{gulati2018modelling} \\
\midrule
n.a. & Emotional Experiences &  & Social Presence & \textbf{Trust} & \cite{lee2022testing} \\
\midrule
n.a. & Expertise, Predictability, Human-likeness, Ease of Use, \textbf{Risk}, Reputation, \textbf{Propensity to Trust Technology} & & & \textbf{Trust} & \cite{nordheim2019initialtrust} \\
\midrule
n.a. & Human-like Cues, Tailored Responses, Task Creativity, Ambiguity &  & Social Presence, Perceived Task Solving Competence & \textbf{Trust toward Chatbot} & \cite{jiang2023make} \\
\midrule
n.a. & Emotion, Emotion Intensity &  &  & Perceived Affect, Perceive Sincerity, \textbf{Perceived Trustworthiness}, Overall Perception & \cite{tastemirova2022microexpressions} \\
\midrule 
HEXACO & Honesty, Emotionality, Extraversion, Agreeableness, Conscientiousness, Openness
Anthropomorphic Design Cues & & Social Presence, Perceived Competence & Positive Consumer Response, \textbf{Trust} & \cite{muller2019chatbot} \\
\midrule
n.a. & \noindent\changed{System Quality} & & \noindent\changed{Subjective Norm, Familiarity, Technology Optimism, Perceived Enjoyment} & \noindent\changed{\textbf{Perceived Trust}} & \cite{liu2021influences} \\
\bottomrule
\end{tabular}
\end{table*}

\subsection{Tools for Privacy, Security and Trust Perception Manipulation}
\label{sec:Tools}

In this section, we specifically highlight the tools used to manipulate privacy, security and trust perceptions. While a majority of studies focused on quantitative methods applying surveys, a significant sample conducted experiments in quantitative or mixed-method settings to understand users' perceptions. 

A majority of these studies relied on commercially available voice-based CAI systems such as Amazon Echo, Google Assistant or Siri~\citep{rajapakshaFieldStudyUsability2021b, choHeyGoogleCan2019b, cowanWhatCanHelp2017a}. Other studies investigated users' perceptions on text-based CAI systems using brokerbots or Facebook e-commerce chatbot~\citep{kasilingamUnderstandingAttitudeIntention2020a,leeBrokerbotCryptocurrencyChatbot2021}

Few studies did not rely on existing chatbots or voice assistants but implemented their own CAI applications using Microsoft's Power Virtual Agent Platform, IBM Watson Platform, specifically developed Amazon Alexa applications, Chatbot Language (CBL), or a conversational agent research toolkit developed by Araujo~\citep{dekkalFactorsAffectingUser2023a, casadei2022chatbots, bruggemeierPerceptionsReactionsConversational2022b, ischenPrivacyConcernsChatbot2020b, gauder2023towards, choWillDeletingHistory2020e, fahnTimeGetConversational2021a}.

More advanced CAI prototypes were developed by studies investigating individual digital study assistants~\citep{konigCriticalSuccessFactors2023a} as well as authentication methods and privacy controls~\citep{renzAuthenticationMethodsVoice2023, mhaidliListenOnlyWhen2020a}. For instance, \cite{renzAuthenticationMethodsVoice2023} implemented a fake voice-banking application for Google Home Mini and four different authentication methods. \cite{mhaidliListenOnlyWhen2020a} developed a smart speaker prototype implementing two privacy controls based on gaze direction and voice volume level and investigated users' perceptions in a lab study. 

Additionally, we found vignette studies to be implemented frequently in trust research~\citep{li2023modeling, wald2021make, toader2019effect, gupta2019ai}. This was prevalent, especially in studies that implemented their own chatbot or voice assistant and common among studies that conducted in-lab experiments. We could also observe that those studies, which use a vignette study did not directly implement a trust scale but focused on either trust measurement from other sources (for instance audio directly)~\citep{li2023s}, or studies which investigated the effect of style and intonation on trust~\citep{tastemirova2022microexpressions}.

Other methods to investigate privacy, security and trust perceptions included the usage of experience sampling methodology. Here,~\cite{malkinPrivacyAttitudesSmart2019} implemented a browser extension to play back historical voice recordings to voice assistant users. Moreover, we need to highlight the rare utilization of long-term studies in the context of CAI. Noteworthy,
\cite{choiUseInternetofThingsSmart2020} conducted a two-month feasibility study with older adults using IoT smart home devices including smart speakers. 

It is to be noted that a few studies did not engage in any interaction with a voice assistant but rather were based on prior experiences with a common assistant or a short video about an interaction was shown before the survey~\citep{muller2019chatbot, purwanto2020interactive, weidmuller2022human, chaSustainabilityCalculusAdopting2021}. Others relied on measurement aids depicting simulated chatbot conversations to manipulate experimental conditions~\citep{lappemanTrustDigitalPrivacy2023a}. Further, scenario-based designs were utilized for example to present virtual prototypes of voice assistants and to investigate users' perceptions concerning privacy controls~\citep{ahmadTangiblePrivacySmart2022}.

\section{Discussion}

We conducted a systematic literature review on empirical methods and privacy, security and trust perception measures in the context of CAI. While the interplay of privacy, security and trust perceptions has been long discussed in the context of new technology, we uniquely contribute to the literature by providing a joint perspective on privacy, security and trust perception measurements and by focusing on CAI systems including both voice- and text-based systems. 

First, our review sheds light on methods used to research users' perceptions in the context of CAI (\textbf{RQ1}). We found that quantitative research methods in particular dominate the field of trust perceptions while privacy and security perceptions are - in addition to quantitative methods - frequently explored through qualitative and mixed-methods studies. Given that privacy and trust perceptions are highly contextual, qualitative and mixed-method studies might be particularly helpful in exploring why people respond a certain way~\citep{Barkhuus2012}. Moreover, we extensively discussed the tools applied by researchers to expose participants to conversational assistants. While we found a significant sample of reviewed papers to implement applications or develop prototypes, the majority relied on peoples' previous experiences with CAI systems or scenario-based experiments to evaluate privacy, security and trust perceptions. Moreover, we found strong reliance on commercially available systems when implementing prototypes which could impose research constraints and hinder research on innovative solutions. This is especially true since the brand or company behind commercially available systems might have a significant influence on privacy and trust perceptions. Thus, our findings suggest that the field could generally profit from the diversification of research methods and research-friendly methods to implement CAI systems.

Second, we were interested in which constructs are assessed and which scales are used when users' perceptions of privacy, security and trust are investigated in the context of CAI systems (\textbf{RQ2 and RQ3}). Our review shows that the most prevalent constructs in this domain are trust, privacy concerns, anthropomorphic aspects and risk perceptions. Nevertheless, our extensive analysis shows that various sub-constructs are commonly assessed such as various types of privacy concerns or trust. While a more fine-grained assessment can enhance our understanding of people's perceptions, a clear distinction between these concepts might not be reflected in the associated scales possibly leading to conflicting research outcomes~\citep{colnagoItConcernPreferencea}. At the same time, constructs are referred to interchangeably increasing the risk of unreliable outcomes. Privacy, security and trust research could benefit from grounding their assessment into already-existing theoretical frameworks and standards on definitions of privacy, security and trust. 

Next, we were interested in the reported reliability and validity scores for privacy, security and trust metrics (\textbf{RQ4}). Therefore, we deployed a previously developed framework for scale validation and found significant differences between the fields of privacy \& security and trust. Our findings reveal that reliability is most frequently assessed when compared to validity, particularly when conceptual models are evaluated using structural equation modelling. Therefore, we urge the research community to follow best practices of scale validation whenever scales are developed, reused or adapted. 

Furthermore, our literature review focused on a joint investigation of privacy, security and trust perception measures (\textbf{RQ5}). Given previous research on the interplay of privacy, security and trust perceptions in online environments, we were interested to which extent these constructs have been jointly researched in the context of CAI. Surprisingly, we found only two papers that were identified by both our literature searches. Despite this finding, we discovered multiple papers jointly investigating privacy, security and trust perceptions, particularly often in the context of CAI adoption. This shows that current research on CAI adoption implicitly assumes a joint influence of these perceptions, yet, lacks explicit evaluation. Moreover, our research shows that a better understanding of the relationship between privacy, security and trust perceptions is required beyond adoption to inform the design of trustworthy and privacy-preserving CAI systems. 

Finally, we were interested in which topics were investigated whenever privacy, security and trust perceptions were researched (\textbf{RQ6}). We found that a majority of papers investigated users' perceptions in relation to CAI adoption. Thereby, researchers relied on a multitude of conceptual frameworks and theories such as TAM or UTAUT. \changed{Despite the varying extensions explored in the reviewed studies, the prevalent use of TAM as an underlying research model is concerning. TAM does not explicitly model factors related to privacy, security and trust and the increased addition of factors makes comparability challenging. Conceptual models that integrate traditional technology acceptance frameworks with privacy and trust factors, such as the privacy-calculus acceptance models~\citep{schomakers2022role}, are essential for advancing research in the field and for the development of more complex and comprehensive frameworks while maintaining comparability.} While investigated privacy and trust constructs were mostly limited to privacy risks and trust in general, they were treated non-uniformly as independent, moderating, mediating or dependent factors. Therefore, it is not surprising that results were found conflicting. Nevertheless, privacy and security risks were commonly treated as inhibitors while trust was expected to drive adoption. Future research could benefit from streamlining investigations and exploring factors beyond privacy risks and trust to better understand their impact on CAI adoption. Moreover, it will be crucial to further investigate factors beyond adoption.

\section{Future Research Directions}

We now want to highlight additional future research directions that have not yet been demonstrated and discussed in the previous section. 

\subsection{Diversification of Application Fields and User Groups}

Through our literature review, we identified application fields and user groups that have been investigated by the research community. We show that the majority of papers investigate CAI systems without a specific application focus. However, as privacy and trust perceptions are highly contextual, the application field can severely impact users' perceptions. Thus, research conducted in a general setting, might not be easily transferable to highly sensitive application fields such as finance. Future research should therefore diversify investigations across application fields to better understand users' privacy, security and trust perceptions in context. Furthermore, future work could investigate application fields that have not yet been covered such as exploring users' perceptions of CAI systems in the public sector or public places. Moreover, we found that comparisons of users' perceptions across multiple application fields is a largely unexplored area. 

Next, our literature review sheds light on investigated user groups and participants. Generally, a strong reliance on crowdworking platforms was observable which can come with reliability and generalizability issues. The research community should critically discuss this trend and rely on traditional research and recruitment methods besides crowdsourcing. On the other hand, research on privacy and security perceptions has frequently focused on technology users while research on trust perception did not. In general, research on trust perceptions could benefit from the diversification of user groups, while the privacy and security field should intensify its efforts in recruiting different types of users. 

\subsection{Need for Long-Term Evaluation}

In our literature research on privacy and security, we discovered only one paper that investigated users' perceptions throughout two months~\citep {choiUseInternetofThingsSmart2020}. Long-term studies may be more complex and time-consuming, yet, can provide valuable insights compared to studies capturing snapshots of users' perceptions. Particularly, as privacy and trust perceptions are known to change over a period of time, long-term studies are vital for enhancing research in this domain. Thereby, researchers in the CAI domain could benefit from other research fields that frequently conduct long-term evaluations such as human-robot-interaction~\citep{fernaeus2010you, de2016long}.  

\subsection{Towards a unified privacy, security and trust scale}

Previous work shows that well-established scales exist to measure privacy, security and trust perceptions~\citep{mayer1995integrative, jianFoundationsEmpiricallyDetermined2000a, lankton2015technology, preibuschGuideMeasuringPrivacy2013}. Yet, our literature research revealed that they are only rarely used to assess users' perceptions in the context of CAI. Therefore, it remains an open question whether they are sufficiently reliable and valid in the context of CAI or whether CAI-specific metrics are needed to evaluate privacy, security and trust perceptions. For example, the usability field has seen the emergence of CAI-specific usability scales to adapt to the unique capabilities of CAI~\citep{zwakmanUsabilityEvaluationArtificial2021c}. Future research could compare and validate multiple privacy, security and trust perception scales in the context of CAI and assess their suitability. 

On a more general note, given the diversity of constructs in privacy, security and trust research, it might be worthwhile to explore adaptable and modular questionnaires to assess perceptions. Similar approaches exist to measure various aspects of usability of traditional interfaces and voice user interfaces~\citep{SchreppUEQ}. While they include aspects of trust, trustworthiness of content and risk handling, privacy and security constructs are not yet covered. A modular questionnaire is especially useful if surveys need to be adapted to particular research questions. As discovered in this literature review, adaptation of scales and various combinations of privacy, security and trust constructs are common when evaluating users' perceptions in the context of CAI. Therefore, future research could assess the suitability of a modular questionnaire for researching privacy, security and trust perceptions, identify relevant constructs and provide additional contextual information to ensure appropriate usage. 

\subsection{Distinction between Privacy and Security Perceptions}

Our literature research revealed that security perceptions are among the less often researched constructs, predominately in the context of security attacks or authentication methods. Moreover, our reviewed papers draw an inconclusive picture of whether people can sufficiently distinguish between privacy and security. Given people's confusion around security and privacy, we argue that a product's security should first and foremost be assessed objectively, e.g. by metrics and certification. We encourage the currently visible trend in that measurements of security perceptions may only be relevant to be assessed for a few applications e.g. when users are actively asked to test secure protection mechanisms or authentication tools. In contrast, we see the need for a clear distinction between security and privacy risks and peoples' perceptions thereof.

\subsection{Impact of Modality}

In our literature review, we explicitly focused on text-based as well as voice-based conversational AI systems to cover the full range of disembodied systems interacting with humans in natural language. Nevertheless, we found that a majority of papers investigated voice-based CAI systems, particularly in the context of privacy and security perceptions. While it is obvious that voice-based systems come with increased privacy risks due to their constant listening capabilities, text-based systems can raise similar concerns regarding data handling and are worthwhile exploring. In addition, we discovered only a few papers that researched the impact of modality on users' privacy, security and trust perceptions. As differences in modalities did not always result in differences in user perceptions, further research is necessary to enhance understanding of the possible transferability of outcomes from one modality to another. Moreover, research on trust perceptions has investigated CAI systems in virtual reality - a modality that was found unexplored in the field of privacy and security. 

\subsection{Impact of Privacy Controls and Authentication Methods}

Our literature review revealed only a few papers that evaluated privacy controls or authentication methods and their influence on users' perceptions. While we acknowledge that papers researching privacy controls for CAI systems, might not investigate privacy, security or trust perceptions and might therefore not be covered in our literature search, the limited number of papers in this domain is still surprising. In particular, none of the papers on privacy controls or authentication methods investigated users' trust. Therefore, it remains unclear to which extent privacy controls and privacy-enhancing features can positively influence users' trust in the context of CAI systems. Given peoples' increasing exposure to CAI systems, we urge the need for research on privacy controls in this domain and their impact on users' behaviour and perceptions.

\subsection{Impact of LLMs}

The vast increase in the use of Large Language Models (LLM) and generative AI for CAI systems is foreseeable to influence users' perceptions. So far, our literature review has revealed only little research on privacy, security and trust perceptions in CAI systems leveraging LLM capabilities. Only one study was found that investigated the usage and perceptions of ChatGPT in an educational context~\citep{tliliWhatIfDevil2023a}. However, it has to be noted that we conducted our literature research in the first half-year of 2023 and considered only peer-reviewed articles. Given the fast-changing and evolving field, it is likely that a literature search conducted at a later point had resulted in a different outcome. On the other hand, the rise of LLMs might make research on the influence of privacy and trust more difficult as these systems are blackbox models with limited transparency, controllability and reproducibility. Thus, studies in which only one variable is controlled might produce different outcomes due to the use of generative models. This makes the assessment of voice assistants or chatbots which use LLMs more difficult. For researchers, the training and fine-tuning of these models also become more expensive as larger amounts of processing power need to be available for fine-tuning and additional training~\citep{kaddour2023challenges}.

\section{Limitations}

While our paper provides valuable insights into the use and development of privacy, security and trust scales in the context of CAI, we acknowledge several limitations of our work. First, from the multitude of names for voice assistants and no common naming convention, it is obvious that the literature review might be incomplete as other terms for CAI might not have been included in our keyword search. While we aimed to cover the most prominent terms through multiple iterations, generative models could potentially help in refining the keyword search to include less often used search terms. Secondly, we limited this work to disembodied conversational AI systems, thereby neglecting embodied systems, in particular robots. This can be a drawback since there has been extensive research on the social interaction between robots and humans. Future work could extend this review to include human-robot interaction and compare findings between embodied and disembodied agents. While we provided a summary of qualitative studies found by our literature search, our work has a strong focus on privacy, security and trust perception scales mostly used in quantitative research and mixed-methods studies. Similarly, behavioral measures or bio-signals were not covered in this work, as this work focused on the evaluation of users' perceptions. Nevertheless, measures of biological signals have been identified throughout this literature review such as \cite{gupta2020measuring}, opening possibilities for future endeavours.

\section{Conclusion}

In this paper, we have conducted a systematic literature review on users' privacy \& security perceptions and trust perceptions in the context of conversational AI. We specifically included both, voice-based as well as text-based systems. Our analysis includes an investigation of meta-information such as author's affiliation, participants, and researched devices as well as a breakdown of researched constructs. Moreover, we provide an overview of topics researched alongside privacy, security and trust perceptions. We found that a majority of studies focus on TAM models when researching privacy and trust perceptions. These are also the studies that include both concepts, privacy \& security and trust at the same time. Further, we identified future research directions specifically urging the community to explicitly investigate the relationship between trust and privacy perceptions. 

\section*{Declaration of competing interest}
The authors declare that they have no known competing financial
interests or personal relationships that could have appeared to influence
the work reported in this paper.

\section*{Data Availability}
Data will be made available on request.

\balance

\section*{Appendix}
\appendix

\section{1}

\newpage
\onecolumn
\scriptsize
\begin{longtable}{p{0.08\textwidth}p{0.05\textwidth}p{0.25\textwidth}p{0.2\textwidth}p{0.05\textwidth}p{0.05\textwidth}p{0.05\textwidth}}
    \caption{\changed{Overview of privacy and security-related constructs used in mixed and quantitative studies. We display whether studies reused scales by showing the corresponding reference. We compute reliability and validity scores based on \cite{rohanSystematicLiteratureReview2023} as described in Section~\ref{sec:ReliabilityValidityScales}. ``n.a'' is used whenever studies did not rely on multi-item scales and thus, reliability and validity assessment was not possible.}} \\
    \toprule
    Construct & Frequency & Reference Paper & Reference Construct & \changed{Reliability Testing} & \changed{Convergent Validity} & \changed{Discriminant Validity}\\
    \midrule 
    
    \multirow{3}{*}{Privacy} & \multirow{3}{*}{3} & Privacy \cite{furiniUsageSmartSpeakers2020b} & not reported & n.a. & n.a. & n.a.\\
    && Privacy \cite{kreyVoiceAssistantsHealthcare2022b} & not reported & n.a. & n.a. & n.a.\\
    && Privacy  \cite{patilCanWeTrust2022a} & \cite{faqihAssessingModeratingEffect2015} & 1.0 & 1.0 & 1.0 \\

    \midrule
    
    Privacy Perception & 1 & Privacy Perception \cite{bruggemeierPerceptionsReactionsConversational2022b} & \cite{casaloRoleSecurityPrivacy2007b} & 0.0 & 0.0 & 0.0\\

    \midrule 
    
    \multirow{40}{*}{\shortstack{Privacy \\ Concerns}} & \multirow{40}{*}{40} & General Privacy Concerns \cite{Mols2021} & \cite{vitakBalancingPrivacyConcerns} & 0.5 & 0.0 & 0.0 \\
    & & Mobile Privacy Concerns \cite{Mols2021}  & \cite{xuMeasuringMobileUsers2012} & 0.5 & 0.0 & 0.0 \\
    & & Privacy Concerns \cite{uysalTrojanHorseUseful2022a} & \cite{smithInformationPrivacyMeasuring1996b} & 1.0 & 1.0 & 1.0 \\
    & & General Privacy Concerns \cite{liaoUnderstandingRolePrivacy2019b} & \cite{vitakDigitalPathHappiness2016} & 0.5 & 0.0 & 0.0 \\
    & & Privacy Concerns \cite{choWillDeletingHistory2020e} & \cite{dinev2006privacy} & 0.5 & 0.0 & 0.0 \\
    & & Privacy Concerns \cite{alghamdiUserStudyEvaluate2022a} & not reported & 0.5 & 0.0 & 0.0 \\
    & & Privacy Concerns \cite{joyInvestigatingUsersPrivacy2022b} & not reported & n.a. & n.a. & n.a. \\
    & & Device Privacy Concerns \cite{lutzPrivacySmartSpeakers2021} & not reported & 0.5 & 0.0 & 0.0 \\
    & & Household Member Privacy Concerns \cite{lutzPrivacySmartSpeakers2021} & not reported  & 0.5 & 0.0 & 0.0 \\
    & & Stranger Privacy Concerns \cite{lutzPrivacySmartSpeakers2021} & not reported & 0.5 & 0.0 & 0.0\\
    & & Company Privacy Concerns \cite{lutzPrivacySmartSpeakers2021} & not reported & 0.5 & 0.0 & 0.0 \\
    & & Contractor Privacy Concerns \cite{lutzPrivacySmartSpeakers2021} & not reported & 0.5 & 0.0 & 0.0 \\
    & & Third-Party Privacy Concerns  \cite{lutzPrivacySmartSpeakers2021} & not reported & 0.5 & 0.0 & 0.0 \\
    & & Government Privacy Concerns \cite{lutzPrivacySmartSpeakers2021} & not reported & 0.5 & 0.0 & 0.0 \\
    & & Privacy Concerns \cite{ngSimulatingEffectsSocial2020a} & \cite{ischenPrivacyConcernsChatbot2020b} & 0.5 & 0.0 & 0.0 \\
    & & Privacy Concerns for Multiple Users \cite{linTransferabilityPrivacyrelatedBehaviours2020a} & not reported & n.a. & n.a. & n.a. \\
    && Privacy Concerns  \cite{dekkalFactorsAffectingUser2023a} & \cite{malhotra2004internet} & 0.5 & 1.0 & 1.0 \\
    && Privacy Concerns  \cite{buteauHeyAlexaWhy2021} & \cite{dinevInternetPrivacyConcerns2005a}  & 1.0 & 0.5 & 1.0 \\
    && Perceived Privacy Concerns  \cite{vimalkumarOkayGoogleWhat2021c} & \cite{alalwanMobileFoodOrdering2020,bansalContextPersonalityMatter2016, smithInformationPrivacyResearch2011,smithInformationPrivacyMeasuring1996b} & 1.0 & 1.0 & 1.0 \\
    && Privacy Concern  \cite{pitardi2021alexa} & \cite{mcleanHeyAlexaExamine2019a} & 1.0 & 1.0 & 1.0 \\
    && Perceived Privacy Concerns  \cite{palEffectTrustIts2022a} & \cite{smithInformationPrivacyResearch2011, alalwanMobileFoodOrdering2020} & 1.0 & 1.0 & 1.0 \\
    && Privacy Concerns \cite{abdiPrivacyNormsSmart2021a} & \cite{malhotra2004internet} & 0.0 & 0.0 & 0.0 \\
    && Privacy Concerns  \cite{choHeyGoogleCan2019b} & \cite{dinev2006privacy} & 0.5 & 0.0 & 0.0 \\
    && VAPA Privacy Concern  \cite{pal2022perceptions} & \cite{kowalczukConsumerAcceptanceSmart2018a} & 1.0 & 1.0 & 1.0 \\
    && Household Member Privacy Concern  \cite{pal2022perceptions} & \cite{kowalczukConsumerAcceptanceSmart2018a} & 1.0 & 1.0 & 1.0 \\
    && Vendor \& Third-Party Privacy Concern  \cite{pal2022perceptions} & \cite{kowalczukConsumerAcceptanceSmart2018a} & 1.0 & 1.0 & 1.0 \\
    && Government Privacy Concern  \cite{pal2022perceptions} & \cite{kowalczukConsumerAcceptanceSmart2018a} & 1.0 & 1.0 & 1.0\\
    && Privacy Concerns  \cite{ischenPrivacyConcernsChatbot2020b} & \cite{xuExaminingFormationIndividual2008} & 0.5 & 0.0 & 0.0 \\
    && Privacy Concerns  \cite{xuSmartSpeakersRequire2022} & \cite{metzger2007making} \cite{quinn2016we}  & 0.5 & 0.0 & 0.0 \\
    && Privacy Concerns  \cite{lappemanTrustDigitalPrivacy2023a} & \cite{malhotra2004internet} & 1.0 & 1.0 & 1.0 \\ 
    && Privacy Concerns  \cite{lv2022research} & \cite{mcleanHeyAlexaExamine2019a} & 1.0 & 1.0 & 1.0\\
    && Privacy Concerns \cite{manikondaWhatPrivacyUser2018a} & not reported & n.a. & n.a. & n.a. \\
    && Privacy Concerns \cite{malkinPrivacyAttitudesSmart2019} & not reported & n.a. & n.a. & n.a. \\
    && Privacy Concerns \cite{abrokwaComparingSecurityPrivacya} & \cite{Gross2021ValidityAR} & 0.5 & 0.0 & 0.0 \\
    && Privacy Concerns \cite{seymourIgnoranceBlissEffect} & \cite{malhotra2004internet} & 0.0 & 0.0 & 0.0 \\
    && \noindent\changed{Privacy Concerns \cite{farooq2022voice}} & \noindent\changed{\cite{kim2011effect}} & 0.5 & 0.0 & 0.0 \\
    && \noindent\changed{Privacy Concerns \cite{reseChatbotsRetailersCustomer2020a}} & \noindent\changed{\cite{rauschnabel2017adoption}} & 1.0 & 1.0 & 1.0 \\
    && \noindent\changed{Privacy Concerns \cite{park2021users}} & \noindent\changed{\cite{smithInformationPrivacyMeasuring1996b}} & 1.0 & 1.0 & 1.0 \\
    && \noindent\changed{Perceived Privacy Concerns \cite{munoz2023voice}} & \noindent\changed{\cite{pavlou2003consumer}} & 0.5 & 0.0 & 0.0 \\
    && \noindent\changed{Perceived Privacy Concerns \cite{kefi2021privacy}} & \noindent\changed{\cite{xuExaminingFormationIndividual2008}} & 0.5 & 1.0 & 1.0 \\

    \midrule 

    \multirow{7}{*}{Other Concerns} & \multirow{7}{*}{7} & Household IPA surveillance concerns \cite{Mols2021} & not reported & 0.5 & 0.0 & 0.0 \\
    & & Household IPA security concerns \cite{Mols2021} & not reported & 0.5 & 0.0 & 0.0  \\
    & & Household IPA platform concerns \cite{Mols2021} & not reported & 0.5 & 0.0 & 0.0 \\
    & & IPA Data Concerns \cite{liaoUnderstandingRolePrivacy2019b} & not reported & 0.5 & 0.0 & 0.0 \\
    & & Mobile Data Concerns \cite{liaoUnderstandingRolePrivacy2019b} & \cite{xuMeasuringMobileUsers2012} & 0.5 & 0.0 & 0.0 \\
    && Privacy and Security Concerns   \cite{alagarsamy2023exploring} & \cite{son2008internet} & 1.0 & 1.0 & 1.0 \\
    && \noindent\changed{Security and Privacy Concerns \cite{shlega2022users}} & \noindent\changed{\cite{van2016development}} & 0.5 & 0.0 & 0.0 \\

    \midrule
    
    \multirow{8}{*}{Control} & \multirow{8}{*}{8} & Perceptions of Control \cite{linTransferabilityPrivacyrelatedBehaviours2020a} & not reported & n.a. & n.a. & n.a. \\
    && Previous Privacy Control Usage \cite{linTransferabilityPrivacyrelatedBehaviours2020a} & not reported & n.a. & n.a. & n.a. \\
    && Smart Assistant Privacy Control Usage \cite{linTransferabilityPrivacyrelatedBehaviours2020a} & not reported & n.a. & n.a. & n.a. \\
    && Privacy Control \cite{alghamdiUserStudyEvaluate2022a} & not reported  & 0.5 & 0.0 & 0.0 \\
    && Controllability   \cite{purwanto2020interactive} & \cite{brill2022siri} & 1.0 & 1.0 & 1.0 \\
    && Perceived Control \cite{pal2020personal} & \cite{keith2013information, mamonov2018impact} & 1.0 & 1.0 & 0.0 \\
    && Perceived Control \cite{ahmadTangiblePrivacySmart2022} & \cite{hinds1998user} & 0.0 & 0.0 & 0.0 \\
    && \noindent\changed{Privacy Boundary Control \cite{kang2023communication}} & \noindent\changed{\cite{childBloggingCommunicationPrivacy2009}} & 0.5 & 0.0 & 0.0 \\

    \midrule

    \multirow{6}{*}{Disclosure} & \multirow{6}{*}{6} &
    Willingness to self-disclose  \cite{kunkel2019let} & \cite{ha2021exploringWillingnessToDisclose} & 0 & 0 & 0\\
    && Willingness to self-disclose  \cite{choi2023inducing} & \cite{ha2021exploringWillingnessToDisclose} & 0.5 & 0 & 0 \\
    && Willingness  \cite{gulati2018modelling} & \cite{pavlou2004building} & 1.0 & 1.0 & 1.0 \\
    && Personal Information Disclosure   \cite{pal2020personal} & & 1.0 & 1.0 & 0.0 \\
    && User self-disclosure   \cite{lappemanTrustDigitalPrivacy2023a} & & 1.0 & 1.0 & 1.0 \\
    && \noindent\changed{Privacy Disclosure \cite{kang2023communication}} & \noindent\changed{\cite{childBloggingCommunicationPrivacy2009}} & 0.5 & 0.0 & 0.0 \\

    \midrule
    \midrule
    \multirow{2}{*}{Security} & \multirow{2}{*}{2} & General Security \cite{mccarthyShoutingLetterboxesStudy2020b} & not reported & n.a. & n.a. & n.a. \\
    && Device Security \cite{kreyVoiceAssistantsHealthcare2022b} & not reported & n.a. & n.a. & n.a. \\

    \midrule 
    
    \multirow{9}{*}{\shortstack{Security \\ Perception}} & \multirow{9}{*}{9} & Perceived Security \cite{renzAuthenticationMethodsVoice2023} & not reported & n.a. & n.a. & n.a. \\
    & & Perceived Security \cite{choWillDeletingHistory2020e} & \cite{salisburyPerceivedSecurityWorld2001} & 0.5 & 0.0 & 0.0 \\
    & & Security Perception \cite{rajapakshaFieldStudyUsability2021b} & not reported & n.a. & n.a. & n.a. \\
    & & Security Perception \cite{bruggemeierPerceptionsReactionsConversational2022b} & \cite{casaloRoleSecurityPrivacy2007b} & 0.0 & 0.0 & 0.0 \\
    && Perceived Security  \cite{buteauHeyAlexaWhy2021} & \cite{flavianConsumerTrustPerceived2006} & 1.0 & 0.5 & 1.0 \\
    && Perceived Security  \cite{fahnTimeGetConversational2021a} & \cite{chengAdoptionInternetBanking2006a} & 0.5 & 0.0 & 0.0 \\
    && Perceived Security  \cite{leeHeyAlexaMagic2020a} & \cite{chengAdoptionInternetBanking2006a, lallmahamoodExaminationIndividualAPerceived1970, oliveiraMobilePaymentUnderstanding2016} & 1.0 & 1.0 & 1.0 \\
    && \noindent\changed{Perceived Security \cite{aw2022alexa}} & \noindent\changed{\cite{chengAdoptionInternetBanking2006a}} & 0.5 & 1.0 & 1.0  \\
    && \noindent\changed{Perceived Security \cite{lee2021continuation}} & \noindent\changed{\cite{chengAdoptionInternetBanking2006a}} & 1.0 & 1.0 & 1.0 \\

    \midrule
    \midrule
    
    \multirow{29}{*}{Risk Perception} & \multirow{29}{*}{29} & Risk Perception \cite{emami-naeiniWhichPrivacySecurity2021b} & not reported & n.a. & n.a. & n.a. \\
    && Perceived Security Risk  \cite{Shofolahan2018} & \cite{hsuExploringFactorsAffecting2018} & 1.0 & 1.0 & 1.0 \\
    && Perceived Privacy Risk  \cite{garciadeblanessebastianApplicationExtensionUTAUT22022} & \cite{featherman2003predicting} & 0.5 & 1.0 & 1.0 \\ 
    && Safety Risk  \cite{patilCanWeTrust2022a} & \cite{supletCustomerPerceptionsPerceived2009} & 1.0 & 1.0 & 1.0 \\
    && Perceived Risk  \cite{beheraCognitiveChatbotPersonalised2021a} & \cite{trivedi2019examining} & 1.0 & 1.0 & 1.0 \\
    && Perceived Privacy Risk  \cite{mcleanHeyAlexaExamine2019a} & AlDebei2014 - Referenced Paper not found & 0.5 & 1.0 & 1.0 \\
    && Perceived Privacy Risk  \cite{aiolfiHowShoppingHabits2023} & \cite{mcleanHeyAlexaExamine2019a} & 1.0 & 1.0 & 1.0 \\
    && Perceived Privacy Risk  \cite{vimalkumarOkayGoogleWhat2021c} & \cite{alalwanMobileFoodOrdering2020, bansalContextPersonalityMatter2016, smithInformationPrivacyResearch2011, smithInformationPrivacyMeasuring1996b} & 1.0 & 1.0 & 1.0 \\
    && Perceived Privacy Risk  \cite{kwangsawadOvercomingCustomerInnovation2022} & \cite{jattamartPerspectivesSocialMedia2020, reseChatbotsRetailersCustomer2020a} & 1.0 & 1.0 & 1.0 \\
    && Perceived Privacy Risk  \cite{chaSustainabilityCalculusAdopting2021} & \cite{treptePrivacyCalculusContextualized2020, krasnovaSelfdisclosurePrivacyCalculus2012, kimWillingnessProvidePersonal2019b} & 1.0 & 1.0 & 1.0 \\
    && Security/Privacy Risk  \cite{kowalczukConsumerAcceptanceSmart2018a} & \cite{yangUserAcceptanceSmart2017} & 1.0 & 1.0 & 1.0 \\
    && Perceived Risk  \cite{kasilingamUnderstandingAttitudeIntention2020a} & \cite{featherman2003predicting} & 1.0 & 1.0 & 1.0 \\
    && Perceived Privacy Risk  \cite{palEffectTrustIts2022a} & \cite{smithInformationPrivacyResearch2011, alalwanMobileFoodOrdering2020} & 1.0 & 1.0 & 1.0 \\
    && Privacy Risk  \cite{guerreiroAdvertisingAcceptanceSmart2022} & \cite{mcleanHeyAlexaExamine2019a} & 1.0 & 1.0 & 1.0 \\
    && Perceived Privacy Risk  \cite{chengHowAIdrivenChatbots2020} & \cite{eeuwenMobileConversationalCommerce2017} & 1.0 & 1.0 & 1.0 \\
    && Security/Privacy Risk  \cite{hanUnderstandingAdoptionIntelligent2018a} & \cite{yangUserAcceptanceSmart2017} & 1.0 & 1.0 & 1.0 \\
    && Perceived Risk   \cite{hasanConsumerTrustPerceived2021} & \cite{zhouImpactPrivacyConcern2011} & 1.0 & 1.0 & 1.0 \\
    && Perceived Risk  \cite{hsu2023semantic} & \cite{xu2009effects} & 1.0 & 1.0 & 1.0  \\ 
    && Risk  \cite{nordheim2019initialtrust} & \cite{corritore2005measuring} & 0.5 & 1.0 & 0.0\\
    && Privacy Risk  \cite{prakash2023determinants} & \cite{featherman2003predicting} & 0.5 & 1.0 & 1.0  \\
    && Risk Perception \cite{ahmadTangiblePrivacySmart2022} & \cite{gulati2018modelling, colesca2009understanding} & 0.0 & 0.0 & 0.0 \\
    && \noindent\changed{Perceived Privacy Risk \cite{kang2023communication}} & \noindent\changed{\cite{xu2011information}} & 0.5 & 0.0 & 0.0 \\
    && \noindent\changed{Perceived Personal Data Collection Risk \cite{patrizi2021talking}} & \noindent\changed{\cite{mcleanHeyAlexaExamine2019a}} & 1.0 & 0.0 & 0.0 \\
    && \noindent\changed{Perceived Personal Data Misuse Risk \cite{patrizi2021talking}} & \noindent\changed{\cite{dinevInternetPrivacyConcerns2005a}} & 1.0 & 0.0 & 0.0 \\
    && \noindent\changed{Privacy Risk \cite{cao2022adoption}} & \noindent\changed{\cite{mcleanHeyAlexaExamine2019a}} & 1.0 & 1.0 & 1.0 \\
    && \noindent\changed{Perceived Privacy Risk \cite{alt2021identifying}} & \noindent\changed{\cite{yang2015understanding}} & 1.0 & 0.0 & 0.0 \\
    && \noindent\changed{Perceived Privacy Risk \cite{maroufkhani2022interactive}} & \noindent\changed{\cite{yangUserAcceptanceSmart2017}} & 1.0 & 1.0 & 1.0 \\
    && \noindent\changed{Perceived Privacy Risk \cite{jain2022interactive}} & \noindent\changed{\cite{yangUserAcceptanceSmart2017}} & 1.0 & 1.0 & 1.0 \\
    && \noindent\changed{Perceived Privacy Risk \cite{song2022should}} & \noindent\changed{\cite{song2022will}} & 1.0 & 1.0 & 1.0 \\
  
    \midrule
    \midrule

    \multirow{34}{*}{\makecell{Other Related \\ Constructs}} & \multirow{34}{*}{34} & Privacy Importance \cite{alghamdiUserStudyEvaluate2022a} & not reported & 0.5 & 0.0 & 0.0 \\
    && Privacy Preferences \cite{alghamdiUserStudyEvaluate2022a} & not reported & 0.5 & 0.0 & 0.0 \\
    && Privacy Preference \cite{ahmadTangiblePrivacySmart2022} & not reported & n.a. & n.a. & n.a. \\
    && Privacy Scores \cite{joyInvestigatingUsersPrivacy2022b} & not reported & n.a. & n.a. & n.a. \\

    && Privacy Self-Efficacy \cite{ahmadTangiblePrivacySmart2022} & \cite{zeissig2017online} & 0.5 & 0.0 & 0.0 \\
    && \noindent\changed{Privacy Self-Efficacy \cite{kang2023communication}} & \noindent\changed{\cite{chen2018revisiting, krasnova2010online}} & 0.5 & 0.0 & 0.0 \\   
    
    && Privacy Setting Review  \cite{xuSmartSpeakersRequire2022} & \cite{childBloggingCommunicationPrivacy2009, John2019, Norton2020} & 0.5 & 0.0 & 0.0 \\
    && Ownership Protection  \cite{xuSmartSpeakersRequire2022} & \cite{childBloggingCommunicationPrivacy2009, John2019, Norton2020} & 0.5 & 0.0 & 0.0 \\
    && \noindent\changed{Security and Privacy Behaviors \cite{shlega2022users}} & \noindent\changed{\cite{haney2019perceptions, yaoPrivacyPerceptionsDesigns2019d, zeng2017end}} & 0.5 & 0.0 & 0.0\\
    && \noindent\changed{Coping Behavior \cite{park2021users}} & \noindent\changed{\cite{son2008internet}} & 1.0 & 1.0 & 1.0 \\

    && Trust in Privacy \cite{cannizzaro2020trust} & not reported & 1.0 & 1.0 & 1.0 \\
    && Trust in Security \cite{cannizzaro2020trust} & not reported & 1.0 & 1.0 & 1.0 \\
   
    && Privacy Awareness \cite{ahmadTangiblePrivacySmart2022} & \cite{zeissig2017online} & 0.5 & 0.0 & 0.0 \\
        
    && \noindent\changed{Privacy Boundary Linkage \cite{kang2023communication}} & \noindent\changed{\cite{childBloggingCommunicationPrivacy2009}} & 0.5 & 0.0 & 0.0 \\
    && \noindent\changed{Perceived Protection of Data \cite{shlega2022users}} & \noindent\changed{not reported} & 0.5 & 0.0 & 0.0 \\
    && Threat to Human Identity \cite{uysalTrojanHorseUseful2022a} & \cite{ferrariBlurringHumanMachine2016, mendeServiceRobotsRising2019} & 1.0 & 1.0 & 1.0 \\
    && IPA Data Confidence \cite{liaoUnderstandingRolePrivacy2019b} & not reported & 0.5 & 0.0 & 0.0 \\
    && Data Protection \cite{kreyVoiceAssistantsHealthcare2022b} & not reported & n.a. & n.a. & n.a.\\    
    
    && Perceived Level of Comfort \cite{tabassumInvestigatingUsersPreferences2019a} & not reported & n.a. & n.a. & n.a.  \\
    && \noindent\changed{Comfort Level \cite{easwara2015privacy}} & \noindent\changed{not reported} & n.a. & n.a. & n.a. \\
    && Level of Comfort with Information Disclosure \cite{ischenPrivacyConcernsChatbot2020b} & \cite{croes2021can, ledbetter2009measuring} & 0.5 & 0.0 & 0.0 \\
    
    && Security Attitudes \cite{abdiPrivacyNormsSmart2021a} & \cite{faklaris2019self} & 0.0 & 0.0 & 0.0 \\
    && Security Attitudes \cite{abrokwaComparingSecurityPrivacya} & \cite{faklaris2019self} & 0.5 & 0.0 & 0.0 \\
    && Security Attitudes \cite{seymourIgnoranceBlissEffect} & \cite{faklaris2019self} & 0.0 & 0.0 & 0.0 \\
    
    && Acceptability \cite{abdiPrivacyNormsSmart2021a} & not reported & 0.0 & 0.0 & 0.0 \\
    && Acceptability \cite{malkinPrivacyAttitudesSmart2019} & not reported & n.a. & n.a. & n.a. \\
    
    && Creepiness \cite{dekkalFactorsAffectingUser2023a} & \cite{langerIntroducingTestingCreepiness2018} & 0.5 & 1.0 & 1.0 \\
    && Recording Sensitivity \cite{tabassumInvestigatingUsersPreferences2019a} & not reported & n.a. & n.a. & n.a. \\
    && Privacy Cynicism  \cite{acikgozRolePrivacyCynicism2022} & \cite{choiRolePrivacyFatigue2018} & 1.0 & 1.0 & 1.0 \\
    && Surveillance Anxiety  \cite{kowalczukConsumerAcceptanceSmart2018a} & \cite{kummerTechnologyinducedAnxietyManifestations2017} & 1.0 & 1.0 & 1.0 \\
    && Skepticism \cite{abrokwaComparingSecurityPrivacya} & \cite{Pew2019} & 0.5 & 0.0 & 0.0 \\
    && Security and Privacy Knowledge \cite{abrokwaComparingSecurityPrivacya} & \cite{Pew2019} & 0.5 & 0.0 & 0.0 \\
    && Perceived Reliability \cite{ahmadTangiblePrivacySmart2022} & \cite{madsen2000measuring} & 0.0 & 0.0 & 0.0 \\
    && Perceived Intrusiveness \cite{pal2022perceptions} & \cite{lucia2021effects} & 1.0 & 1.0 & 1.0 \\

\bottomrule
\label{tab:TypePrivacyConstructs}
\end{longtable}

\pagebreak

\scriptsize
\begin{longtable}{p{0.08\textwidth}p{0.05\textwidth}p{0.25\textwidth}p{0.2\textwidth}p{0.05\textwidth}p{0.05\textwidth}p{0.05\textwidth}}
    \caption{\changed{Overview of trust-related constructs used in mixed and quantitative studies. We display whether studies reused scales by showing the corresponding reference. We compute reliability and validity scores based on \cite{rohanSystematicLiteratureReview2023} as described in Section~\ref{sec:ReliabilityValidityScales}. ``n.a'' is used whenever studies did not rely on multi-item scales and thus, reliability and validity assessment was not possible.}} \\
    \toprule
    Construct & Frequency & Reference Paper & Reference Construct & \changed{Reliability Testing} & \changed{Convergent Validity} & \changed{Discriminant Validity}\\
    \midrule 
        
    \multirow{55}{*}{Trust} & \multirow{55}{*}{55} & Trust in AIA \cite{uysalTrojanHorseUseful2022a} & \cite{leachGroupVirtueImportance2007} & 1.0 & 1.0 & 1.0 \\
    && Trust toward Alexa \cite{choWillDeletingHistory2020e} & \cite{kohEffectsSpecializationComputers2010} & 0.5 & 0.0 & 0.0 \\
    && Trust \cite{seymourIgnoranceBlissEffect} & \cite{cannizzaro2020trust} & 0.0 & 0.0 & 0.0 \\
    && Trust \cite{ngSimulatingEffectsSocial2020a} & \cite{nordheimTrustChatbotsCustomer} & 0.5 & 0.0 & 0.0 \\
    && Trust  \cite{garciadeblanessebastianApplicationExtensionUTAUT22022} & \cite{lu2011dynamics} & 0.5 & 1.0 & 1.0 \\
    && Propensity to Trust  \cite{patilCanWeTrust2022a} & \cite{ashleighNewPropensityTrust2012, mcknight2002impact, mayerEffectPerformanceAppraisal1999, liWhyWeTrust2008a} & 1.0 & 1.0 & 1.0 \\
    && Perceived Trust Belief  \cite{patilCanWeTrust2022a} & \cite{akterDevelopmentValidationInstrument2013} & 1.0 & 1.0 & 1.0\\
    && Perceived Trust  \cite{beheraCognitiveChatbotPersonalised2021a} & \cite{kasilingamUnderstandingAttitudeIntention2020a} & 1.0 & 1.0 & 1.0 \\
    && Trust  \cite{dekkalFactorsAffectingUser2023a} & \cite{khalidTrustVirtualAgent2018} & 0.5 & 1.0 & 1.0\\
    && Trust  \cite{vimalkumarOkayGoogleWhat2021c} & \cite{alalwanMobileFoodOrdering2020, bansalContextPersonalityMatter2016, smithInformationPrivacyResearch2011, smithInformationPrivacyMeasuring1996b} & 1.0 & 1.0 & 1.0 \\
    && Trust  \cite{acikgozRolePrivacyCynicism2022} & \cite{kimWillingnessProvidePersonal2019b} & 1.0 & 1.0 & 1.0 \\ 
    && Trust  \cite{pitardi2021alexa} & \cite{chattaraman2019should, hassaneinManipulatingPerceivedSocial2007, yeEnhancingCustomerTrust2019} & 1.0 & 1.0 & 1.0\\
    && Trust  \cite{vanbusselAnalyzingDeterminantsAccept2022b} & \cite{chandraEvaluatingRoleTrust2010} & 1.0 & 0.5 & 1.0 \\
    && Trust  \cite{fahnTimeGetConversational2021a} & \cite{davisUserAcceptanceInformation1993} & 0.5 & 0.0 & 0.0 \\
    && TS-Trust  \cite{fahnTimeGetConversational2021a} & \cite{jianFoundationsEmpiricallyDetermined2000a}& 0.5 & 0.0 & 0.0 \\
    && Mistrust  \cite{fahnTimeGetConversational2021a} & \cite{jianFoundationsEmpiricallyDetermined2000a} & 0.5 & 0.0 & 0.0 \\
    && Trust  \cite{kasilingamUnderstandingAttitudeIntention2020a} & \cite{chongPredictingConsumerDecisions2012, tsuweiWhatDrivesMalaysian2009} & 1.0 & 1.0 & 1.0 \\
    && Trust  \cite{palEffectTrustIts2022a} & \cite{ alalwanMobileFoodOrdering2020, yeEnhancingCustomerTrust2019} & 1.0 & 1.0 & 1.0 \\
    && Trust  \cite{guerreiroAdvertisingAcceptanceSmart2022} & \cite{kaaria2017technology} & 1.0 & 1.0 & 1.0 \\
    && Trust (General after Mayer) \cite{cannizzaro2020trust} & \cite{mayer1995integrative} & 0.0 & 0.0 & 0.0\\
    && Trust (General after Mayer) \cite{ bhattacherjee2002individualTrust} & \cite{mayer1995integrative} & 1.0 & 1.0 & 1.0\\
    && Trust (General after Mayer) \cite{mari2021role} & \cite{mayer1995integrative} & 0.5 & 0.0 & 0.0\\
    && Trust (General) \cite{lin2023measuringtrustgame} & not reported \cite{yamagishi1994trust} & n.a. & n.a. & n.a.\\
    && Trust (General) \cite{hsu2023semantic} & \cite{dwyer2007trust, metzger2007making} & 1.0 & 1.0 & 1.0 \\
    && Trust (General) \cite{jiang2023make} & \cite{gefen2003trust} & 1.0 & 1.0 & 1.0\\
    && Trust (General) \cite{choung2023trust} & \changed{\cite{schmidt2020transparency}} & 0.0 & 0.0 & 0.0\\
    && Trust (General) \cite{lopez2023enhancing} & \cite{lee1992trust} & 0.5 & 0.0 & 0.0\\
    && Trust (General) \cite{hasanConsumerTrustPerceived2021} & \cite{kaaria2017technology} & 1.0 & 1.0 & 1.0 \\
    && Trust (General) \cite{moradinezhad2021investigating} & \cite{jianFoundationsEmpiricallyDetermined2000a} & 0.0 & 0.0 & 0.0 \\
    && Trust (General) \cite{wald2021make} & \cite{bickmore2001relational} & 0.0 & 0.0 & 0.0\\
    && Trust (General) \cite{pitardi2021alexa} & \cite{chattaraman2019should,hassanein2009cross} & 1.0 & 1.0 & 1.0\\
    && Trust (General) \cite{lee2021role} & \cite{kim2010empirical, pham2015effects} & 0.5 & 1.0 & 1.0 \\
    && Trust (General) \cite{gupta2020measuring} & \cite{jianFoundationsEmpiricallyDetermined2000a} & 0.0 & 0.0 & 0.0\\
    && Trust (General) \cite{toader2019effect} & \cite{gupta2009task} & 1.0 & 1.0 & 1.0\\

    && Perceived Trust \cite{liu2021influences} & \cite{kim2011effect, shuhaiber2019understanding} & 1.0 & 1.0 & 1.0\\
    && Perceived Trust \cite{pal2020personal} & \cite{dinev2006privacy, malhotra2004internet} & 1.0 & 1.0 & 0.0\\
    && Perceived Trust \cite{purwanto2020interactive} & \cite{brill2022siri, ejdys2018building, ejdys2019role} & 1.0 & 1.0 & 1.0 \\
    && Perceived Trust \cite{ahmadTangiblePrivacySmart2022} & \cite{jianFoundationsEmpiricallyDetermined2000a} & 0.0 & 0.0 & 0.0 \\
    && Perceived Trust \cite{munoz2023voice} & \cite{corritore2003line} & 0.5 & 0.0 & 0.0 \\
    
    && Brand Trust \cite{lappemanTrustDigitalPrivacy2023a} & \cite{bruner2005marketing} & 1.0 & 1.0 & 1.0 \\
    && Brand Trust \cite{lv2022research} & \cite{nordheimTrustChatbotsCustomer} & 1.0 & 1.0 & 0.0 \\
    && Emotional Trust \cite{lappemanTrustDigitalPrivacy2023a} & \cite{komiak2006effects} & 1.0 & 1.0 & 1.0  \\
    && Emotional Trust \cite{pal2022perceptions} & \cite{komiak2006effects} & 1.0 & 1.0 & 1.0  \\
    && Emotional Trust \cite{chen2021anthropomorphism} & \cite{komiak2006effects} & 1.0 & 1.0 & 1.0  \\
    
    && Cognitive Trust \cite{lappemanTrustDigitalPrivacy2023a} & not reported & 1.0 & 1.0 & 1.0  \\
    && Cognitive Trust \cite{pal2022perceptions} & \cite{komiak2006effects} & 1.0 & 1.0 & 1.0  \\
    && Cognitive Trust \cite{chen2021anthropomorphism} & \cite{komiak2006effects} & 1.0 & 1.0 & 1.0  \\
    
    && Trusting Intention \cite{kunkel2019let} & \cite{mcknight2011trust} & 0.0 & 0.0 & 0.0 \\
    && Trusting Intention \cite{prakash2023determinants} & \cite{lankton2015technology} & 0.5 & 1.0 & 1.0 \\
   
    && Human-Like Trustworthiness \cite{weidmuller2022human} & \cite{mayer1995integrative, lankton2015technology} & 0.5 & 1.0 & 1.0 \\
    && Machine-Like Trustworthiness \cite{weidmuller2022human} & \cite{mayer1995integrative, lankton2015technology} & 0.5 & 1.0 & 1.0 \\

    && Institution-based Trust \cite{kunkel2019let} & \cite{mcknight2002impact} & 0.0 & 0.0 & 0.0\\
    && Disposition to Trust \cite{kunkel2019let} & \cite{mcknight2002impact} & 0.0 & 0.0 & 0.0 \\
    && Perceived Appropriateness Trust \cite{casadei2022chatbots} & \cite{davis1989perceivedTAM, nordheim2019initialtrust} & n.a. & n.a. & n.a. \\
    && Perceived Trustworthiness \cite{tastemirova2022microexpressions} & \cite{latoschik2017effect, ho2010revisiting} & 0.5 & 0.0 & 0.0 \\

    \midrule
    
    \multirowcell{11}{Technology \& \\ Trust} & \multirow{11}{*}{11} & 
    Human-Computer Trust \cite{pesonen2021you} & \cite{gulati2019design} & 0.5 & 0.0 & 0.0  \\
    && Trust in Chatbots \cite{cyr2009exploring} & not reported & 1.0 & 1.0 & 0.0 \\
    && Trust in Chatbots \cite{everard2005presentation} & not reported & n.a. & n.a. & n.a. \\
    && Trust in Chatbots \cite{alagarsamy2023exploring} & not reported & 1.0 & 1.0 & 1.0 . \\
    && Trust in Automation \cite{li2023s} & \cite{jianFoundationsEmpiricallyDetermined2000a} & 0.0 & 0.0 & 0.0\\
    && Trust in Automation \cite{li2023modeling} & \cite{jianFoundationsEmpiricallyDetermined2000a}  & n.a. & n.a. & n.a.\\
    && Trust in Automation \cite{merritt2013trust} & \cite{jianFoundationsEmpiricallyDetermined2000a} & 0.0 & 0.0 & 0.0 \\
    && Trust in Automation \cite{weitz2019you} & \cite{jianFoundationsEmpiricallyDetermined2000a} & 0 & 0 & 0\\
    && Trust in AI \cite{salah2023chatting} & \cite{gulati2019design} & 1.0 & 1.0 & 1.0\\
    && Trust in Technology \cite{lv2022research} & & 1.0 & 1.0 & 0.0\\
    && System Trust Scale  \cite{gupta2019ai} & \cite{jianFoundationsEmpiricallyDetermined2000a} & n.a. & n.a. & n.a. \\

    \multirow{25}{*}{Trust (Mayer)} & \multirow{25}{*}{25} & 
    Benevolence \cite{goodman2023s} & \cite{elkins2013sound, harwood2017internet, mcknight2011trust} & 1.0 & 1.0 & 1.0 \\
    && Benevolence \cite{weidmuller2022human} & \cite{elkins2013sound, harwood2017internet, mcknight2011trust} & 0.5 & 1.0 & 1.0 \\
    && Benevolence \cite{hu2021dual} & \cite{elkins2013sound, harwood2017internet, mcknight2011trust} & 1.0 & 1.0 & 1.0 \\
    && Benevolence \cite{mari2021role} & \cite{elkins2013sound, harwood2017internet, mcknight2011trust} & 0.5 & 0.0 & 0.0\\
    && Benevolence \cite{gulati2018modelling} & \cite{elkins2013sound, harwood2017internet, mcknight2011trust} & 1.0 & 1.0 & 1.0 \\ \\

    && Integrity \cite{lappemanTrustDigitalPrivacy2023a} & \cite{elkins2013sound}& 1.0 & 1.0 & 1.0 \\
    && Integrity \cite{weidmuller2022human} & \cite{elkins2013sound}& 0.5 & 1.0 & 1.0\\
    && Integrity \cite{goodman2023s} & \cite{elkins2013sound}& 1.0 & 1.0 & 1.0\\
    && Integrity \cite{hu2021dual} & \cite{elkins2013sound} & 1.0 & 1.0 & 1.0\\
    && Integrity \cite{mari2021role} & \cite{elkins2013sound} & 0.5 & 0.0 & 0.0\\
    
    && Trust Belief  \cite{prakash2023determinants} & \cite{lankton2015technology}& 0.5 & 1.0 & 1.0 \\
    && Trust Belief  \cite{jiang2023make} & \cite{lankton2015technology}& 1.0 & 1.0 & 1.0 \\
    && Trust Belief  \cite{kunkel2019let} & \cite{lankton2015technology} & 0.0 & 0.0 & 0.0\\
    && Competence \cite{lappemanTrustDigitalPrivacy2023a} & \cite{skjuve2019measuring, mcknight2002impact} & 1.0 & 1.0 & 1.0 \\
    && Competence \cite{weidmuller2022human} & \cite{skjuve2019measuring, mcknight2002impact} & 0.5 & 1.0 & 1.0\\
    && Competence \cite{wald2021make} & \cite{skjuve2019measuring, mcknight2002impact} & 0.0 & 0.0 & 0.0\\
    && Competence \cite{hu2021dual} & \cite{skjuve2019measuring, mcknight2002impact} & 1.0 & 1.0 & 1.0\\
    && Competence \cite{mari2021role} & \cite{skjuve2019measuring, mcknight2002impact} & 0.5 & 0.0 & 0.0\\
    && Competence \cite{gulati2018modelling} & \cite{skjuve2019measuring, mcknight2002impact} & 1.0 & 1.0 & 1.0\\

    && Technical Competence \cite{lee2022testing} & \cite{madsen2000measuring} & 0.0 & 1.0 & 0.0 \\
    && Chatbot Competence  & \cite{cho2006mechanism} & 1.0 & 1.0 & 1.0 \\
    && Chatbot Competence  & \cite{holzwarth2006influence} & 0.5 & 0.0 & 1.0 \\
    && Perceived Task Solving Competence & \cite{erskine2019location} & 1.0 & 1.0 & 1.0\\
    && Perceived Task Solving Competence & \cite{fiske2018model, aaker} & 0.0 & 0.0 & 0.0 \\

    && Perceived Performance \cite{purwanto2020interactive} & \cite{davis1989perceivedTAM, xiao2002impact, malhotra2004internet, kim2008trust}& 1.0 & 1.0 & 1.0 \\
    && Professionalism \cite{lv2022research} & \cite{nordheimTrustChatbotsCustomer} & 1.0 & 1.0 & 0.0\\
    && Ability  \cite{goodman2023s} & \cite{elkins2013sound} & 1.0 & 1.0 & 1.0\\

    && Perceived Benefits \cite{pal2020personal} & & 1.0 & 1.0 & 0.0\\
    && Reputation \cite{nordheim2019initialtrust}  & \cite{corritore2005measuring}& n.a. & n.a. & n.a. \\

    \multirowcell{13}{Other constructs \\ related to trust} & \multirow{13}{*}{13} &
    Reliability  \cite{lee2022testing} & \cite{madsen2000measuring, lankton2015technology} & 0.0 & 1.0 & 0.0\\
    && Reliability  \cite{weidmuller2022human} & \cite{madsen2000measuring, lankton2015technology} & 0.5 & 1.0 & 1.0\\
    && Understandability \cite{lee2022testing} & & 0.0 & 1.0 & 0.0\\
    && Previous Chatbot Usage \cite{jiang2023make} & & 1.0 & 1.0 & 1.0 \\
    && Previous Chatbot Usage \cite{wald2021make} & & 0.0 & 0.0 & 0.0\\
    && Tailored response  \cite{jiang2023make} & \cite{schuetzler2020impact} & 1.0 & 1.0 & 1.0  \\
    && Communication Delay  & \cite{schanke2021estimating} &  0.0 & 0.0 & 0.0 \\
    && Communication Delay  & \cite{liu2011bringing}  &  0.0 & 0.0 & 0.0\\
    && Communal Relationship  \cite{cheng2022human} & \cite{li2019service, wang2017smile} & 1.0 & 1.0 & 1.0 \\

    \midrule
    \midrule

    \multirow{3}{*}{Interactivity} & \multirow{3}{*}{3} &
    Synchronicity \cite{purwanto2020interactive} & \cite{brill2022siri}& 1.0 & 1.0 & 1.0  \\
    && Bidirectionality \cite{purwanto2020interactive} & \cite{yoo2010role, brill2022siri}& 1.0 & 1.0 & 1.0\\
    && Perceived Interactivity \cite{prakash2023determinants} & \cite{go2019humanizing} & 0.5 & 1.0 & 1.0\\

    \multirowcell{29}{\shortstack{Affect,\\ Humanness, \\ Antropomorphism}} & \multirow{29}{*}{29} &
    Voice Humanness  \cite{hu2021dual} & \cite{go2019humanizing} & 1.0 & 1.0 & 1.0\\
    &&Understanding Humanness  \cite{hu2021dual} & \cite{go2019humanizing} & 1.0 & 1.0 & 1.0 \\
    &&Affinity for Technology  \cite{gupta2022trust} & \cite{franke2019personal} & 0.0 & 0.0 & 0.0 \\
    &&Perceived Warmth  \cite{cheng2022human} & \cite{kirmani2017doing} & 1.0 & 1.0 & 1.0 \\  
    &&Virtual assistant warmth  \cite{toader2019effect} & \cite{aaker} & 1.0 & 1.0 & 1.0 \\
    &&Perceived Humanity   \cite{zhang2021motivation} & \cite{venkatesh2000theoretical} & 1.0 & 1.0 & 0.0 \\
    &&Perceived Social Presence  \cite{zhang2021motivation} & \cite{venkatesh2000theoretical} & 1.0 & 1.0 & 0.0  \\
    &&Perceived Social Interactivity  \cite{zhang2021motivation} & \cite{venkatesh2000theoretical} & 1.0 & 1.0 & 0.0  \\
    &&Antropomorphism   \cite{chen2021anthropomorphism} & \cite{waytz2010sees, bartneck2009measurement} & 1.0 & 1.0 & 1.0\\
    &&Antropomorphism   \cite{wald2021make} & \cite{waytz2010sees, bartneck2009measurement} & 0.0 & 0.0 & 0.0 \\
    &&Antropomorphic design cues  \cite{toader2019effect} & \cite{nowak2005influence} & 1.0 & 1.0 & 1.0\\
    &&Social Presence  \cite{lee2022testing} & \cite{gefen2003trust, gray1977social, biocca2003toward, king2006meta, bruwer2008psychometric}& 0.0 & 1.0 & 0.0 \\
    &&Social Presence  \cite{kunkel2019let} & \cite{gefen2003trust, gray1977social, biocca2003toward, king2006meta, bruwer2008psychometric} & 0.0 & 0.0 & 0.0\\
    &&Social Presence  \cite{prakash2023determinants} & \cite{gefen2003trust, gray1977social, biocca2003toward, king2006meta, bruwer2008psychometric}& 0.5 & 1.0 & 1.0 \\
    &&Social Presence  \cite{jiang2023make} & \cite{gefen2003trust, gray1977social, biocca2003toward, king2006meta, bruwer2008psychometric} & 1.0 & 1.0 & 1.0\\
    &&Social Presence  \cite{pitardi2021alexa} & \cite{gefen2003trust, gray1977social, biocca2003toward, king2006meta, bruwer2008psychometric} & 1.0 & 1.0 & 1.0 \\
    &&Social Presence  \cite{toader2019effect} & \cite{gefen2003trust, gray1977social, biocca2003toward, king2006meta, bruwer2008psychometric} & 1.0 & 1.0 & 1.0 \\
    
    &&Social Cognition \cite{pitardi2021alexa} & \cite{fiske2012sage} & 1.0 & 1.0 & 1.0\\
    &&Interpersonal Attraction \cite{chen2021anthropomorphism} & \cite{hanUnderstandingAdoptionIntelligent2018a} & 1.0 & 1.0 & 1.0 \\
    &&Humanlike traits \cite{hsu2023semantic} & \cite{lewis2015investigating} & 1.0 & 1.0 & 1.0\\
    &&Humanlike Cues  \cite{li2021anthropomorphism} & & 0 & 0 &  \\
    &&Humanlikeness  \cite{lv2022research} & \cite{ho2010revisiting} & 1.0 & 1.0 & 0\\ 
    &&Humanlikeness  \cite{weidmuller2022human} & \cite{ho2010revisiting}  & 0.5 & 1.0 & 1.0\\
    &&Humanlikeness  \cite{nordheim2019initialtrust} & \cite{ho2010revisiting} &  n.a. & n.a. & n.a. \\
    
    &&Helpfulness  \cite{weidmuller2022human} & \cite{lankton2015technology} & 0.5 & 1.0 & 1.0 \\
    &&Faith \cite{lee2022testing} & \cite{madsen2000measuring} & 0 & 1.0 & 0.0\\
    &&Personal Attachment\cite{lee2022testing} & \cite{madsen2000measuring}& 0.0 & 1.0 & 0.0 \\
    &&Familiarity  \cite{liu2021influences} & & 1.0 & 1.0 & 1.0 \\

    \midrule
    
    \multirowcell{15}{Character \& \\ Personality} & \multirow{15}{*}{15} &
    Self-Esteem \cite{salah2023chatting} & \cite{gnambs2018structure} & 1.0 & 1.0 & 1.0\\
    && Disposition to Trust \cite{alagarsamy2023exploring} & \cite{tan2005consumer} & 1.0 & 1.0 & 1.0\\
    && Propensity to Trust Technology \cite{prakash2023determinants} & \cite{lankton2015technology, mayer1995integrative} & 0.5 & 1.0 & 1.0\\
    
    &&Psychological Well-Being  \cite{salah2023chatting} & \cite{babnik2022ryff}& 1.0 & 1.0 & 1.0 \\
    &&Personality  \cite{lv2022research} & not reported & 1.0 & 1.0 & 0.0\\
    &&Conformity  \cite{schreuter2021trust} & \cite{mehrabian1995basic} & 0.0 & 0.0 & 0.0 \\
    &&Ambiguity Tolerance  \cite{jiang2023make} & \cite{budner19621962} & 1.0 & 1.0 & 1.0\\
    &&Technology Optimism  \cite{liu2021influences} & \cite{shuhaiber2019understanding} & 1.0 & 1.0 & 1.0\\ 
    &&Extraversion  \cite{muller2019chatbot} & \cite{wang2016empirical} & 0.5 & 0.0 & 0.0 \\
    &&Agreeableness \cite{muller2019chatbot} & \cite{wang2016empirical}& 0.5 & 0.0 & 0.0 \\
    &&Emotionality  \cite{muller2019chatbot} & \cite{wang2016empirical}& 0.5 & 0.0 & 0\\
    &&Honesty-Humility \cite{muller2019chatbot} & \cite{wang2016empirical}& 0.5 & 0.0 & 0.0\\
    &&Honesty \cite{gulati2018modelling} & \cite{mcknight2002impact} & 1.0 & 1.0 & 1.0\\
    
    &&Social Disclosure \cite{toader2019effect} & \cite{cho2006mechanism} & 1.0 & 1.0 & 1.0 \\

    \midrule
    \multirow{13}{*}{Other Constructs} & \multirow{13}{*}{13} &
    Personalized Services \cite{pal2020personal} & \cite{xu2011personalization} & 1.0 & 1.0 & 1.0\\
    && Follow Advice \cite{kunkel2019let} & not reported & n.a. & n.a. & n.a. \\
    && Follow Advice \cite{lin2023measuringtrustgame} & not reported & n.a. & n.a. & n.a.\\
    &&Give Information  \cite{kunkel2019let} & not reported & 0.0 & 0.0 & 0.0\\
    &&Job Anxiety  \cite{salah2023chatting} & \cite{wang2022development} & 1.0 & 1.0 & 1.0 \\
    &&Reciprocity \cite{gulati2018modelling} & \cite{kankanhalli2005contributing}& 1.0 & 1.0 & 1.0 \\
    && Technology Fear  \cite{alagarsamy2023exploring} & \cite{cabrera2021identifying} & 1.0 & 1.0 & 1.0\\
    &&Decision Satisfaction \cite{mari2021role} & \cite{fitzsimons2000consumer, fitzsimons1997decision} & 0.5 & 0.0 & 0.0 \\
    &&Ubiquitiy \cite{compeau1995computer} & & 1.0 & 1.0 & 1.0\\
    &&Positive Consumer Response \cite{toader2019effect} & \cite{cyr2004design} & 1.0 & 1.0 & 1.0\\
    &&Perceived Stereotyping \cite{salah2023chatting} & \cite{schepman2023general} & 1.0 & 1.0 & 1.0\\
    &&Perceived Contingency  \cite{prakash2023determinants} & \cite{go2019humanizing}  & 0.5 & 1.0 & 1.0\\
    &&Perceived Complementary \cite{pal2020personal} & \cite{lee2014early} & 1.0 & 1.0 & 0.0 \\
    \bottomrule

\label{tab:TypeTrustConstructs}
\end{longtable}
\twocolumn

\begin{table}[ht]
  \caption{Research Methods in our sample}
  \label{tab:ResearchMethodsExtensive}
  \scriptsize
  \begin{tabular}{p{0.3\columnwidth}p{0.25\columnwidth}p{0.25\columnwidth}}
    \toprule
    Research Method & Privacy \& Security - Frequency (Percentage) & Trust - Frequency (Percentage)\\
    \midrule 
    Quantitative & \noindent\changed{60 (60\%)} & 47 (81\%) \\
    \midrule
    Survey & \noindent\changed{45 (45\%)} & 22 (38\%) \\
    Experiment and Survey & \noindent\changed{12 (12\%)} & 11 (19\%) \\
    Vignette Study & 3 (3\%) & 14 (24\%) \\  
    Trust Game & - & 2 (3\%) \\
    \midrule
    Mixed & \noindent\changed{19 (19\%)} & 6 (10\%)\\
    \midrule 
    Experiment and Survey and Interview & 5 (5\%)&\\
    Interview and Survey & \noindent\changed{3 (3\%)} & 2 (3\%)\\
    Focus group and Survey & 2 (2\%) & \\
    Survey and Analysis of posted reviews & 2 (2\%) & \\
    Experiment and Interview and Focus Group and Observation & 1 (1\%) &\\
    Experiment and Interview and Datalogs and Drawing & 1 (1\%) &\\
    Experiment and Survey and Card Sorting & 1 (1\%) & \\
    Experiment and Mixed Survey & 1 (1\%) & 3 (5\%) \\
    Experience Sampling Method and Survey & 1 (1\%) &\\ 
    Delphi Study and Survey & 1 (1\%) &\\ 
    Design Science Research and Interview and Survey & 1 (1\%) & \\
    Mixed Survey & & 1 (2\%)\\
    \midrule 
    Qualitative & \noindent\changed{21 (21\%)}& 5 (9\%) \\
    \midrule 
    Interview & \noindent\changed{15 (15\%)} & 3 (5\%) \\ 
    \noindent\changed{Focus Group} & \noindent\changed{1 (1\%)} & \\
    Interview and Focus Group & 1 (1\%) & \\
    Interview and Diary Study & 1 (1\%) & \\ 
    Interview and Datalogs & 1 (1\%) & \\ 
    Interview and Drawing & 1 (1\%) & \\
    Focus Group and Co-Design & 1 (1\%) & \\
    Interview with Coding & & 1 (2\%) \\
    Collaborative Design  & & 1 (2\%) \\
    \bottomrule
\end{tabular}
\end{table}

\bibliographystyle{cas-model2-names}

\bibliography{cas-dc-template}

\begin{thebibliography}{345}
\expandafter\ifx\csname natexlab\endcsname\relax\def\natexlab#1{#1}\fi
\providecommand{\url}[1]{\texttt{#1}}
\providecommand{\href}[2]{#2}
\providecommand{\path}[1]{#1}
\providecommand{\DOIprefix}{doi:}
\providecommand{\ArXivprefix}{arXiv:}
\providecommand{\URLprefix}{URL: }
\providecommand{\Pubmedprefix}{pmid:}
\providecommand{\doi}[1]{\href{http://dx.doi.org/#1}{\path{#1}}}
\providecommand{\Pubmed}[1]{\href{pmid:#1}{\path{#1}}}
\providecommand{\bibinfo}[2]{#2}
\ifx\xfnm\relax \def\xfnm[#1]{\unskip,\space#1}\fi
\bibitem[{Aaker et~al.(2010)Aaker, Vohs and Mogilner}]{aaker}
\bibinfo{author}{Aaker, J.}, \bibinfo{author}{Vohs, K.D.}, \bibinfo{author}{Mogilner, C.}, \bibinfo{year}{2010}.
\newblock \bibinfo{title}{Nonprofits are seen as warm and for-profits as competent: Firm stereotypes matter}.
\newblock \bibinfo{journal}{Journal of consumer research} \bibinfo{volume}{37}, \bibinfo{pages}{224--237}.
\bibitem[{Abdi et~al.(2021)Abdi, Zhan, Ramokapane and Such}]{abdiPrivacyNormsSmart2021a}
\bibinfo{author}{Abdi, N.}, \bibinfo{author}{Zhan, X.}, \bibinfo{author}{Ramokapane, K.M.}, \bibinfo{author}{Such, J.}, \bibinfo{year}{2021}.
\newblock \bibinfo{title}{Privacy {{Norms}} for {{Smart Home Personal Assistants}}}, in: \bibinfo{booktitle}{Proceedings of the 2021 {{CHI Conference}} on {{Human Factors}} in {{Computing Systems}}}, \bibinfo{publisher}{{ACM}}, \bibinfo{address}{{Yokohama Japan}}. pp. \bibinfo{pages}{1--14}.
\newblock \DOIprefix\doi{10.1145/3411764.3445122}.
\bibitem[{Abrokwa et~al.()Abrokwa, Das, Akgul and Mazurek}]{abrokwaComparingSecurityPrivacya}
\bibinfo{author}{Abrokwa, D.}, \bibinfo{author}{Das, S.}, \bibinfo{author}{Akgul, O.}, \bibinfo{author}{Mazurek, M.L.}, .
\newblock \bibinfo{title}{Comparing {{Security}} and {{Privacy Attitudes}} {{Among U}}.{{S}}. {{Users}} of {{Different Smartphone}} and {{Smart-Speaker Platforms}}} .
\bibitem[{Acikgoz and Vega(2022)}]{acikgozRolePrivacyCynicism2022}
\bibinfo{author}{Acikgoz, F.}, \bibinfo{author}{Vega, R.P.}, \bibinfo{year}{2022}.
\newblock \bibinfo{title}{The {{Role}} of {{Privacy Cynicism}} in {{Consumer Habits}} with {{Voice Assistants}}: {{A Technology Acceptance Model Perspective}}}.
\newblock \bibinfo{journal}{International Journal of Human{\textendash}Computer Interaction} \bibinfo{volume}{38}, \bibinfo{pages}{1138--1152}.
\newblock \DOIprefix\doi{10.1080/10447318.2021.1987677}.
\bibitem[{Ackerman(2004)}]{Ackerman2004}
\bibinfo{author}{Ackerman, M.S.}, \bibinfo{year}{2004}.
\newblock \bibinfo{title}{Privacy in pervasive environments: next generation labeling protocols}.
\newblock \bibinfo{journal}{Personal and Ubiquitous Computing} \bibinfo{volume}{8}, \bibinfo{pages}{430--439}.
\newblock \URLprefix \url{https://doi.org/10.1007/s00779-004-0305-8}, \DOIprefix\doi{10.1007/s00779-004-0305-8}.
\bibitem[{Ahmad et~al.(2022)Ahmad, Akter, Buher, Farzan, Kapadia and Lee}]{ahmadTangiblePrivacySmart2022}
\bibinfo{author}{Ahmad, I.}, \bibinfo{author}{Akter, T.}, \bibinfo{author}{Buher, Z.}, \bibinfo{author}{Farzan, R.}, \bibinfo{author}{Kapadia, A.}, \bibinfo{author}{Lee, A.J.}, \bibinfo{year}{2022}.
\newblock \bibinfo{title}{Tangible {{Privacy}} for {{Smart Voice Assistants}}: {{Bystanders}}' {{Perceptions}} of {{Physical Device Controls}}}.
\newblock \bibinfo{journal}{Proceedings of the ACM on Human-Computer Interaction} \bibinfo{volume}{6}, \bibinfo{pages}{1--31}.
\newblock \DOIprefix\doi{10.1145/3555089}.
\bibitem[{Ahmad et~al.(2020)Ahmad, Farzan, Kapadia and Lee}]{ahmadTangiblePrivacyUserCentric2020b}
\bibinfo{author}{Ahmad, I.}, \bibinfo{author}{Farzan, R.}, \bibinfo{author}{Kapadia, A.}, \bibinfo{author}{Lee, A.J.}, \bibinfo{year}{2020}.
\newblock \bibinfo{title}{Tangible {{Privacy}}: {{Towards User-Centric Sensor Designs}} for {{Bystander Privacy}}}.
\newblock \bibinfo{journal}{Proceedings of the ACM on Human-Computer Interaction} \bibinfo{volume}{4}, \bibinfo{pages}{1--28}.
\newblock \DOIprefix\doi{10.1145/3415187}.
\bibitem[{Aiolfi(2023)}]{aiolfiHowShoppingHabits2023}
\bibinfo{author}{Aiolfi, S.}, \bibinfo{year}{2023}.
\newblock \bibinfo{title}{How shopping habits change with artificial intelligence: Smart speakers' usage intention}.
\newblock \bibinfo{journal}{International Journal of Retail \& Distribution Management} \bibinfo{volume}{ahead-of-print}.
\newblock \DOIprefix\doi{10.1108/IJRDM-11-2022-0441}.
\bibitem[{Akter et~al.(2013)Akter, D'Ambra and Ray}]{akterDevelopmentValidationInstrument2013}
\bibinfo{author}{Akter, S.}, \bibinfo{author}{D'Ambra, J.}, \bibinfo{author}{Ray, P.}, \bibinfo{year}{2013}.
\newblock \bibinfo{title}{Development and validation of an instrument to measure user perceived service quality of {{mHealth}}}.
\newblock \bibinfo{journal}{Information \& Management} \bibinfo{volume}{50}, \bibinfo{pages}{181--195}.
\newblock \DOIprefix\doi{10.1016/j.im.2013.03.001}.
\bibitem[{{Al-Ameen} et~al.(2021){Al-Ameen}, Chauhan, Ahsan and Kocabas}]{al-ameenLookUserPrivacy2021a}
\bibinfo{author}{{Al-Ameen}, M.N.}, \bibinfo{author}{Chauhan, A.}, \bibinfo{author}{Ahsan, M.M.}, \bibinfo{author}{Kocabas, H.}, \bibinfo{year}{2021}.
\newblock \bibinfo{title}{A look into user's privacy perceptions and data practices of {{IoT}} devices}.
\newblock \bibinfo{journal}{Information \& Computer Security} \bibinfo{volume}{29}, \bibinfo{pages}{573--588}.
\newblock \DOIprefix\doi{10.1108/ICS-08-2020-0134}.
\bibitem[{Alagarsamy and Mehrolia(2023)}]{alagarsamy2023exploring}
\bibinfo{author}{Alagarsamy, S.}, \bibinfo{author}{Mehrolia, S.}, \bibinfo{year}{2023}.
\newblock \bibinfo{title}{Exploring chatbot trust: Antecedents and behavioural outcomes}.
\newblock \bibinfo{journal}{Heliyon} \bibinfo{volume}{9}.
\bibitem[{Alalwan(2020)}]{alalwanMobileFoodOrdering2020}
\bibinfo{author}{Alalwan, A.A.}, \bibinfo{year}{2020}.
\newblock \bibinfo{title}{Mobile food ordering apps: {{An}} empirical study of the factors affecting customer e-satisfaction and continued intention to reuse}.
\newblock \bibinfo{journal}{International Journal of Information Management} \bibinfo{volume}{50}, \bibinfo{pages}{28--44}.
\newblock \DOIprefix\doi{10.1016/j.ijinfomgt.2019.04.008}.
\bibitem[{Alghamdi et~al.(2022)Alghamdi, Alsoubai, Akter, Alghamdi and Wisniewski}]{alghamdiUserStudyEvaluate2022a}
\bibinfo{author}{Alghamdi, L.}, \bibinfo{author}{Alsoubai, A.}, \bibinfo{author}{Akter, M.}, \bibinfo{author}{Alghamdi, F.}, \bibinfo{author}{Wisniewski, P.}, \bibinfo{year}{2022}.
\newblock \bibinfo{title}{A user study to evaluate a web-based prototype for smart home internet of things device management}, in: \bibinfo{booktitle}{International Conference on Human-Computer Interaction}, \bibinfo{organization}{Springer}. pp. \bibinfo{pages}{383--405}.
\newblock \DOIprefix\doi{10.1007/978-3-031-05563-8_24}.
\bibitem[{Alt and Ibolya(2021)}]{alt2021identifying}
\bibinfo{author}{Alt, M.A.}, \bibinfo{author}{Ibolya, V.}, \bibinfo{year}{2021}.
\newblock \bibinfo{title}{Identifying relevant segments of potential banking chatbot users based on technology adoption behavior}.
\newblock \bibinfo{journal}{Market-Tr{\v{z}}i{\v{s}}te} \bibinfo{volume}{33}, \bibinfo{pages}{165--183}.
\bibitem[{Ammari et~al.(2019)Ammari, Kaye, Tsai and Bentley}]{ammariMusicSearchIoT2019a}
\bibinfo{author}{Ammari, T.}, \bibinfo{author}{Kaye, J.}, \bibinfo{author}{Tsai, J.Y.}, \bibinfo{author}{Bentley, F.}, \bibinfo{year}{2019}.
\newblock \bibinfo{title}{Music, {{Search}}, and {{IoT}}: {{How People}} ({{Really}}) {{Use Voice Assistants}}}.
\newblock \bibinfo{journal}{ACM Transactions on Computer-Human Interaction} \bibinfo{volume}{26}, \bibinfo{pages}{1--28}.
\newblock \DOIprefix\doi{10.1145/3311956}.
\bibitem[{Ashleigh et~al.(2012)Ashleigh, Higgs and Dulewicz}]{ashleighNewPropensityTrust2012}
\bibinfo{author}{Ashleigh, M.J.}, \bibinfo{author}{Higgs, M.}, \bibinfo{author}{Dulewicz, V.}, \bibinfo{year}{2012}.
\newblock \bibinfo{title}{A new propensity to trust scale and its relationship with individual well-being: Implications for {{HRM}} policies and practices}.
\newblock \bibinfo{journal}{Human Resource Management Journal} \bibinfo{volume}{22}, \bibinfo{pages}{360--376}.
\newblock \DOIprefix\doi{10.1111/1748-8583.12007}.
\bibitem[{Aw et~al.(2022)Aw, Tan, Cham, Raman and Ooi}]{aw2022alexa}
\bibinfo{author}{Aw, E.C.X.}, \bibinfo{author}{Tan, G.W.H.}, \bibinfo{author}{Cham, T.H.}, \bibinfo{author}{Raman, R.}, \bibinfo{author}{Ooi, K.B.}, \bibinfo{year}{2022}.
\newblock \bibinfo{title}{Alexa, what's on my shopping list? transforming customer experience with digital voice assistants}.
\newblock \bibinfo{journal}{Technological Forecasting and Social Change} \bibinfo{volume}{180}, \bibinfo{pages}{121711}.
\bibitem[{Ayalon and Toch(2019)}]{Ayalon2019}
\bibinfo{author}{Ayalon, O.}, \bibinfo{author}{Toch, E.}, \bibinfo{year}{2019}.
\newblock \bibinfo{title}{Evaluating $\{$Users’$\}$ perceptions about a $\{$System’s$\}$ privacy: Differentiating social and institutional aspects}, in: \bibinfo{booktitle}{Fifteenth Symposium on Usable Privacy and Security (SOUPS 2019)}, pp. \bibinfo{pages}{41--59}.
\bibitem[{Babnik et~al.(2022)Babnik, Benko and von Humboldt}]{babnik2022ryff}
\bibinfo{author}{Babnik, K.}, \bibinfo{author}{Benko, E.}, \bibinfo{author}{von Humboldt, S.}, \bibinfo{year}{2022}.
\newblock \bibinfo{title}{Ryff’s psychological well-being scale}, in: \bibinfo{booktitle}{Encyclopedia of gerontology and population aging}. \bibinfo{publisher}{Springer}, pp. \bibinfo{pages}{4344--4349}.
\bibitem[{Bansal et~al.(2016)Bansal, Zahedi and Gefen}]{bansalContextPersonalityMatter2016}
\bibinfo{author}{Bansal, G.}, \bibinfo{author}{Zahedi, F.M.}, \bibinfo{author}{Gefen, D.}, \bibinfo{year}{2016}.
\newblock \bibinfo{title}{Do context and personality matter? {{Trust}} and privacy concerns in disclosing private information online}.
\newblock \bibinfo{journal}{Information \& Management} \bibinfo{volume}{53}, \bibinfo{pages}{1--21}.
\newblock \DOIprefix\doi{10.1016/j.im.2015.08.001}.
\bibitem[{{Bar-Ilan}(2018)}]{bar-ilanTaleThreeDatabases2018}
\bibinfo{author}{{Bar-Ilan}, J.}, \bibinfo{year}{2018}.
\newblock \bibinfo{title}{Tale of {{Three Databases}}: {{The Implication}} of {{Coverage Demonstrated}} for a {{Sample Query}}}.
\newblock \bibinfo{journal}{Frontiers in Research Metrics and Analytics} \bibinfo{volume}{3}.
\bibitem[{Barkhuus(2012)}]{Barkhuus2012}
\bibinfo{author}{Barkhuus, L.}, \bibinfo{year}{2012}.
\newblock \bibinfo{title}{The mismeasurement of privacy: Using contextual integrity to reconsider privacy in hci}, in: \bibinfo{booktitle}{Proceedings of the SIGCHI Conference on Human Factors in Computing Systems}, \bibinfo{publisher}{Association for Computing Machinery}, \bibinfo{address}{New York, NY, USA}. p. \bibinfo{pages}{367–376}.
\newblock \URLprefix \url{https://doi.org/10.1145/2207676.2207727}, \DOIprefix\doi{10.1145/2207676.2207727}.
\bibitem[{Bartneck et~al.(2009)Bartneck, Kuli{\'c}, Croft and Zoghbi}]{bartneck2009measurement}
\bibinfo{author}{Bartneck, C.}, \bibinfo{author}{Kuli{\'c}, D.}, \bibinfo{author}{Croft, E.}, \bibinfo{author}{Zoghbi, S.}, \bibinfo{year}{2009}.
\newblock \bibinfo{title}{Measurement instruments for the anthropomorphism, animacy, likeability, perceived intelligence, and perceived safety of robots}.
\newblock \bibinfo{journal}{International journal of social robotics} \bibinfo{volume}{1}, \bibinfo{pages}{71--81}.
\bibitem[{Bauer and Freitag(2018)}]{bauer2018measuring}
\bibinfo{author}{Bauer, P.C.}, \bibinfo{author}{Freitag, M.}, \bibinfo{year}{2018}.
\newblock \bibinfo{title}{Measuring trust}.
\newblock \bibinfo{journal}{The Oxford handbook of social and political trust} \bibinfo{volume}{15}.
\bibitem[{Behera et~al.(2021)Behera, Bala and Ray}]{beheraCognitiveChatbotPersonalised2021a}
\bibinfo{author}{Behera, R.K.}, \bibinfo{author}{Bala, P.K.}, \bibinfo{author}{Ray, A.}, \bibinfo{year}{2021}.
\newblock \bibinfo{title}{Cognitive {{Chatbot}} for {{Personalised Contextual Customer Service}}: {{Behind}} the {{Scene}} and beyond the {{Hype}}}.
\newblock \bibinfo{journal}{Information Systems Frontiers} \DOIprefix\doi{10.1007/s10796-021-10168-y}.
\bibitem[{Berg et~al.(1995)Berg, Dickhaut and McCabe}]{berg1995trustgame}
\bibinfo{author}{Berg, J.}, \bibinfo{author}{Dickhaut, J.}, \bibinfo{author}{McCabe, K.}, \bibinfo{year}{1995}.
\newblock \bibinfo{title}{Trust, reciprocity, and social history}.
\newblock \bibinfo{journal}{Games and economic behavior} \bibinfo{volume}{10}, \bibinfo{pages}{122--142}.
\bibitem[{Bhattacherjee(2002)}]{bhattacherjee2002individualTrust}
\bibinfo{author}{Bhattacherjee, A.}, \bibinfo{year}{2002}.
\newblock \bibinfo{title}{Individual trust in online firms: Scale development and initial test}.
\newblock \bibinfo{journal}{Journal of management information systems} \bibinfo{volume}{19}, \bibinfo{pages}{211--241}.
\bibitem[{Bickmore and Cassell(2001)}]{bickmore2001relational}
\bibinfo{author}{Bickmore, T.}, \bibinfo{author}{Cassell, J.}, \bibinfo{year}{2001}.
\newblock \bibinfo{title}{Relational agents: a model and implementation of building user trust}, in: \bibinfo{booktitle}{Proceedings of the SIGCHI conference on Human factors in computing systems}, pp. \bibinfo{pages}{396--403}.
\bibitem[{Biocca et~al.(2003)Biocca, Harms and Burgoon}]{biocca2003toward}
\bibinfo{author}{Biocca, F.}, \bibinfo{author}{Harms, C.}, \bibinfo{author}{Burgoon, J.K.}, \bibinfo{year}{2003}.
\newblock \bibinfo{title}{Toward a more robust theory and measure of social presence: Review and suggested criteria}.
\newblock \bibinfo{journal}{Presence: Teleoperators \& virtual environments} \bibinfo{volume}{12}, \bibinfo{pages}{456--480}.
\bibitem[{Bolton et~al.(2021)Bolton, Dargahi, Belguith, {Al-Rakhami} and Sodhro}]{boltonSecurityPrivacyChallenges2021}
\bibinfo{author}{Bolton, T.}, \bibinfo{author}{Dargahi, T.}, \bibinfo{author}{Belguith, S.}, \bibinfo{author}{{Al-Rakhami}, M.S.}, \bibinfo{author}{Sodhro, A.H.}, \bibinfo{year}{2021}.
\newblock \bibinfo{title}{On the {{Security}} and {{Privacy Challenges}} of {{Virtual Assistants}}}.
\newblock \bibinfo{journal}{Sensors} \bibinfo{volume}{21}, \bibinfo{pages}{2312}.
\newblock \DOIprefix\doi{10.3390/s21072312}.
\bibitem[{Borsci et~al.(2022)Borsci, Malizia, Schmettow, {van der Velde}, Tariverdiyeva, Balaji and Chamberlain}]{borsciChatbotUsabilityScale2022b}
\bibinfo{author}{Borsci, S.}, \bibinfo{author}{Malizia, A.}, \bibinfo{author}{Schmettow, M.}, \bibinfo{author}{{van der Velde}, F.}, \bibinfo{author}{Tariverdiyeva, G.}, \bibinfo{author}{Balaji, D.}, \bibinfo{author}{Chamberlain, A.}, \bibinfo{year}{2022}.
\newblock \bibinfo{title}{The {{Chatbot Usability Scale}}: The {{Design}} and {{Pilot}} of a {{Usability Scale}} for {{Interaction}} with {{AI-Based Conversational Agents}}}.
\newblock \bibinfo{journal}{Personal and Ubiquitous Computing} \bibinfo{volume}{26}, \bibinfo{pages}{95--119}.
\newblock \DOIprefix\doi{10.1007/s00779-021-01582-9}.
\bibitem[{Brandimarte et~al.(2013)Brandimarte, Acquisti and Loewenstein}]{Brandimarte2013}
\bibinfo{author}{Brandimarte, L.}, \bibinfo{author}{Acquisti, A.}, \bibinfo{author}{Loewenstein, G.}, \bibinfo{year}{2013}.
\newblock \bibinfo{title}{Misplaced {Confidences}: {Privacy} and the {Control} {Paradox}}.
\newblock \bibinfo{journal}{Social Psychological and Personality Science} \bibinfo{volume}{4}, \bibinfo{pages}{340--347}.
\newblock \URLprefix \url{https://doi.org/10.1177/1948550612455931}, \DOIprefix\doi{10.1177/1948550612455931}. \bibinfo{note}{publisher: SAGE Publications Inc}.
\bibitem[{Brill et~al.(2022)Brill, Munoz and Miller}]{brill2022siri}
\bibinfo{author}{Brill, T.M.}, \bibinfo{author}{Munoz, L.}, \bibinfo{author}{Miller, R.J.}, \bibinfo{year}{2022}.
\newblock \bibinfo{title}{Siri, alexa, and other digital assistants: a study of customer satisfaction with artificial intelligence applications}, in: \bibinfo{booktitle}{The Role of Smart Technologies in Decision Making}. \bibinfo{publisher}{Routledge}, pp. \bibinfo{pages}{35--70}.
\bibitem[{Br{\"u}ggemeier and Lalone(2022)}]{bruggemeierPerceptionsReactionsConversational2022b}
\bibinfo{author}{Br{\"u}ggemeier, B.}, \bibinfo{author}{Lalone, P.}, \bibinfo{year}{2022}.
\newblock \bibinfo{title}{Perceptions and reactions to conversational privacy initiated by a conversational user interface}.
\newblock \bibinfo{journal}{Computer Speech \& Language} \bibinfo{volume}{71}, \bibinfo{pages}{101269}.
\newblock \DOIprefix\doi{10.1016/j.csl.2021.101269}.
\bibitem[{Bruner et~al.(2005)Bruner, Hensel and James}]{bruner2005marketing}
\bibinfo{author}{Bruner, G.C.}, \bibinfo{author}{Hensel, P.J.}, \bibinfo{author}{James, K.E.}, \bibinfo{year}{2005}.
\newblock \bibinfo{title}{Marketing scales handbook}.
\bibitem[{Bruwer et~al.(2008)Bruwer, Emsley, Kidd, Lochner and Seedat}]{bruwer2008psychometric}
\bibinfo{author}{Bruwer, B.}, \bibinfo{author}{Emsley, R.}, \bibinfo{author}{Kidd, M.}, \bibinfo{author}{Lochner, C.}, \bibinfo{author}{Seedat, S.}, \bibinfo{year}{2008}.
\newblock \bibinfo{title}{Psychometric properties of the multidimensional scale of perceived social support in youth}.
\newblock \bibinfo{journal}{Comprehensive psychiatry} \bibinfo{volume}{49}, \bibinfo{pages}{195--201}.
\bibitem[{Budner(1962)}]{budner19621962}
\bibinfo{author}{Budner, S.}, \bibinfo{year}{1962}.
\newblock \bibinfo{title}{Intolerance of ambiguity as a personality variable}.
\newblock \bibinfo{journal}{Journal ofPersonalityAssessment,(1)} , \bibinfo{pages}{29--50}.
\bibitem[{Buteau and Lee(2021)}]{buteauHeyAlexaWhy2021}
\bibinfo{author}{Buteau, E.}, \bibinfo{author}{Lee, J.}, \bibinfo{year}{2021}.
\newblock \bibinfo{title}{Hey {{Alexa}}, why do we use voice assistants? {{The}} driving factors of voice assistant technology use}.
\newblock \bibinfo{journal}{Communication Research Reports} \bibinfo{volume}{38}, \bibinfo{pages}{336--345}.
\newblock \DOIprefix\doi{10.1080/08824096.2021.1980380}.
\bibitem[{Cabrera-S{\'a}nchez et~al.(2021)Cabrera-S{\'a}nchez, Villarejo-Ramos, Li{\'e}bana-Cabanillas and Shaikh}]{cabrera2021identifying}
\bibinfo{author}{Cabrera-S{\'a}nchez, J.P.}, \bibinfo{author}{Villarejo-Ramos, {\'A}.F.}, \bibinfo{author}{Li{\'e}bana-Cabanillas, F.}, \bibinfo{author}{Shaikh, A.A.}, \bibinfo{year}{2021}.
\newblock \bibinfo{title}{Identifying relevant segments of ai applications adopters--expanding the utaut2’s variables}.
\newblock \bibinfo{journal}{Telematics and Informatics} \bibinfo{volume}{58}, \bibinfo{pages}{101529}.
\bibitem[{Cannizzaro et~al.(2020)Cannizzaro, Procter, Ma and Maple}]{cannizzaro2020trust}
\bibinfo{author}{Cannizzaro, S.}, \bibinfo{author}{Procter, R.}, \bibinfo{author}{Ma, S.}, \bibinfo{author}{Maple, C.}, \bibinfo{year}{2020}.
\newblock \bibinfo{title}{Trust in the smart home: Findings from a nationally representative survey in the uk}.
\newblock \bibinfo{journal}{Plos one} \bibinfo{volume}{15}, \bibinfo{pages}{e0231615}.
\bibitem[{Cao et~al.(2022)Cao, Sun, Goh, Wang and Kuiavska}]{cao2022adoption}
\bibinfo{author}{Cao, D.}, \bibinfo{author}{Sun, Y.}, \bibinfo{author}{Goh, E.}, \bibinfo{author}{Wang, R.}, \bibinfo{author}{Kuiavska, K.}, \bibinfo{year}{2022}.
\newblock \bibinfo{title}{Adoption of smart voice assistants technology among airbnb guests: a revised self-efficacy-based value adoption model (svam)}.
\newblock \bibinfo{journal}{International Journal of Hospitality Management} \bibinfo{volume}{101}, \bibinfo{pages}{103124}.
\bibitem[{Carlos~Roca et~al.(2009)Carlos~Roca, Jos{\'e}~Garc{\'i}a and Jos{\'e} De La~Vega}]{carlosrocaImportancePerceivedTrust2009}
\bibinfo{author}{Carlos~Roca, J.}, \bibinfo{author}{Jos{\'e}~Garc{\'i}a, J.}, \bibinfo{author}{Jos{\'e} De La~Vega, J.}, \bibinfo{year}{2009}.
\newblock \bibinfo{title}{The importance of perceived trust, security and privacy in online trading systems}.
\newblock \bibinfo{journal}{Information Management \& Computer Security} \bibinfo{volume}{17}, \bibinfo{pages}{96--113}.
\newblock \DOIprefix\doi{10.1108/09685220910963983}.
\bibitem[{Casadei et~al.(2022)Casadei, Schl{\"o}gl and Bergmann}]{casadei2022chatbots}
\bibinfo{author}{Casadei, A.}, \bibinfo{author}{Schl{\"o}gl, S.}, \bibinfo{author}{Bergmann, M.}, \bibinfo{year}{2022}.
\newblock \bibinfo{title}{Chatbots for robotic process automation: Investigating perceived trust and user satisfaction}, in: \bibinfo{booktitle}{2022 IEEE 3rd International Conference on Human-Machine Systems (ICHMS)}, \bibinfo{organization}{IEEE}. pp. \bibinfo{pages}{1--6}.
\bibitem[{Casal{\'o} et~al.(2007)Casal{\'o}, Flavi{\'a}n and Guinal{\'i}u}]{casaloRoleSecurityPrivacy2007b}
\bibinfo{author}{Casal{\'o}, L.V.}, \bibinfo{author}{Flavi{\'a}n, C.}, \bibinfo{author}{Guinal{\'i}u, M.}, \bibinfo{year}{2007}.
\newblock \bibinfo{title}{The role of security, privacy, usability and reputation in the development of online banking}.
\newblock \bibinfo{journal}{Online Information Review} \bibinfo{volume}{31}, \bibinfo{pages}{583--603}.
\newblock \DOIprefix\doi{10.1108/14684520710832315}.
\bibitem[{Center(2019)}]{Pew2019}
\bibinfo{author}{Center, P.R.}, \bibinfo{year}{2019}.
\newblock \bibinfo{title}{American trends panel wave 49}.
\newblock \bibinfo{howpublished}{[Online] \url{https://www.pewresearch.org/internet/dataset/american-trends-panel-wave-49/}}.
\bibitem[{Cha et~al.(2021)Cha, Wi, Park and Kim}]{chaSustainabilityCalculusAdopting2021}
\bibinfo{author}{Cha, H.S.}, \bibinfo{author}{Wi, J.H.}, \bibinfo{author}{Park, C.}, \bibinfo{author}{Kim, T.}, \bibinfo{year}{2021}.
\newblock \bibinfo{title}{Sustainability {{Calculus}} in {{Adopting Smart Speakers}}{\textemdash}{{Personalized Services}} and {{Privacy Risks}}}.
\newblock \bibinfo{journal}{Sustainability} \bibinfo{volume}{13}, \bibinfo{pages}{602}.
\newblock \DOIprefix\doi{10.3390/su13020602}.
\bibitem[{Chandra et~al.(2010)Chandra, Srivastava and Theng}]{chandraEvaluatingRoleTrust2010}
\bibinfo{author}{Chandra, S.}, \bibinfo{author}{Srivastava, S.C.}, \bibinfo{author}{Theng, Y.L.}, \bibinfo{year}{2010}.
\newblock \bibinfo{title}{Evaluating the {{Role}} of {{Trust}} in {{Consumer Adoption}} of {{Mobile Payment Systems}}: {{An Empirical Analysis}}}.
\newblock \bibinfo{journal}{Communications of the Association for Information Systems} \bibinfo{volume}{27}.
\newblock \DOIprefix\doi{10.17705/1CAIS.02729}.
\bibitem[{Chattaraman et~al.(2019)Chattaraman, Kwon, Gilbert and Ross}]{chattaraman2019should}
\bibinfo{author}{Chattaraman, V.}, \bibinfo{author}{Kwon, W.S.}, \bibinfo{author}{Gilbert, J.E.}, \bibinfo{author}{Ross, K.}, \bibinfo{year}{2019}.
\newblock \bibinfo{title}{Should ai-based, conversational digital assistants employ social-or task-oriented interaction style? a task-competency and reciprocity perspective for older adults}.
\newblock \bibinfo{journal}{Computers in Human Behavior} \bibinfo{volume}{90}, \bibinfo{pages}{315--330}.
\bibitem[{Chen(2018)}]{chen2018revisiting}
\bibinfo{author}{Chen, H.T.}, \bibinfo{year}{2018}.
\newblock \bibinfo{title}{Revisiting the privacy paradox on social media with an extended privacy calculus model: The effect of privacy concerns, privacy self-efficacy, and social capital on privacy management}.
\newblock \bibinfo{journal}{American behavioral scientist} \bibinfo{volume}{62}, \bibinfo{pages}{1392--1412}.
\bibitem[{Chen and Park(2021)}]{chen2021anthropomorphism}
\bibinfo{author}{Chen, Q.Q.}, \bibinfo{author}{Park, H.J.}, \bibinfo{year}{2021}.
\newblock \bibinfo{title}{How anthropomorphism affects trust in intelligent personal assistants}.
\newblock \bibinfo{journal}{Industrial Management \& Data Systems} \bibinfo{volume}{121}, \bibinfo{pages}{2722--2737}.
\bibitem[{Cheng et~al.(2006)Cheng, Lam and Yeung}]{chengAdoptionInternetBanking2006a}
\bibinfo{author}{Cheng, T.C.E.}, \bibinfo{author}{Lam, D.Y.C.}, \bibinfo{author}{Yeung, A.C.L.}, \bibinfo{year}{2006}.
\newblock \bibinfo{title}{Adoption of internet banking: {{An}} empirical study in {{Hong Kong}}}.
\newblock \bibinfo{journal}{Decision Support Systems} \bibinfo{volume}{42}, \bibinfo{pages}{1558--1572}.
\newblock \DOIprefix\doi{10.1016/j.dss.2006.01.002}.
\bibitem[{Cheng et~al.(2022)Cheng, Zhang, Cohen and Mou}]{cheng2022human}
\bibinfo{author}{Cheng, X.}, \bibinfo{author}{Zhang, X.}, \bibinfo{author}{Cohen, J.}, \bibinfo{author}{Mou, J.}, \bibinfo{year}{2022}.
\newblock \bibinfo{title}{Human vs. ai: Understanding the impact of anthropomorphism on consumer response to chatbots from the perspective of trust and relationship norms}.
\newblock \bibinfo{journal}{Information Processing \& Management} \bibinfo{volume}{59}, \bibinfo{pages}{102940}.
\bibitem[{Cheng and Jiang(2020)}]{chengHowAIdrivenChatbots2020}
\bibinfo{author}{Cheng, Y.}, \bibinfo{author}{Jiang, H.}, \bibinfo{year}{2020}.
\newblock \bibinfo{title}{How {{Do AI-driven Chatbots Impact User Experience}}? {{Examining Gratifications}}, {{Perceived Privacy Risk}}, {{Satisfaction}}, {{Loyalty}}, and {{Continued Use}}}.
\newblock \bibinfo{journal}{Journal of Broadcasting \& Electronic Media} \bibinfo{volume}{64}, \bibinfo{pages}{592--614}.
\newblock \DOIprefix\doi{10.1080/08838151.2020.1834296}.
\bibitem[{Child et~al.(2009)Child, Pearson and Petronio}]{childBloggingCommunicationPrivacy2009}
\bibinfo{author}{Child, J.T.}, \bibinfo{author}{Pearson, J.C.}, \bibinfo{author}{Petronio, S.}, \bibinfo{year}{2009}.
\newblock \bibinfo{title}{Blogging, communication, and privacy management: {{Development}} of the {{Blogging Privacy Management Measure}}}.
\newblock \bibinfo{journal}{Journal of the American Society for Information Science and Technology} \bibinfo{volume}{60}, \bibinfo{pages}{2079--2094}.
\newblock \DOIprefix\doi{10.1002/asi.21122}.
\bibitem[{Chita-Tegmark et~al.(2021)Chita-Tegmark, Law, Rabb and Scheutz}]{chita2021can}
\bibinfo{author}{Chita-Tegmark, M.}, \bibinfo{author}{Law, T.}, \bibinfo{author}{Rabb, N.}, \bibinfo{author}{Scheutz, M.}, \bibinfo{year}{2021}.
\newblock \bibinfo{title}{Can you trust your trust measure?}, in: \bibinfo{booktitle}{Proceedings of the 2021 ACM/IEEE international conference on human-robot interaction}, pp. \bibinfo{pages}{92--100}.
\bibitem[{Cho(2019)}]{choHeyGoogleCan2019b}
\bibinfo{author}{Cho, E.}, \bibinfo{year}{2019}.
\newblock \bibinfo{title}{Hey {{Google}}, {{Can I Ask You Something}} in {{Private}}?}, in: \bibinfo{booktitle}{Proceedings of the 2019 {{CHI Conference}} on {{Human Factors}} in {{Computing Systems}}}, \bibinfo{publisher}{{ACM}}, \bibinfo{address}{{Glasgow Scotland Uk}}. pp. \bibinfo{pages}{1--9}.
\newblock \DOIprefix\doi{10.1145/3290605.3300488}.
\bibitem[{Cho et~al.(2020)Cho, Sundar, Abdullah and Motalebi}]{choWillDeletingHistory2020e}
\bibinfo{author}{Cho, E.}, \bibinfo{author}{Sundar, S.S.}, \bibinfo{author}{Abdullah, S.}, \bibinfo{author}{Motalebi, N.}, \bibinfo{year}{2020}.
\newblock \bibinfo{title}{Will {{Deleting History Make Alexa More Trustworthy}}?: {{Effects}} of {{Privacy}} and {{Content Customization}} on {{User Experience}} of {{Smart Speakers}}}, in: \bibinfo{booktitle}{Proceedings of the 2020 {{CHI Conference}} on {{Human Factors}} in {{Computing Systems}}}, \bibinfo{publisher}{{ACM}}, \bibinfo{address}{{Honolulu HI USA}}. pp. \bibinfo{pages}{1--13}.
\newblock \DOIprefix\doi{10.1145/3313831.3376551}.
\bibitem[{Cho(2006)}]{cho2006mechanism}
\bibinfo{author}{Cho, J.}, \bibinfo{year}{2006}.
\newblock \bibinfo{title}{The mechanism of trust and distrust formation and their relational outcomes}.
\newblock \bibinfo{journal}{Journal of retailing} \bibinfo{volume}{82}, \bibinfo{pages}{25--35}.
\bibitem[{Choi et~al.(2018)Choi, Park and Jung}]{choiRolePrivacyFatigue2018}
\bibinfo{author}{Choi, H.}, \bibinfo{author}{Park, J.}, \bibinfo{author}{Jung, Y.}, \bibinfo{year}{2018}.
\newblock \bibinfo{title}{The role of privacy fatigue in online privacy behavior}.
\newblock \bibinfo{journal}{Computers in Human Behavior} \bibinfo{volume}{81}, \bibinfo{pages}{42--51}.
\newblock \DOIprefix\doi{10.1016/j.chb.2017.12.001}.
\bibitem[{Choi and Zhou(2023)}]{choi2023inducing}
\bibinfo{author}{Choi, S.}, \bibinfo{author}{Zhou, J.}, \bibinfo{year}{2023}.
\newblock \bibinfo{title}{Inducing consumers’ self-disclosure through the fit between chatbot’s interaction styles and regulatory focus}.
\newblock \bibinfo{journal}{Journal of Business Research} \bibinfo{volume}{166}, \bibinfo{pages}{114127}.
\bibitem[{Choi et~al.(2020)Choi, Thompson and Demiris}]{choiUseInternetofThingsSmart2020}
\bibinfo{author}{Choi, Y.K.}, \bibinfo{author}{Thompson, H.J.}, \bibinfo{author}{Demiris, G.}, \bibinfo{year}{2020}.
\newblock \bibinfo{title}{Use of an {{Internet-of-Things Smart Home System}} for {{Healthy Aging}} in {{Older Adults}} in {{Residential Settings}}: {{Pilot Feasibility Study}}}.
\newblock \bibinfo{journal}{JMIR Aging} \bibinfo{volume}{3}, \bibinfo{pages}{e21964}.
\newblock \DOIprefix\doi{10.2196/21964}.
\bibitem[{Chong et~al.(2012)Chong, Chan and Ooi}]{chongPredictingConsumerDecisions2012}
\bibinfo{author}{Chong, A.Y.L.}, \bibinfo{author}{Chan, F.T.S.}, \bibinfo{author}{Ooi, K.B.}, \bibinfo{year}{2012}.
\newblock \bibinfo{title}{Predicting consumer decisions to adopt mobile commerce: {{Cross}} country empirical examination between {{China}} and {{Malaysia}}}.
\newblock \bibinfo{journal}{Decision Support Systems} \bibinfo{volume}{53}, \bibinfo{pages}{34--43}.
\newblock \DOIprefix\doi{10.1016/j.dss.2011.12.001}.
\bibitem[{Choung et~al.(2023)Choung, David and Ross}]{choung2023trust}
\bibinfo{author}{Choung, H.}, \bibinfo{author}{David, P.}, \bibinfo{author}{Ross, A.}, \bibinfo{year}{2023}.
\newblock \bibinfo{title}{Trust in ai and its role in the acceptance of ai technologies}.
\newblock \bibinfo{journal}{International Journal of Human--Computer Interaction} \bibinfo{volume}{39}, \bibinfo{pages}{1727--1739}.
\bibitem[{Colesca(2009)}]{colesca2009understanding}
\bibinfo{author}{Colesca, S.E.}, \bibinfo{year}{2009}.
\newblock \bibinfo{title}{Understanding trust in e-government}.
\newblock \bibinfo{journal}{Engineering Economics} \bibinfo{volume}{63}.
\bibitem[{Colnago et~al.(2022)Colnago, Cranor, Acquisti and Stanton}]{colnagoItConcernPreferencea}
\bibinfo{author}{Colnago, J.}, \bibinfo{author}{Cranor, L.F.}, \bibinfo{author}{Acquisti, A.}, \bibinfo{author}{Stanton, K.H.}, \bibinfo{year}{2022}.
\newblock \bibinfo{title}{Is it a concern or a preference? an investigation into the ability of privacy scales to capture and distinguish granular privacy constructs}, in: \bibinfo{booktitle}{Eighteenth Symposium on Usable Privacy and Security (SOUPS 2022)}, \bibinfo{publisher}{USENIX Association}, \bibinfo{address}{Boston, MA}. pp. \bibinfo{pages}{331--346}.
\newblock \URLprefix \url{https://www.usenix.org/conference/soups2022/presentation/colnago}.
\bibitem[{Compeau and Higgins(1995)}]{compeau1995computer}
\bibinfo{author}{Compeau, D.R.}, \bibinfo{author}{Higgins, C.A.}, \bibinfo{year}{1995}.
\newblock \bibinfo{title}{Computer self-efficacy: Development of a measure and initial test}.
\newblock \bibinfo{journal}{MIS quarterly} , \bibinfo{pages}{189--211}.
\bibitem[{Corritore et~al.(2003)Corritore, Kracher and Wiedenbeck}]{corritore2003line}
\bibinfo{author}{Corritore, C.L.}, \bibinfo{author}{Kracher, B.}, \bibinfo{author}{Wiedenbeck, S.}, \bibinfo{year}{2003}.
\newblock \bibinfo{title}{On-line trust: concepts, evolving themes, a model}.
\newblock \bibinfo{journal}{International journal of human-computer studies} \bibinfo{volume}{58}, \bibinfo{pages}{737--758}.
\bibitem[{Corritore et~al.(2005)Corritore, Marble, Wiedenbeck, Kracher and Chandran}]{corritore2005measuring}
\bibinfo{author}{Corritore, C.L.}, \bibinfo{author}{Marble, R.P.}, \bibinfo{author}{Wiedenbeck, S.}, \bibinfo{author}{Kracher, B.}, \bibinfo{author}{Chandran, A.}, \bibinfo{year}{2005}.
\newblock \bibinfo{title}{Measuring online trust of websites: Credibility, perceived ease of use, and risk}.
\newblock \bibinfo{journal}{AMCIS 2005 Proceedings} .
\bibitem[{Cowan et~al.(2017)Cowan, Pantidi, Coyle, Morrissey, Clarke, {Al-Shehri}, Earley and Bandeira}]{cowanWhatCanHelp2017a}
\bibinfo{author}{Cowan, B.R.}, \bibinfo{author}{Pantidi, N.}, \bibinfo{author}{Coyle, D.}, \bibinfo{author}{Morrissey, K.}, \bibinfo{author}{Clarke, P.}, \bibinfo{author}{{Al-Shehri}, S.}, \bibinfo{author}{Earley, D.}, \bibinfo{author}{Bandeira, N.}, \bibinfo{year}{2017}.
\newblock \bibinfo{title}{"{{What}} can i help you with?": Infrequent users' experiences of intelligent personal assistants}, in: \bibinfo{booktitle}{Proceedings of the 19th {{International Conference}} on {{Human-Computer Interaction}} with {{Mobile Devices}} and {{Services}}}, \bibinfo{publisher}{{Association for Computing Machinery}}, \bibinfo{address}{{New York, NY, USA}}. pp. \bibinfo{pages}{1--12}.
\newblock \DOIprefix\doi{10.1145/3098279.3098539}.
\bibitem[{Croes and Antheunis(2021)}]{croes2021can}
\bibinfo{author}{Croes, E.A.}, \bibinfo{author}{Antheunis, M.L.}, \bibinfo{year}{2021}.
\newblock \bibinfo{title}{Can we be friends with mitsuku? a longitudinal study on the process of relationship formation between humans and a social chatbot}.
\newblock \bibinfo{journal}{Journal of Social and Personal Relationships} \bibinfo{volume}{38}, \bibinfo{pages}{279--300}.
\bibitem[{Culnan and Williams(2009)}]{Culnan2009}
\bibinfo{author}{Culnan, M.}, \bibinfo{author}{Williams, C.}, \bibinfo{year}{2009}.
\newblock \bibinfo{title}{How {Ethics} {Can} {Enhance} {Organizational} {Privacy}: {Lessons} from the {ChoicePoint} and {TJX} {Data} {Breaches}}.
\newblock \bibinfo{journal}{Management Information Systems Quarterly} \bibinfo{volume}{33}, \bibinfo{pages}{673--687}.
\newblock \URLprefix \url{https://aisel.aisnet.org/misq/vol33/iss4/6}.
\bibitem[{Cyr et~al.(2004)Cyr, Bonanni and Ilsever}]{cyr2004design}
\bibinfo{author}{Cyr, D.}, \bibinfo{author}{Bonanni, C.}, \bibinfo{author}{Ilsever, J.}, \bibinfo{year}{2004}.
\newblock \bibinfo{title}{Design and e-loyalty across cultures in electronic commerce}, in: \bibinfo{booktitle}{Proceedings of the 6th international conference on Electronic commerce}, pp. \bibinfo{pages}{351--360}.
\bibitem[{Cyr et~al.(2009)Cyr, Head, Larios and Pan}]{cyr2009exploring}
\bibinfo{author}{Cyr, D.}, \bibinfo{author}{Head, M.}, \bibinfo{author}{Larios, H.}, \bibinfo{author}{Pan, B.}, \bibinfo{year}{2009}.
\newblock \bibinfo{title}{Exploring human images in website design: a multi-method approach}.
\newblock \bibinfo{journal}{MIS quarterly} , \bibinfo{pages}{539--566}.
\bibitem[{Davis(1989)}]{davis1989perceivedTAM}
\bibinfo{author}{Davis, F.D.}, \bibinfo{year}{1989}.
\newblock \bibinfo{title}{Perceived usefulness, perceived ease of use, and user acceptance of information technology}.
\newblock \bibinfo{journal}{MIS quarterly} , \bibinfo{pages}{319--340}.
\bibitem[{Davis(1993)}]{davisUserAcceptanceInformation1993}
\bibinfo{author}{Davis, F.D.}, \bibinfo{year}{1993}.
\newblock \bibinfo{title}{User acceptance of information technology: System characteristics, user perceptions and behavioral impacts}.
\newblock \bibinfo{journal}{International Journal of Man-Machine Studies} \bibinfo{volume}{38}, \bibinfo{pages}{475--487}.
\newblock \DOIprefix\doi{10.1006/imms.1993.1022}.
\bibitem[{{de Barcelos Silva} et~al.(2020){de Barcelos Silva}, Gomes, {da Costa}, {da Rosa Righi}, Barbosa, Pessin, De~Doncker and Federizzi}]{debarcelossilvaIntelligentPersonalAssistants2020a}
\bibinfo{author}{{de Barcelos Silva}, A.}, \bibinfo{author}{Gomes, M.M.}, \bibinfo{author}{{da Costa}, C.A.}, \bibinfo{author}{{da Rosa Righi}, R.}, \bibinfo{author}{Barbosa, J.L.V.}, \bibinfo{author}{Pessin, G.}, \bibinfo{author}{De~Doncker, G.}, \bibinfo{author}{Federizzi, G.}, \bibinfo{year}{2020}.
\newblock \bibinfo{title}{Intelligent personal assistants: {{A}} systematic literature review}.
\newblock \bibinfo{journal}{Expert Systems with Applications} \bibinfo{volume}{147}, \bibinfo{pages}{113193}.
\newblock \DOIprefix\doi{10.1016/j.eswa.2020.113193}.
\bibitem[{De~Graaf et~al.(2016)De~Graaf, Ben~Allouch and van Dijk}]{de2016long}
\bibinfo{author}{De~Graaf, M.M.}, \bibinfo{author}{Ben~Allouch, S.}, \bibinfo{author}{van Dijk, J.A.}, \bibinfo{year}{2016}.
\newblock \bibinfo{title}{Long-term evaluation of a social robot in real homes}.
\newblock \bibinfo{journal}{Interaction studies} \bibinfo{volume}{17}, \bibinfo{pages}{462--491}.
\bibitem[{Dekkal et~al.(2023)Dekkal, Arcand, Prom~Tep, Rajaobelina and Ricard}]{dekkalFactorsAffectingUser2023a}
\bibinfo{author}{Dekkal, M.}, \bibinfo{author}{Arcand, M.}, \bibinfo{author}{Prom~Tep, S.}, \bibinfo{author}{Rajaobelina, L.}, \bibinfo{author}{Ricard, L.}, \bibinfo{year}{2023}.
\newblock \bibinfo{title}{Factors affecting user trust and intention in adopting chatbots: The moderating role of technology anxiety in insurtech}.
\newblock \bibinfo{journal}{Journal of Financial Services Marketing} \DOIprefix\doi{10.1057/s41264-023-00230-y}.
\bibitem[{Dinev et~al.(2006)Dinev, Bellotto, Hart, Russo, Serra and Colautti}]{dinev2006privacy}
\bibinfo{author}{Dinev, T.}, \bibinfo{author}{Bellotto, M.}, \bibinfo{author}{Hart, P.}, \bibinfo{author}{Russo, V.}, \bibinfo{author}{Serra, I.}, \bibinfo{author}{Colautti, C.}, \bibinfo{year}{2006}.
\newblock \bibinfo{title}{Privacy calculus model in e-commerce--a study of italy and the united states}.
\newblock \bibinfo{journal}{European Journal of Information Systems} \bibinfo{volume}{15}, \bibinfo{pages}{389--402}.
\bibitem[{Dinev and Hart(2005)}]{dinevInternetPrivacyConcerns2005a}
\bibinfo{author}{Dinev, T.}, \bibinfo{author}{Hart, P.}, \bibinfo{year}{2005}.
\newblock \bibinfo{title}{Internet {{Privacy Concerns}} and {{Social Awareness}} as {{Determinants}} of {{Intention}} to {{Transact}}}.
\newblock \bibinfo{journal}{International Journal of Electronic Commerce} \bibinfo{volume}{10}, \bibinfo{pages}{7--29}.
\newblock \DOIprefix\doi{10.2753/JEC1086-4415100201}.
\bibitem[{Distler et~al.(2021)Distler, Fassl, Habib, Krombholz, Lenzini, Lallemand, Cranor and Koenig}]{distlerSystematicLiteratureReview2021}
\bibinfo{author}{Distler, V.}, \bibinfo{author}{Fassl, M.}, \bibinfo{author}{Habib, H.}, \bibinfo{author}{Krombholz, K.}, \bibinfo{author}{Lenzini, G.}, \bibinfo{author}{Lallemand, C.}, \bibinfo{author}{Cranor, L.F.}, \bibinfo{author}{Koenig, V.}, \bibinfo{year}{2021}.
\newblock \bibinfo{title}{A {{Systematic Literature Review}} of {{Empirical Methods}} and {{Risk Representation}} in {{Usable Privacy}} and {{Security Research}}}.
\newblock \bibinfo{journal}{ACM Transactions on Computer-Human Interaction} \bibinfo{volume}{28}, \bibinfo{pages}{1--50}.
\newblock \DOIprefix\doi{10.1145/3469845}.
\bibitem[{Durall~Gazulla et~al.(2023)Durall~Gazulla, Martins and Fern{\'a}ndez-Ferrer}]{durall2023designing}
\bibinfo{author}{Durall~Gazulla, E.}, \bibinfo{author}{Martins, L.}, \bibinfo{author}{Fern{\'a}ndez-Ferrer, M.}, \bibinfo{year}{2023}.
\newblock \bibinfo{title}{Designing learning technology collaboratively: Analysis of a chatbot co-design}.
\newblock \bibinfo{journal}{Education and Information Technologies} \bibinfo{volume}{28}, \bibinfo{pages}{109--134}.
\bibitem[{Dwyer et~al.(2007)Dwyer, Hiltz and Passerini}]{dwyer2007trust}
\bibinfo{author}{Dwyer, C.}, \bibinfo{author}{Hiltz, S.}, \bibinfo{author}{Passerini, K.}, \bibinfo{year}{2007}.
\newblock \bibinfo{title}{Trust and privacy concern within social networking sites: A comparison of facebook and myspace}.
\newblock \bibinfo{journal}{AMCIS 2007 proceedings} , \bibinfo{pages}{339}.
\bibitem[{Dzindolet et~al.(2003)Dzindolet, Peterson, Pomranky, Pierce and Beck}]{dzindolet2003roleoftrustinautomationreliance}
\bibinfo{author}{Dzindolet, M.T.}, \bibinfo{author}{Peterson, S.A.}, \bibinfo{author}{Pomranky, R.A.}, \bibinfo{author}{Pierce, L.G.}, \bibinfo{author}{Beck, H.P.}, \bibinfo{year}{2003}.
\newblock \bibinfo{title}{The role of trust in automation reliance}.
\newblock \bibinfo{journal}{International journal of human-computer studies} \bibinfo{volume}{58}, \bibinfo{pages}{697--718}.
\bibitem[{Easwara~Moorthy and Vu(2015)}]{easwara2015privacy}
\bibinfo{author}{Easwara~Moorthy, A.}, \bibinfo{author}{Vu, K.P.L.}, \bibinfo{year}{2015}.
\newblock \bibinfo{title}{Privacy concerns for use of voice activated personal assistant in the public space}.
\newblock \bibinfo{journal}{International Journal of Human-Computer Interaction} \bibinfo{volume}{31}, \bibinfo{pages}{307--335}.
\bibitem[{Edu et~al.(2021)Edu, Such and {Suarez-Tangil}}]{eduSmartHomePersonal2021}
\bibinfo{author}{Edu, J.S.}, \bibinfo{author}{Such, J.M.}, \bibinfo{author}{{Suarez-Tangil}, G.}, \bibinfo{year}{2021}.
\newblock \bibinfo{title}{Smart {{Home Personal Assistants}}: {{A Security}} and {{Privacy Review}}}.
\newblock \bibinfo{journal}{ACM Computing Surveys} \bibinfo{volume}{53}, \bibinfo{pages}{1--36}.
\newblock \DOIprefix\doi{10.1145/3412383}, \href{http://arxiv.org/abs/1903.05593}{\tt arXiv:1903.05593}.
\bibitem[{van Eeuwen(2017)}]{eeuwenMobileConversationalCommerce2017}
\bibinfo{author}{van Eeuwen, M.}, \bibinfo{year}{2017}.
\newblock \bibinfo{title}{Mobile Conversational Commerce: Messenger Chatbots as the next Interface between Businesses and Consumers}.
\newblock Master's thesis. University of Twente.
\bibitem[{Ejdys(2018)}]{ejdys2018building}
\bibinfo{author}{Ejdys, J.}, \bibinfo{year}{2018}.
\newblock \bibinfo{title}{Building technology trust in ict application at a university}.
\newblock \bibinfo{journal}{International Journal of Emerging Markets} \bibinfo{volume}{13}, \bibinfo{pages}{980--997}.
\bibitem[{Ejdys et~al.(2019)Ejdys, Ginevicius, Rozsa and Janoskova}]{ejdys2019role}
\bibinfo{author}{Ejdys, J.}, \bibinfo{author}{Ginevicius, R.}, \bibinfo{author}{Rozsa, Z.}, \bibinfo{author}{Janoskova, K.}, \bibinfo{year}{2019}.
\newblock \bibinfo{title}{The role of perceived risk and security level in building trust in e-government solutions}.
\newblock \bibinfo{journal}{E+ M Ekonomie a Management} \bibinfo{volume}{22}, \bibinfo{pages}{220--235}.
\bibitem[{Elkins and Derrick(2013)}]{elkins2013sound}
\bibinfo{author}{Elkins, A.C.}, \bibinfo{author}{Derrick, D.C.}, \bibinfo{year}{2013}.
\newblock \bibinfo{title}{The sound of trust: voice as a measurement of trust during interactions with embodied conversational agents}.
\newblock \bibinfo{journal}{Group decision and negotiation} \bibinfo{volume}{22}, \bibinfo{pages}{897--913}.
\bibitem[{{Emami-Naeini} et~al.(2021){Emami-Naeini}, Dheenadhayalan, Agarwal and Cranor}]{emami-naeiniWhichPrivacySecurity2021b}
\bibinfo{author}{{Emami-Naeini}, P.}, \bibinfo{author}{Dheenadhayalan, J.}, \bibinfo{author}{Agarwal, Y.}, \bibinfo{author}{Cranor, L.F.}, \bibinfo{year}{2021}.
\newblock \bibinfo{title}{Which {{Privacy}} and {{Security Attributes Most Impact Consumers}}' {{Risk Perception}} and {{Willingness}} to {{Purchase IoT Devices}}?}, in: \bibinfo{booktitle}{2021 {{IEEE Symposium}} on {{Security}} and {{Privacy}} ({{SP}})}, pp. \bibinfo{pages}{519--536}.
\newblock \DOIprefix\doi{10.1109/SP40001.2021.00112}.
\bibitem[{Erskine et~al.(2019)Erskine, Khojah and McDaniel}]{erskine2019location}
\bibinfo{author}{Erskine, M.A.}, \bibinfo{author}{Khojah, M.}, \bibinfo{author}{McDaniel, A.E.}, \bibinfo{year}{2019}.
\newblock \bibinfo{title}{Location selection using heat maps: Relative advantage, task-technology fit, and decision-making performance}.
\newblock \bibinfo{journal}{Computers in Human Behavior} \bibinfo{volume}{101}, \bibinfo{pages}{151--162}.
\bibitem[{Everard and Galletta(2005)}]{everard2005presentation}
\bibinfo{author}{Everard, A.}, \bibinfo{author}{Galletta, D.F.}, \bibinfo{year}{2005}.
\newblock \bibinfo{title}{How presentation flaws affect perceived site quality, trust, and intention to purchase from an online store}.
\newblock \bibinfo{journal}{Journal of management information systems} \bibinfo{volume}{22}, \bibinfo{pages}{56--95}.
\bibitem[{Fahn and Riener(2021)}]{fahnTimeGetConversational2021a}
\bibinfo{author}{Fahn, V.}, \bibinfo{author}{Riener, A.}, \bibinfo{year}{2021}.
\newblock \bibinfo{title}{Time to {{Get Conversational}}: {{Assessment}} of the {{Potential}} of {{Conversational User Interfaces}} for {{Mobile Banking}}}, in: \bibinfo{booktitle}{Mensch Und {{Computer}} 2021}, \bibinfo{publisher}{{ACM}}, \bibinfo{address}{{Ingolstadt Germany}}. pp. \bibinfo{pages}{34--43}.
\newblock \DOIprefix\doi{10.1145/3473856.3473872}.
\bibitem[{Fakhimi et~al.(2023)Fakhimi, Garry and Biggemann}]{fakhimi2023effectsQualitative}
\bibinfo{author}{Fakhimi, A.}, \bibinfo{author}{Garry, T.}, \bibinfo{author}{Biggemann, S.}, \bibinfo{year}{2023}.
\newblock \bibinfo{title}{The effects of anthropomorphised virtual conversational assistants on consumer engagement and trust during service encounters}.
\newblock \bibinfo{journal}{Australasian Marketing Journal} , \bibinfo{pages}{14413582231181140}.
\bibitem[{Faklaris et~al.(2019)Faklaris, Dabbish and Hong}]{faklaris2019self}
\bibinfo{author}{Faklaris, C.}, \bibinfo{author}{Dabbish, L.A.}, \bibinfo{author}{Hong, J.I.}, \bibinfo{year}{2019}.
\newblock \bibinfo{title}{A $\{$Self-Report$\}$ measure of $\{$End-User$\}$ security attitudes ($\{$$\{$$\{$$\{$$\{$SA-6$\}$$\}$$\}$$\}$$\}$)}, in: \bibinfo{booktitle}{Fifteenth Symposium on Usable Privacy and Security (SOUPS 2019)}, pp. \bibinfo{pages}{61--77}.
\bibitem[{Faqih and Jaradat(2015)}]{faqihAssessingModeratingEffect2015}
\bibinfo{author}{Faqih, K.M.S.}, \bibinfo{author}{Jaradat, M.I.R.M.}, \bibinfo{year}{2015}.
\newblock \bibinfo{title}{Assessing the moderating effect of gender differences and individualism-collectivism at individual-level on the adoption of mobile commerce technology: {{TAM3}} perspective}.
\newblock \bibinfo{journal}{Journal of Retailing and Consumer Services} \bibinfo{volume}{22}, \bibinfo{pages}{37--52}.
\newblock \DOIprefix\doi{10.1016/j.jretconser.2014.09.006}.
\bibitem[{Farooq et~al.(2022)Farooq, Jeske, van Schaik and Moran}]{farooq2022voice}
\bibinfo{author}{Farooq, A.}, \bibinfo{author}{Jeske, D.}, \bibinfo{author}{van Schaik, P.}, \bibinfo{author}{Moran, M.}, \bibinfo{year}{2022}.
\newblock \bibinfo{title}{Voice assistants:(physical) device use perceptions, acceptance, and privacy concerns}, in: \bibinfo{booktitle}{Conference on e-Business, e-Services and e-Society}, \bibinfo{organization}{Springer}. pp. \bibinfo{pages}{485--498}.
\bibitem[{Featherman and Pavlou(2003)}]{featherman2003predicting}
\bibinfo{author}{Featherman, M.S.}, \bibinfo{author}{Pavlou, P.A.}, \bibinfo{year}{2003}.
\newblock \bibinfo{title}{Predicting e-services adoption: a perceived risk facets perspective}.
\newblock \bibinfo{journal}{International journal of human-computer studies} \bibinfo{volume}{59}, \bibinfo{pages}{451--474}.
\bibitem[{Fernaeus et~al.(2010)Fernaeus, H{\aa}kansson, Jacobsson and Ljungblad}]{fernaeus2010you}
\bibinfo{author}{Fernaeus, Y.}, \bibinfo{author}{H{\aa}kansson, M.}, \bibinfo{author}{Jacobsson, M.}, \bibinfo{author}{Ljungblad, S.}, \bibinfo{year}{2010}.
\newblock \bibinfo{title}{How do you play with a robotic toy animal? a long-term study of pleo}, in: \bibinfo{booktitle}{Proceedings of the 9th international Conference on interaction Design and Children}, pp. \bibinfo{pages}{39--48}.
\bibitem[{Ferrari et~al.(2016)Ferrari, Paladino and Jetten}]{ferrariBlurringHumanMachine2016}
\bibinfo{author}{Ferrari, F.}, \bibinfo{author}{Paladino, M.}, \bibinfo{author}{Jetten, J.}, \bibinfo{year}{2016}.
\newblock \bibinfo{title}{Blurring {{Human}}{\textendash}{{Machine Distinctions}}: {{Anthropomorphic Appearance}} in {{Social Robots}} as a {{Threat}} to {{Human Distinctiveness}}}.
\newblock \bibinfo{journal}{International Journal of Social Robotics} \bibinfo{volume}{8}.
\newblock \DOIprefix\doi{10.1007/s12369-016-0338-y}.
\bibitem[{Fiske et~al.(2018)Fiske, Cuddy, Glick and Xu}]{fiske2018model}
\bibinfo{author}{Fiske, S.T.}, \bibinfo{author}{Cuddy, A.J.}, \bibinfo{author}{Glick, P.}, \bibinfo{author}{Xu, J.}, \bibinfo{year}{2018}.
\newblock \bibinfo{title}{A model of (often mixed) stereotype content: Competence and warmth respectively follow from perceived status and competition}, in: \bibinfo{booktitle}{Social cognition}. \bibinfo{publisher}{Routledge}, pp. \bibinfo{pages}{162--214}.
\bibitem[{Fiske and Macrae(2012)}]{fiske2012sage}
\bibinfo{author}{Fiske, S.T.}, \bibinfo{author}{Macrae, C.N.}, \bibinfo{year}{2012}.
\newblock \bibinfo{title}{The SAGE handbook of social cognition}.
\newblock \bibinfo{publisher}{Sage}.
\bibitem[{Fitzsimons(2000)}]{fitzsimons2000consumer}
\bibinfo{author}{Fitzsimons, G.J.}, \bibinfo{year}{2000}.
\newblock \bibinfo{title}{Consumer response to stockouts}.
\newblock \bibinfo{journal}{Journal of consumer research} \bibinfo{volume}{27}, \bibinfo{pages}{249--266}.
\bibitem[{Fitzsimons et~al.(1997)Fitzsimons, Greenleaf and Lehmann}]{fitzsimons1997decision}
\bibinfo{author}{Fitzsimons, G.J.}, \bibinfo{author}{Greenleaf, E.A.}, \bibinfo{author}{Lehmann, D.R.}, \bibinfo{year}{1997}.
\newblock \bibinfo{title}{Decision and consumption satisfaction: Implications for channel relations}.
\newblock \bibinfo{journal}{Marketing Studies Center Working Paper Series} \bibinfo{volume}{313}.
\bibitem[{Flavi{\'a}n and Guinal{\'i}u(2006)}]{flavianConsumerTrustPerceived2006}
\bibinfo{author}{Flavi{\'a}n, C.}, \bibinfo{author}{Guinal{\'i}u, M.}, \bibinfo{year}{2006}.
\newblock \bibinfo{title}{Consumer trust, perceived security and privacy policy: {{Three}} basic elements of loyalty to a web site}.
\newblock \bibinfo{journal}{Industrial Management \& Data Systems} \bibinfo{volume}{106}, \bibinfo{pages}{601--620}.
\newblock \DOIprefix\doi{10.1108/02635570610666403}.
\bibitem[{Franke et~al.(2019)Franke, Attig and Wessel}]{franke2019personal}
\bibinfo{author}{Franke, T.}, \bibinfo{author}{Attig, C.}, \bibinfo{author}{Wessel, D.}, \bibinfo{year}{2019}.
\newblock \bibinfo{title}{A personal resource for technology interaction: development and validation of the affinity for technology interaction (ati) scale}.
\newblock \bibinfo{journal}{International Journal of Human--Computer Interaction} \bibinfo{volume}{35}, \bibinfo{pages}{456--467}.
\bibitem[{Furini et~al.(2020)Furini, Mirri, Montangero and Prandi}]{furiniUsageSmartSpeakers2020b}
\bibinfo{author}{Furini, M.}, \bibinfo{author}{Mirri, S.}, \bibinfo{author}{Montangero, M.}, \bibinfo{author}{Prandi, C.}, \bibinfo{year}{2020}.
\newblock \bibinfo{title}{On the {{Usage}} of {{Smart Speakers During}} the {{Covid-19 Coronavirus Lockdown}}}, in: \bibinfo{booktitle}{Proceedings of the 6th {{EAI International Conference}} on {{Smart Objects}} and {{Technologies}} for {{Social Good}}}, \bibinfo{publisher}{{ACM}}, \bibinfo{address}{{Antwerp Belgium}}. pp. \bibinfo{pages}{187--192}.
\newblock \DOIprefix\doi{10.1145/3411170.3411260}.
\bibitem[{{Garc{\'i}a de Blanes Sebasti{\'a}n} et~al.(2022){Garc{\'i}a de Blanes Sebasti{\'a}n}, Sarmiento~Guede and Antonovica}]{garciadeblanessebastianApplicationExtensionUTAUT22022}
\bibinfo{author}{{Garc{\'i}a de Blanes Sebasti{\'a}n}, M.}, \bibinfo{author}{Sarmiento~Guede, J.R.}, \bibinfo{author}{Antonovica, A.}, \bibinfo{year}{2022}.
\newblock \bibinfo{title}{Application and extension of the {{UTAUT2}} model for determining behavioral intention factors in use of the artificial intelligence virtual assistants}.
\newblock \bibinfo{journal}{Frontiers in Psychology} \bibinfo{volume}{13}.
\bibitem[{Gauder et~al.(2023)Gauder, Pepino, Riera, Brussino, Vidal, Gravano and Ferrer}]{gauder2023towards}
\bibinfo{author}{Gauder, L.}, \bibinfo{author}{Pepino, L.}, \bibinfo{author}{Riera, P.}, \bibinfo{author}{Brussino, S.}, \bibinfo{author}{Vidal, J.}, \bibinfo{author}{Gravano, A.}, \bibinfo{author}{Ferrer, L.}, \bibinfo{year}{2023}.
\newblock \bibinfo{title}{Towards detecting the level of trust in the skills of a virtual assistant from the user’s speech}.
\newblock \bibinfo{journal}{Computer Speech \& Language} \bibinfo{volume}{80}, \bibinfo{pages}{101487}.
\bibitem[{Gefen et~al.(2003)Gefen, Karahanna and Straub}]{gefen2003trust}
\bibinfo{author}{Gefen, D.}, \bibinfo{author}{Karahanna, E.}, \bibinfo{author}{Straub, D.W.}, \bibinfo{year}{2003}.
\newblock \bibinfo{title}{Trust and tam in online shopping: An integrated model}.
\newblock \bibinfo{journal}{MIS quarterly} , \bibinfo{pages}{51--90}.
\bibitem[{Gnambs et~al.(2018)Gnambs, Scharl and Schroeders}]{gnambs2018structure}
\bibinfo{author}{Gnambs, T.}, \bibinfo{author}{Scharl, A.}, \bibinfo{author}{Schroeders, U.}, \bibinfo{year}{2018}.
\newblock \bibinfo{title}{The structure of the rosenberg self-esteem scale}.
\newblock \bibinfo{journal}{Zeitschrift f{\"u}r Psychologie} .
\bibitem[{Go and Sundar(2019)}]{go2019humanizing}
\bibinfo{author}{Go, E.}, \bibinfo{author}{Sundar, S.S.}, \bibinfo{year}{2019}.
\newblock \bibinfo{title}{Humanizing chatbots: The effects of visual, identity and conversational cues on humanness perceptions}.
\newblock \bibinfo{journal}{Computers in Human Behavior} \bibinfo{volume}{97}, \bibinfo{pages}{304--316}.
\bibitem[{Goodman and Mayhorn(2023)}]{goodman2023s}
\bibinfo{author}{Goodman, K.L.}, \bibinfo{author}{Mayhorn, C.B.}, \bibinfo{year}{2023}.
\newblock \bibinfo{title}{It's not what you say but how you say it: Examining the influence of perceived voice assistant gender and pitch on trust and reliance}.
\newblock \bibinfo{journal}{Applied Ergonomics} \bibinfo{volume}{106}, \bibinfo{pages}{103864}.
\bibitem[{Gray(1977)}]{gray1977social}
\bibinfo{author}{Gray, G.}, \bibinfo{year}{1977}.
\newblock \bibinfo{title}{The social psychology of telecommunications: by john short, ederyn williams, and bruce christie 195 pages, john wiley and sons, london, new york, sydney, toronto, 1976}.
\bibitem[{Gross(2021)}]{Gross2021ValidityAR}
\bibinfo{author}{Gross, T.}, \bibinfo{year}{2021}.
\newblock \bibinfo{title}{Validity and reliability of the scale internet users’ information privacy concerns (iuipc)}.
\newblock \bibinfo{journal}{Proceedings on Privacy Enhancing Technologies} \bibinfo{volume}{2021}, \bibinfo{pages}{235 -- 258}.
\bibitem[{Guerreiro et~al.(2022)Guerreiro, Loureiro and Ribeiro}]{guerreiroAdvertisingAcceptanceSmart2022}
\bibinfo{author}{Guerreiro, J.}, \bibinfo{author}{Loureiro, S.M.C.}, \bibinfo{author}{Ribeiro, C.}, \bibinfo{year}{2022}.
\newblock \bibinfo{title}{Advertising acceptance via smart speakers}.
\newblock \bibinfo{journal}{Spanish Journal of Marketing - ESIC} \bibinfo{volume}{26}, \bibinfo{pages}{286--308}.
\newblock \DOIprefix\doi{10.1108/SJME-02-2022-0028}.
\bibitem[{Gulati et~al.(2018)Gulati, Sousa and Lamas}]{gulati2018modelling}
\bibinfo{author}{Gulati, S.}, \bibinfo{author}{Sousa, S.}, \bibinfo{author}{Lamas, D.}, \bibinfo{year}{2018}.
\newblock \bibinfo{title}{Modelling trust in human-like technologies}, in: \bibinfo{booktitle}{Proceedings of the 9th Indian Conference on Human-Computer Interaction}, pp. \bibinfo{pages}{1--10}.
\bibitem[{Gulati et~al.(2019)Gulati, Sousa and Lamas}]{gulati2019design}
\bibinfo{author}{Gulati, S.}, \bibinfo{author}{Sousa, S.}, \bibinfo{author}{Lamas, D.}, \bibinfo{year}{2019}.
\newblock \bibinfo{title}{Design, development and evaluation of a human-computer trust scale}.
\newblock \bibinfo{journal}{Behaviour \& Information Technology} \bibinfo{volume}{38}, \bibinfo{pages}{1004--1015}.
\bibitem[{Gupta et~al.(2022)Gupta, Basu, Ghantasala, Qiu and Gadiraju}]{gupta2022trust}
\bibinfo{author}{Gupta, A.}, \bibinfo{author}{Basu, D.}, \bibinfo{author}{Ghantasala, R.}, \bibinfo{author}{Qiu, S.}, \bibinfo{author}{Gadiraju, U.}, \bibinfo{year}{2022}.
\newblock \bibinfo{title}{To trust or not to trust: How a conversational interface affects trust in a decision support system}, in: \bibinfo{booktitle}{Proceedings of the ACM Web Conference 2022}, pp. \bibinfo{pages}{3531--3540}.
\bibitem[{Gupta et~al.(2019)Gupta, Hajika, Pai, Duenser, Lochner and Billinghurst}]{gupta2019ai}
\bibinfo{author}{Gupta, K.}, \bibinfo{author}{Hajika, R.}, \bibinfo{author}{Pai, Y.S.}, \bibinfo{author}{Duenser, A.}, \bibinfo{author}{Lochner, M.}, \bibinfo{author}{Billinghurst, M.}, \bibinfo{year}{2019}.
\newblock \bibinfo{title}{In ai we trust: Investigating the relationship between biosignals, trust and cognitive load in vr}, in: \bibinfo{booktitle}{Proceedings of the 25th ACM Symposium on Virtual Reality Software and Technology}, pp. \bibinfo{pages}{1--10}.
\bibitem[{Gupta et~al.(2020)Gupta, Hajika, Pai, Duenser, Lochner and Billinghurst}]{gupta2020measuring}
\bibinfo{author}{Gupta, K.}, \bibinfo{author}{Hajika, R.}, \bibinfo{author}{Pai, Y.S.}, \bibinfo{author}{Duenser, A.}, \bibinfo{author}{Lochner, M.}, \bibinfo{author}{Billinghurst, M.}, \bibinfo{year}{2020}.
\newblock \bibinfo{title}{Measuring human trust in a virtual assistant using physiological sensing in virtual reality}, in: \bibinfo{booktitle}{2020 IEEE Conference on virtual reality and 3D user interfaces (VR)}, \bibinfo{organization}{IEEE}. pp. \bibinfo{pages}{756--765}.
\bibitem[{Gupta et~al.(2009)Gupta, Yadav and Varadarajan}]{gupta2009task}
\bibinfo{author}{Gupta, P.}, \bibinfo{author}{Yadav, M.S.}, \bibinfo{author}{Varadarajan, R.}, \bibinfo{year}{2009}.
\newblock \bibinfo{title}{How task-facilitative interactive tools foster buyers’ trust in online retailers: a process view of trust development in the electronic marketplace}.
\newblock \bibinfo{journal}{Journal of Retailing} \bibinfo{volume}{85}, \bibinfo{pages}{159--176}.
\bibitem[{Gusenbauer and Haddaway(2020)}]{gusenbauerWhichAcademicSearch2020}
\bibinfo{author}{Gusenbauer, M.}, \bibinfo{author}{Haddaway, N.}, \bibinfo{year}{2020}.
\newblock \bibinfo{title}{Which {{Academic Search Systems}} are {{Suitable}} for {{Systematic Reviews}} or {{Meta}}-{{Analyses}}? {{Evaluating Retrieval Qualities}} of {{Google Scholar}}, {{PubMed}} and 26 other {{Resources}} [{{OPEN ACCESS}}]}.
\newblock \bibinfo{journal}{Research Synthesis Methods} \bibinfo{volume}{11}, \bibinfo{pages}{181--217}.
\newblock \DOIprefix\doi{10.1002/jrsm.1378}.
\bibitem[{Ha et~al.(2021)Ha, Chen, Uy and Capistrano}]{ha2021exploringWillingnessToDisclose}
\bibinfo{author}{Ha, Q.A.}, \bibinfo{author}{Chen, J.V.}, \bibinfo{author}{Uy, H.U.}, \bibinfo{author}{Capistrano, E.P.}, \bibinfo{year}{2021}.
\newblock \bibinfo{title}{Exploring the privacy concerns in using intelligent virtual assistants under perspectives of information sensitivity and anthropomorphism}.
\newblock \bibinfo{journal}{International journal of human--computer interaction} \bibinfo{volume}{37}, \bibinfo{pages}{512--527}.
\bibitem[{Hair et~al.(2021)Hair, Hult, Ringle, Sarstedt, Danks and Ray}]{hairPartialLeastSquares2021}
\bibinfo{author}{Hair, J.F.}, \bibinfo{author}{Hult, G.T.M.}, \bibinfo{author}{Ringle, C.M.}, \bibinfo{author}{Sarstedt, M.}, \bibinfo{author}{Danks, N.P.}, \bibinfo{author}{Ray, S.}, \bibinfo{year}{2021}.
\newblock \bibinfo{title}{Partial {{Least Squares Structural Equation Modeling}} ({{PLS-SEM}}) {{Using R}}: {{A Workbook}}}.
\newblock Classroom {{Companion}}: {{Business}}, \bibinfo{publisher}{{Springer International Publishing}}, \bibinfo{address}{{Cham}}.
\newblock \DOIprefix\doi{10.1007/978-3-030-80519-7}.
\bibitem[{Han and Yang(2018)}]{hanUnderstandingAdoptionIntelligent2018a}
\bibinfo{author}{Han, S.}, \bibinfo{author}{Yang, H.}, \bibinfo{year}{2018}.
\newblock \bibinfo{title}{Understanding adoption of intelligent personal assistants: {{A}} parasocial relationship perspective}.
\newblock \bibinfo{journal}{Industrial Management \& Data Systems} \bibinfo{volume}{118}, \bibinfo{pages}{618--636}.
\newblock \DOIprefix\doi{10.1108/IMDS-05-2017-0214}.
\bibitem[{Haney et~al.(2019)Haney, Furman, Theofanos and Fahl}]{haney2019perceptions}
\bibinfo{author}{Haney, J.}, \bibinfo{author}{Furman, S.M.}, \bibinfo{author}{Theofanos, M.}, \bibinfo{author}{Fahl, Y.A.}, \bibinfo{year}{2019}.
\newblock \bibinfo{title}{Perceptions of smart home privacy and security responsibility, concerns, and mitigations} .
\bibitem[{Harkous et~al.(2016)Harkous, Fawaz, Shin and Aberer}]{harkous2016pribots}
\bibinfo{author}{Harkous, H.}, \bibinfo{author}{Fawaz, K.}, \bibinfo{author}{Shin, K.G.}, \bibinfo{author}{Aberer, K.}, \bibinfo{year}{2016}.
\newblock \bibinfo{title}{$\{$PriBots$\}$: Conversational privacy with chatbots}, in: \bibinfo{booktitle}{Twelfth Symposium on Usable Privacy and Security (SOUPS 2016)}.
\bibitem[{Harwood and Garry(2017)}]{harwood2017internet}
\bibinfo{author}{Harwood, T.}, \bibinfo{author}{Garry, T.}, \bibinfo{year}{2017}.
\newblock \bibinfo{title}{Internet of things: understanding trust in techno-service systems}.
\newblock \bibinfo{journal}{Journal of Service Management} \bibinfo{volume}{28}, \bibinfo{pages}{442--475}.
\bibitem[{Hasan et~al.(2021)Hasan, Shams and Rahman}]{hasanConsumerTrustPerceived2021}
\bibinfo{author}{Hasan, R.}, \bibinfo{author}{Shams, R.}, \bibinfo{author}{Rahman, M.}, \bibinfo{year}{2021}.
\newblock \bibinfo{title}{Consumer trust and perceived risk for voice-controlled artificial intelligence: {{The}} case of {{Siri}}}.
\newblock \bibinfo{journal}{Journal of Business Research} \bibinfo{volume}{131}, \bibinfo{pages}{591--597}.
\newblock \DOIprefix\doi{10.1016/j.jbusres.2020.12.012}.
\bibitem[{Hassanein and Head(2007)}]{hassaneinManipulatingPerceivedSocial2007}
\bibinfo{author}{Hassanein, K.}, \bibinfo{author}{Head, M.}, \bibinfo{year}{2007}.
\newblock \bibinfo{title}{Manipulating perceived social presence through the web interface and its impact on attitude towards online shopping}.
\newblock \bibinfo{journal}{International Journal of Human-Computer Studies} \bibinfo{volume}{65}, \bibinfo{pages}{689--708}.
\newblock \DOIprefix\doi{10.1016/j.ijhcs.2006.11.018}.
\bibitem[{Hassanein et~al.(2009)Hassanein, Head and Ju}]{hassanein2009cross}
\bibinfo{author}{Hassanein, K.}, \bibinfo{author}{Head, M.}, \bibinfo{author}{Ju, C.}, \bibinfo{year}{2009}.
\newblock \bibinfo{title}{A cross-cultural comparison of the impact of social presence on website trust, usefulness and enjoyment}.
\newblock \bibinfo{journal}{International Journal of Electronic Business} \bibinfo{volume}{7}, \bibinfo{pages}{625--641}.
\bibitem[{Hinds(1998)}]{hinds1998user}
\bibinfo{author}{Hinds, P.J.}, \bibinfo{year}{1998}.
\newblock \bibinfo{title}{User control and its many facets: A study of perceived control in human-computer interaction}.
\newblock \bibinfo{publisher}{Hewlett Packard Laboratories California}.
\bibitem[{Ho and MacDorman(2010)}]{ho2010revisiting}
\bibinfo{author}{Ho, C.C.}, \bibinfo{author}{MacDorman, K.F.}, \bibinfo{year}{2010}.
\newblock \bibinfo{title}{Revisiting the uncanny valley theory: Developing and validating an alternative to the godspeed indices}.
\newblock \bibinfo{journal}{Computers in Human Behavior} \bibinfo{volume}{26}, \bibinfo{pages}{1508--1518}.
\bibitem[{Hoff and Bashir(2015)}]{hoff2015trust}
\bibinfo{author}{Hoff, K.A.}, \bibinfo{author}{Bashir, M.}, \bibinfo{year}{2015}.
\newblock \bibinfo{title}{Trust in automation: Integrating empirical evidence on factors that influence trust}.
\newblock \bibinfo{journal}{Human factors} \bibinfo{volume}{57}, \bibinfo{pages}{407--434}.
\bibitem[{Holzwarth et~al.(2006)Holzwarth, Janiszewski and Neumann}]{holzwarth2006influence}
\bibinfo{author}{Holzwarth, M.}, \bibinfo{author}{Janiszewski, C.}, \bibinfo{author}{Neumann, M.M.}, \bibinfo{year}{2006}.
\newblock \bibinfo{title}{The influence of avatars on online consumer shopping behavior}.
\newblock \bibinfo{journal}{Journal of marketing} \bibinfo{volume}{70}, \bibinfo{pages}{19--36}.
\bibitem[{Hornung and Smolnik(2022)}]{hornungAIInvadingWorkplace2022b}
\bibinfo{author}{Hornung, O.}, \bibinfo{author}{Smolnik, S.}, \bibinfo{year}{2022}.
\newblock \bibinfo{title}{{{AI}} invading the workplace: Negative emotions towards the organizational use of personal virtual assistants}.
\newblock \bibinfo{journal}{Electronic Markets} \bibinfo{volume}{32}, \bibinfo{pages}{123--138}.
\newblock \DOIprefix\doi{10.1007/s12525-021-00493-0}.
\bibitem[{Hsu and Lin(2018)}]{hsuExploringFactorsAffecting2018}
\bibinfo{author}{Hsu, C.L.}, \bibinfo{author}{Lin, J.C.C.}, \bibinfo{year}{2018}.
\newblock \bibinfo{title}{Exploring {{Factors Affecting}} the {{Adoption}} of {{Internet}} of {{Things Services}}}.
\newblock \bibinfo{journal}{Journal of Computer Information Systems} \bibinfo{volume}{58}, \bibinfo{pages}{49--57}.
\newblock \DOIprefix\doi{10.1080/08874417.2016.1186524}.
\bibitem[{Hsu and Lee(2023)}]{hsu2023semantic}
\bibinfo{author}{Hsu, W.C.}, \bibinfo{author}{Lee, M.H.}, \bibinfo{year}{2023}.
\newblock \bibinfo{title}{Semantic technology and anthropomorphism: Exploring the impacts of voice assistant personality on user trust, perceived risk, and attitude}.
\newblock \bibinfo{journal}{Journal of Global Information Management (JGIM)} \bibinfo{volume}{31}, \bibinfo{pages}{1--21}.
\bibitem[{Hu et~al.(2021)Hu, Lu et~al.}]{hu2021dual}
\bibinfo{author}{Hu, P.}, \bibinfo{author}{Lu, Y.}, et~al., \bibinfo{year}{2021}.
\newblock \bibinfo{title}{Dual humanness and trust in conversational ai: A person-centered approach}.
\newblock \bibinfo{journal}{Computers in Human Behavior} \bibinfo{volume}{119}, \bibinfo{pages}{106727}.
\bibitem[{Ischen et~al.(2020)Ischen, Araujo, Voorveld, Van~Noort and Smit}]{ischenPrivacyConcernsChatbot2020b}
\bibinfo{author}{Ischen, C.}, \bibinfo{author}{Araujo, T.}, \bibinfo{author}{Voorveld, H.}, \bibinfo{author}{Van~Noort, G.}, \bibinfo{author}{Smit, E.}, \bibinfo{year}{2020}.
\newblock \bibinfo{title}{Privacy {{Concerns}} in {{Chatbot Interactions}}}, in: \bibinfo{editor}{F{\o}lstad, A.}, \bibinfo{editor}{Araujo, T.}, \bibinfo{editor}{Papadopoulos, S.}, \bibinfo{editor}{Law, E.L.C.}, \bibinfo{editor}{Granmo, O.C.}, \bibinfo{editor}{Luger, E.}, \bibinfo{editor}{Brandtzaeg, P.B.} (Eds.), \bibinfo{booktitle}{Chatbot {{Research}} and {{Design}}}. \bibinfo{publisher}{{Springer International Publishing}}, \bibinfo{address}{{Cham}}. volume \bibinfo{volume}{11970}, pp. \bibinfo{pages}{34--48}.
\newblock \DOIprefix\doi{10.1007/978-3-030-39540-7_3}.
\bibitem[{Ivarsson and Lindwall(2023)}]{ivarsson2023suspicious}
\bibinfo{author}{Ivarsson, J.}, \bibinfo{author}{Lindwall, O.}, \bibinfo{year}{2023}.
\newblock \bibinfo{title}{Suspicious minds: the problem of trust and conversational agents}.
\newblock \bibinfo{journal}{Computer Supported Cooperative Work (CSCW)} , \bibinfo{pages}{1--27}.
\bibitem[{Jain et~al.(2022)Jain, Basu, Dwivedi and Kaur}]{jain2022interactive}
\bibinfo{author}{Jain, S.}, \bibinfo{author}{Basu, S.}, \bibinfo{author}{Dwivedi, Y.K.}, \bibinfo{author}{Kaur, S.}, \bibinfo{year}{2022}.
\newblock \bibinfo{title}{Interactive voice assistants--does brand credibility assuage privacy risks?}
\newblock \bibinfo{journal}{Journal of Business Research} \bibinfo{volume}{139}, \bibinfo{pages}{701--717}.
\bibitem[{Jameel and Karem(2022)}]{jameel2022perceived}
\bibinfo{author}{Jameel, A.S.}, \bibinfo{author}{Karem, M.A.}, \bibinfo{year}{2022}.
\newblock \bibinfo{title}{Perceived trust and enjoyment: Predicting behavioural intention to use mobile payment systems}, in: \bibinfo{booktitle}{2022 International Conference on Intelligent Technology, System and Service for Internet of Everything (ITSS-IoE)}, \bibinfo{organization}{IEEE}. pp. \bibinfo{pages}{1--6}.
\bibitem[{Jattamart and Leelasantitham(2020)}]{jattamartPerspectivesSocialMedia2020}
\bibinfo{author}{Jattamart, A.}, \bibinfo{author}{Leelasantitham, A.}, \bibinfo{year}{2020}.
\newblock \bibinfo{title}{Perspectives to social media usage of depressed patients and caregivers affecting to change the health behavior of patients in terms of information and perceived privacy risks}.
\newblock \bibinfo{journal}{Heliyon} \bibinfo{volume}{6}, \bibinfo{pages}{e04244}.
\newblock \DOIprefix\doi{10.1016/j.heliyon.2020.e04244}.
\bibitem[{Jian et~al.(2000)Jian, Bisantz and Drury}]{jianFoundationsEmpiricallyDetermined2000a}
\bibinfo{author}{Jian, J.Y.}, \bibinfo{author}{Bisantz, A.M.}, \bibinfo{author}{Drury, C.G.}, \bibinfo{year}{2000}.
\newblock \bibinfo{title}{Foundations for an {{Empirically Determined Scale}} of {{Trust}} in {{Automated Systems}}}.
\newblock \bibinfo{journal}{International Journal of Cognitive Ergonomics} \bibinfo{volume}{4}, \bibinfo{pages}{53--71}.
\newblock \DOIprefix\doi{10.1207/S15327566IJCE0401_04}.
\bibitem[{Jiang et~al.(2023)Jiang, Yang and Zheng}]{jiang2023make}
\bibinfo{author}{Jiang, Y.}, \bibinfo{author}{Yang, X.}, \bibinfo{author}{Zheng, T.}, \bibinfo{year}{2023}.
\newblock \bibinfo{title}{Make chatbots more adaptive: Dual pathways linking human-like cues and tailored response to trust in interactions with chatbots}.
\newblock \bibinfo{journal}{Computers in Human Behavior} \bibinfo{volume}{138}, \bibinfo{pages}{107485}.
\bibitem[{John(2022)}]{John2019}
\bibinfo{author}{John, A.S.}, \bibinfo{year}{2022}.
\newblock \bibinfo{title}{How to {{Set Up}} a {{Smart Speaker}} for {{Privacy}}}.
\newblock \bibinfo{howpublished}{[Accessed Online 13-12-2023] \url{https://www.consumerreports.org/electronics-computers/privacy/smart-speaker-privacy-settings-a8054333211/}}.
\bibitem[{Joy et~al.(2022)Joy, Kotevska and {Al-Masri}}]{joyInvestigatingUsersPrivacy2022b}
\bibinfo{author}{Joy, D.}, \bibinfo{author}{Kotevska, O.}, \bibinfo{author}{{Al-Masri}, E.}, \bibinfo{year}{2022}.
\newblock \bibinfo{title}{Investigating {{Users}}' {{Privacy Concerns}} of {{Internet}} of {{Things}} ({{IoT}}) {{Smart Devices}}}, in: \bibinfo{booktitle}{2022 {{IEEE}} 4th {{Eurasia Conference}} on {{IOT}}, {{Communication}} and {{Engineering}} ({{ECICE}})}, pp. \bibinfo{pages}{70--76}.
\newblock \DOIprefix\doi{10.1109/ECICE55674.2022.10042926}.
\bibitem[{K{\"a}{\"a}ri{\"a}(2017)}]{kaaria2017technology}
\bibinfo{author}{K{\"a}{\"a}ri{\"a}, A.}, \bibinfo{year}{2017}.
\newblock \bibinfo{title}{Technology acceptance of voice assistants: anthropomorphism as factor}.
\newblock Master's thesis. University of Jyv{\"a}skyl{\"a}.
\bibitem[{Kaddour et~al.(2023)Kaddour, Harris, Mozes, Bradley, Raileanu and McHardy}]{kaddour2023challenges}
\bibinfo{author}{Kaddour, J.}, \bibinfo{author}{Harris, J.}, \bibinfo{author}{Mozes, M.}, \bibinfo{author}{Bradley, H.}, \bibinfo{author}{Raileanu, R.}, \bibinfo{author}{McHardy, R.}, \bibinfo{year}{2023}.
\newblock \bibinfo{title}{Challenges and applications of large language models}.
\newblock \bibinfo{journal}{arXiv preprint arXiv:2307.10169} .
\bibitem[{Kang and Oh(2023)}]{kang2023communication}
\bibinfo{author}{Kang, H.}, \bibinfo{author}{Oh, J.}, \bibinfo{year}{2023}.
\newblock \bibinfo{title}{Communication privacy management for smart speaker use: Integrating the role of privacy self-efficacy and the multidimensional view}.
\newblock \bibinfo{journal}{New Media \& Society} \bibinfo{volume}{25}, \bibinfo{pages}{1153--1175}.
\bibitem[{Kankanhalli et~al.(2005)Kankanhalli, Tan and Wei}]{kankanhalli2005contributing}
\bibinfo{author}{Kankanhalli, A.}, \bibinfo{author}{Tan, B.C.}, \bibinfo{author}{Wei, K.K.}, \bibinfo{year}{2005}.
\newblock \bibinfo{title}{Contributing knowledge to electronic knowledge repositories: An empirical investigation}.
\newblock \bibinfo{journal}{MIS quarterly} , \bibinfo{pages}{113--143}.
\bibitem[{Kasilingam(2020)}]{kasilingamUnderstandingAttitudeIntention2020a}
\bibinfo{author}{Kasilingam, D.L.}, \bibinfo{year}{2020}.
\newblock \bibinfo{title}{Understanding the attitude and intention to use smartphone chatbots for shopping}.
\newblock \bibinfo{journal}{Technology in Society} \bibinfo{volume}{62}, \bibinfo{pages}{101280}.
\newblock \DOIprefix\doi{10.1016/j.techsoc.2020.101280}.
\bibitem[{Kefi et~al.(2021)Kefi, Besson, Sokolova and Aouina-Mejri}]{kefi2021privacy}
\bibinfo{author}{Kefi, H.}, \bibinfo{author}{Besson, E.}, \bibinfo{author}{Sokolova, K.}, \bibinfo{author}{Aouina-Mejri, C.}, \bibinfo{year}{2021}.
\newblock \bibinfo{title}{Privacy and intelligent virtual assistants usage across generations}.
\newblock \bibinfo{journal}{Syst{\`e}mes d’information et management} \bibinfo{volume}{26}, \bibinfo{pages}{43--76}.
\bibitem[{Keith et~al.(2013)Keith, Thompson, Hale, Lowry and Greer}]{keith2013information}
\bibinfo{author}{Keith, M.J.}, \bibinfo{author}{Thompson, S.C.}, \bibinfo{author}{Hale, J.}, \bibinfo{author}{Lowry, P.B.}, \bibinfo{author}{Greer, C.}, \bibinfo{year}{2013}.
\newblock \bibinfo{title}{Information disclosure on mobile devices: Re-examining privacy calculus with actual user behavior}.
\newblock \bibinfo{journal}{International journal of human-computer studies} \bibinfo{volume}{71}, \bibinfo{pages}{1163--1173}.
\bibitem[{Khalid et~al.(2018)Khalid, Shiung, Sheng and Helander}]{khalidTrustVirtualAgent2018}
\bibinfo{author}{Khalid, H.M.}, \bibinfo{author}{Shiung, L.W.}, \bibinfo{author}{Sheng, V.B.}, \bibinfo{author}{Helander, M.G.}, \bibinfo{year}{2018}.
\newblock \bibinfo{title}{Trust of {{Virtual Agent}} in {{Multi Actor Interactions}}}.
\newblock \bibinfo{journal}{Journal of Robotics, Networking and Artificial Life} \bibinfo{volume}{4}, \bibinfo{pages}{295--298}.
\newblock \DOIprefix\doi{10.2991/jrnal.2018.4.4.8}.
\bibitem[{Kim et~al.(2010)Kim, Mirusmonov and Lee}]{kim2010empirical}
\bibinfo{author}{Kim, C.}, \bibinfo{author}{Mirusmonov, M.}, \bibinfo{author}{Lee, I.}, \bibinfo{year}{2010}.
\newblock \bibinfo{title}{An empirical examination of factors influencing the intention to use mobile payment}.
\newblock \bibinfo{journal}{Computers in human behavior} \bibinfo{volume}{26}, \bibinfo{pages}{310--322}.
\bibitem[{Kim et~al.(2019)Kim, Park, Park and Ahn}]{kimWillingnessProvidePersonal2019b}
\bibinfo{author}{Kim, D.}, \bibinfo{author}{Park, K.}, \bibinfo{author}{Park, Y.}, \bibinfo{author}{Ahn, J.H.}, \bibinfo{year}{2019}.
\newblock \bibinfo{title}{Willingness to provide personal information: {{Perspective}} of privacy calculus in {{IoT}} services}.
\newblock \bibinfo{journal}{Computers in Human Behavior} \bibinfo{volume}{92}, \bibinfo{pages}{273--281}.
\newblock \DOIprefix\doi{10.1016/j.chb.2018.11.022}.
\bibitem[{Kim et~al.(2008)Kim, Ferrin and Rao}]{kim2008trust}
\bibinfo{author}{Kim, D.J.}, \bibinfo{author}{Ferrin, D.L.}, \bibinfo{author}{Rao, H.R.}, \bibinfo{year}{2008}.
\newblock \bibinfo{title}{A trust-based consumer decision-making model in electronic commerce: The role of trust, perceived risk, and their antecedents}.
\newblock \bibinfo{journal}{Decision support systems} \bibinfo{volume}{44}, \bibinfo{pages}{544--564}.
\bibitem[{Kim et~al.(2011)Kim, Chung and Lee}]{kim2011effect}
\bibinfo{author}{Kim, M.J.}, \bibinfo{author}{Chung, N.}, \bibinfo{author}{Lee, C.K.}, \bibinfo{year}{2011}.
\newblock \bibinfo{title}{The effect of perceived trust on electronic commerce: Shopping online for tourism products and services in south korea}.
\newblock \bibinfo{journal}{Tourism Management} \bibinfo{volume}{32}, \bibinfo{pages}{256--265}.
\bibitem[{King and He(2006)}]{king2006meta}
\bibinfo{author}{King, W.R.}, \bibinfo{author}{He, J.}, \bibinfo{year}{2006}.
\newblock \bibinfo{title}{A meta-analysis of the technology acceptance model}.
\newblock \bibinfo{journal}{Information \& management} \bibinfo{volume}{43}, \bibinfo{pages}{740--755}.
\bibitem[{Kirmani et~al.(2017)Kirmani, Hamilton, Thompson and Lantzy}]{kirmani2017doing}
\bibinfo{author}{Kirmani, A.}, \bibinfo{author}{Hamilton, R.W.}, \bibinfo{author}{Thompson, D.V.}, \bibinfo{author}{Lantzy, S.}, \bibinfo{year}{2017}.
\newblock \bibinfo{title}{Doing well versus doing good: The differential effect of underdog positioning on moral and competent service providers}.
\newblock \bibinfo{journal}{Journal of Marketing} \bibinfo{volume}{81}, \bibinfo{pages}{103--117}.
\bibitem[{Kitkowska et~al.(2023)Kitkowska, Karegar and W{\"a}stlund}]{kitkowskaShareProtectUnderstanding2023}
\bibinfo{author}{Kitkowska, A.}, \bibinfo{author}{Karegar, F.}, \bibinfo{author}{W{\"a}stlund, E.}, \bibinfo{year}{2023}.
\newblock \bibinfo{title}{Share or {{Protect}}: {{Understanding}} the {{Interplay}} of {{Trust}}, {{Privacy Concerns}}, and {{Data Sharing Purposes}} in {{Health}} and {{Well-Being Apps}}}, in: \bibinfo{booktitle}{Proceedings of the 15th {{Biannual Conference}} of the {{Italian SIGCHI Chapter}}}, \bibinfo{publisher}{{ACM}}, \bibinfo{address}{{Torino Italy}}. pp. \bibinfo{pages}{1--14}.
\newblock \DOIprefix\doi{10.1145/3605390.3605417}.
\bibitem[{Koh and Sundar(2010)}]{kohEffectsSpecializationComputers2010}
\bibinfo{author}{Koh, Y.J.}, \bibinfo{author}{Sundar, S.S.}, \bibinfo{year}{2010}.
\newblock \bibinfo{title}{Effects of specialization in computers, web sites, and web agents on e-commerce trust}.
\newblock \bibinfo{journal}{International Journal of Human-Computer Studies} \bibinfo{volume}{68}, \bibinfo{pages}{899--912}.
\newblock \DOIprefix\doi{10.1016/j.ijhcs.2010.08.002}.
\bibitem[{Komiak and Benbasat(2006)}]{komiak2006effects}
\bibinfo{author}{Komiak, S.Y.}, \bibinfo{author}{Benbasat, I.}, \bibinfo{year}{2006}.
\newblock \bibinfo{title}{The effects of personalization and familiarity on trust and adoption of recommendation agents}.
\newblock \bibinfo{journal}{MIS quarterly} , \bibinfo{pages}{941--960}.
\bibitem[{K{\"o}nig et~al.(2023)K{\"o}nig, Karrenbauer and Breitner}]{konigCriticalSuccessFactors2023a}
\bibinfo{author}{K{\"o}nig, C.M.}, \bibinfo{author}{Karrenbauer, C.}, \bibinfo{author}{Breitner, M.H.}, \bibinfo{year}{2023}.
\newblock \bibinfo{title}{Critical success factors and challenges for individual digital study assistants in higher education: {{A}} mixed methods analysis}.
\newblock \bibinfo{journal}{Education and Information Technologies} \bibinfo{volume}{28}, \bibinfo{pages}{4475--4503}.
\newblock \DOIprefix\doi{10.1007/s10639-022-11394-w}.
\bibitem[{Kosa(2010)}]{kosa2010vampire}
\bibinfo{author}{Kosa, T.A.}, \bibinfo{year}{2010}.
\newblock \bibinfo{title}{Vampire bats: Trust in privacy}, in: \bibinfo{booktitle}{2010 Eighth International Conference on Privacy, Security and Trust}, \bibinfo{organization}{IEEE}. pp. \bibinfo{pages}{96--102}.
\bibitem[{Kowalczuk(2018)}]{kowalczukConsumerAcceptanceSmart2018a}
\bibinfo{author}{Kowalczuk, P.}, \bibinfo{year}{2018}.
\newblock \bibinfo{title}{Consumer acceptance of smart speakers: A mixed methods approach}.
\newblock \bibinfo{journal}{Journal of Research in Interactive Marketing} \bibinfo{volume}{12}, \bibinfo{pages}{418--431}.
\newblock \DOIprefix\doi{10.1108/JRIM-01-2018-0022}.
\bibitem[{Krasnova et~al.(2010)Krasnova, Spiekermann, Koroleva and Hildebrand}]{krasnova2010online}
\bibinfo{author}{Krasnova, H.}, \bibinfo{author}{Spiekermann, S.}, \bibinfo{author}{Koroleva, K.}, \bibinfo{author}{Hildebrand, T.}, \bibinfo{year}{2010}.
\newblock \bibinfo{title}{Online social networks: Why we disclose}.
\newblock \bibinfo{journal}{Journal of information technology} \bibinfo{volume}{25}, \bibinfo{pages}{109--125}.
\bibitem[{Krasnova et~al.(2012)Krasnova, Veltri and G{\"u}nther}]{krasnovaSelfdisclosurePrivacyCalculus2012}
\bibinfo{author}{Krasnova, H.}, \bibinfo{author}{Veltri, N.F.}, \bibinfo{author}{G{\"u}nther, O.}, \bibinfo{year}{2012}.
\newblock \bibinfo{title}{Self-disclosure and {{Privacy Calculus}} on {{Social Networking Sites}}: {{The Role}} of {{Culture}}}.
\newblock \bibinfo{journal}{Business \& Information Systems Engineering} \bibinfo{volume}{4}, \bibinfo{pages}{127--135}.
\newblock \DOIprefix\doi{10.1007/s12599-012-0216-6}.
\bibitem[{Krey and Ramirez~Garcia(2022)}]{kreyVoiceAssistantsHealthcare2022b}
\bibinfo{author}{Krey, M.}, \bibinfo{author}{Ramirez~Garcia, R.}, \bibinfo{year}{2022}.
\newblock \bibinfo{title}{Voice {{Assistants}} in {{Healthcare}}: {{The Patient}}'s {{Perception}}}, in: \bibinfo{booktitle}{2022 8th {{International Conference}} on {{Information Management}} ({{ICIM}})}, pp. \bibinfo{pages}{120--130}.
\newblock \DOIprefix\doi{10.1109/ICIM56520.2022.00029}.
\bibitem[{Kummer et~al.(2017)Kummer, Recker and Bick}]{kummerTechnologyinducedAnxietyManifestations2017}
\bibinfo{author}{Kummer, T.F.}, \bibinfo{author}{Recker, J.}, \bibinfo{author}{Bick, M.}, \bibinfo{year}{2017}.
\newblock \bibinfo{title}{Technology-induced anxiety: {{Manifestations}}, cultural influences, and its effect on the adoption of sensor-based technology in {{German}} and {{Australian}} hospitals}.
\newblock \bibinfo{journal}{Information \& Management} \bibinfo{volume}{54}, \bibinfo{pages}{73--89}.
\newblock \DOIprefix\doi{10.1016/j.im.2016.04.002}.
\bibitem[{Kunkel et~al.(2019)Kunkel, Donkers, Michael, Barbu and Ziegler}]{kunkel2019let}
\bibinfo{author}{Kunkel, J.}, \bibinfo{author}{Donkers, T.}, \bibinfo{author}{Michael, L.}, \bibinfo{author}{Barbu, C.M.}, \bibinfo{author}{Ziegler, J.}, \bibinfo{year}{2019}.
\newblock \bibinfo{title}{Let me explain: Impact of personal and impersonal explanations on trust in recommender systems}, in: \bibinfo{booktitle}{Proceedings of the 2019 CHI conference on human factors in computing systems}, pp. \bibinfo{pages}{1--12}.
\bibitem[{Kwangsawad and Jattamart(2022)}]{kwangsawadOvercomingCustomerInnovation2022}
\bibinfo{author}{Kwangsawad, A.}, \bibinfo{author}{Jattamart, A.}, \bibinfo{year}{2022}.
\newblock \bibinfo{title}{Overcoming customer innovation resistance to the sustainable adoption of chatbot services: {{A}} community-enterprise perspective in {{Thailand}}}.
\newblock \bibinfo{journal}{Journal of Innovation \& Knowledge} \bibinfo{volume}{7}, \bibinfo{pages}{100211}.
\newblock \DOIprefix\doi{10.1016/j.jik.2022.100211}.
\bibitem[{Lallmahamood(2007)}]{lallmahamoodExaminationIndividualAPerceived1970}
\bibinfo{author}{Lallmahamood, M.}, \bibinfo{year}{2007}.
\newblock \bibinfo{title}{An examination of individuals perceived security and privacy of the internet in malaysia and the influence ofthis on their intention to use e-commerce: Using anextension of the technology acceptance model}.
\newblock \bibinfo{journal}{The Journal of Internet Banking and Commerce} \bibinfo{volume}{12}, \bibinfo{pages}{1--26}.
\newblock \URLprefix \url{https://api.semanticscholar.org/CorpusID:167934491}.
\bibitem[{Langer and K{\"o}nig(2018)}]{langerIntroducingTestingCreepiness2018}
\bibinfo{author}{Langer, M.}, \bibinfo{author}{K{\"o}nig, C.J.}, \bibinfo{year}{2018}.
\newblock \bibinfo{title}{Introducing and {{Testing}} the {{Creepiness}} of {{Situation Scale}} ({{CRoSS}})}.
\newblock \bibinfo{journal}{Frontiers in Psychology} \bibinfo{volume}{9}.
\bibitem[{Lankton et~al.(2015)Lankton, McKnight and Tripp}]{lankton2015technology}
\bibinfo{author}{Lankton, N.K.}, \bibinfo{author}{McKnight, D.H.}, \bibinfo{author}{Tripp, J.}, \bibinfo{year}{2015}.
\newblock \bibinfo{title}{Technology, humanness, and trust: Rethinking trust in technology}.
\newblock \bibinfo{journal}{Journal of the Association for Information Systems} \bibinfo{volume}{16}, \bibinfo{pages}{1}.
\bibitem[{Lappeman et~al.(2023)Lappeman, Marlie, Johnson and Poggenpoel}]{lappemanTrustDigitalPrivacy2023a}
\bibinfo{author}{Lappeman, J.}, \bibinfo{author}{Marlie, S.}, \bibinfo{author}{Johnson, T.}, \bibinfo{author}{Poggenpoel, S.}, \bibinfo{year}{2023}.
\newblock \bibinfo{title}{Trust and digital privacy: Willingness to disclose personal information to banking chatbot services}.
\newblock \bibinfo{journal}{Journal of Financial Services Marketing} \bibinfo{volume}{28}, \bibinfo{pages}{337--357}.
\newblock \DOIprefix\doi{10.1057/s41264-022-00154-z}.
\bibitem[{Latoschik et~al.(2017)Latoschik, Roth, Gall, Achenbach, Waltemate and Botsch}]{latoschik2017effect}
\bibinfo{author}{Latoschik, M.E.}, \bibinfo{author}{Roth, D.}, \bibinfo{author}{Gall, D.}, \bibinfo{author}{Achenbach, J.}, \bibinfo{author}{Waltemate, T.}, \bibinfo{author}{Botsch, M.}, \bibinfo{year}{2017}.
\newblock \bibinfo{title}{The effect of avatar realism in immersive social virtual realities}, in: \bibinfo{booktitle}{Proceedings of the 23rd ACM symposium on virtual reality software and technology}, pp. \bibinfo{pages}{1--10}.
\bibitem[{Lau et~al.(2018)Lau, Zimmerman and Schaub}]{lauAlexaAreYou2018c}
\bibinfo{author}{Lau, J.}, \bibinfo{author}{Zimmerman, B.}, \bibinfo{author}{Schaub, F.}, \bibinfo{year}{2018}.
\newblock \bibinfo{title}{Alexa, {{Are You Listening}}?: {{Privacy Perceptions}}, {{Concerns}} and {{Privacy-seeking Behaviors}} with {{Smart Speakers}}}.
\newblock \bibinfo{journal}{Proceedings of the ACM on Human-Computer Interaction} \bibinfo{volume}{2}, \bibinfo{pages}{1--31}.
\newblock \DOIprefix\doi{10.1145/3274371}.
\bibitem[{Laufer and Wolfe(1977)}]{Laufer1977}
\bibinfo{author}{Laufer, R.S.}, \bibinfo{author}{Wolfe, M.}, \bibinfo{year}{1977}.
\newblock \bibinfo{title}{Privacy as a concept and a social issue: A multidimensional developmental theory}.
\newblock \bibinfo{journal}{Journal of Social Issues} \bibinfo{volume}{33}, \bibinfo{pages}{22--42}.
\newblock \DOIprefix\doi{https://doi.org/10.1111/j.1540-4560.1977.tb01880.x}.
\bibitem[{Lazar et~al.(2017)Lazar, Feng and Hochheiser}]{lazarResearchMethodsHumanComputer2017}
\bibinfo{author}{Lazar, J.}, \bibinfo{author}{Feng, J.H.}, \bibinfo{author}{Hochheiser, H.}, \bibinfo{year}{2017}.
\newblock \bibinfo{title}{Research {{Methods}} in {{Human-Computer Interaction}}}.
\newblock \bibinfo{publisher}{{Morgan Kaufmann}}.
\bibitem[{Leach et~al.(2007)Leach, Ellemers and Barreto}]{leachGroupVirtueImportance2007}
\bibinfo{author}{Leach, C.W.}, \bibinfo{author}{Ellemers, N.}, \bibinfo{author}{Barreto, M.}, \bibinfo{year}{2007}.
\newblock \bibinfo{title}{Group virtue: {{The}} importance of morality (vs. competence and sociability) in the positive evaluation of in-groups}.
\newblock \bibinfo{journal}{Journal of Personality and Social Psychology} \bibinfo{volume}{93}, \bibinfo{pages}{234--249}.
\newblock \DOIprefix\doi{10.1037/0022-3514.93.2.234}.
\bibitem[{Ledbetter(2009)}]{ledbetter2009measuring}
\bibinfo{author}{Ledbetter, A.M.}, \bibinfo{year}{2009}.
\newblock \bibinfo{title}{Measuring online communication attitude: Instrument development and validation}.
\newblock \bibinfo{journal}{Communication Monographs} \bibinfo{volume}{76}, \bibinfo{pages}{463--486}.
\bibitem[{Lee et~al.(2019)Lee, Kim and Choi}]{lee2019adoption}
\bibinfo{author}{Lee, J.}, \bibinfo{author}{Kim, J.}, \bibinfo{author}{Choi, J.Y.}, \bibinfo{year}{2019}.
\newblock \bibinfo{title}{The adoption of virtual reality devices: The technology acceptance model integrating enjoyment, social interaction, and strength of the social ties}.
\newblock \bibinfo{journal}{Telematics and Informatics} \bibinfo{volume}{39}, \bibinfo{pages}{37--48}.
\bibitem[{Lee and Moray(1992)}]{lee1992trust}
\bibinfo{author}{Lee, J.}, \bibinfo{author}{Moray, N.}, \bibinfo{year}{1992}.
\newblock \bibinfo{title}{Trust, control strategies and allocation of function in human-machine systems}.
\newblock \bibinfo{journal}{Ergonomics} \bibinfo{volume}{35}, \bibinfo{pages}{1243--1270}.
\bibitem[{Lee and See(2004)}]{lee2004trust}
\bibinfo{author}{Lee, J.D.}, \bibinfo{author}{See, K.A.}, \bibinfo{year}{2004}.
\newblock \bibinfo{title}{Trust in automation: Designing for appropriate reliance}.
\newblock \bibinfo{journal}{Human factors} \bibinfo{volume}{46}, \bibinfo{pages}{50--80}.
\bibitem[{Lee and Ashton(2004)}]{lee2004psychometricHEXACO}
\bibinfo{author}{Lee, K.}, \bibinfo{author}{Ashton, M.C.}, \bibinfo{year}{2004}.
\newblock \bibinfo{title}{Psychometric properties of the hexaco personality inventory}.
\newblock \bibinfo{journal}{Multivariate behavioral research} \bibinfo{volume}{39}, \bibinfo{pages}{329--358}.
\bibitem[{Lee et~al.(2020)Lee, Lee and Sheehan}]{leeHeyAlexaMagic2020a}
\bibinfo{author}{Lee, K.}, \bibinfo{author}{Lee, K.Y.}, \bibinfo{author}{Sheehan, L.}, \bibinfo{year}{2020}.
\newblock \bibinfo{title}{Hey {{Alexa}}! {{A Magic Spell}} of {{Social Glue}}?: {{Sharing}} a {{Smart Voice Assistant Speaker}} and {{Its Impact}} on {{Users}}' {{Perception}} of {{Group Harmony}}}.
\newblock \bibinfo{journal}{Information Systems Frontiers} \bibinfo{volume}{22}, \bibinfo{pages}{563--583}.
\newblock \DOIprefix\doi{10.1007/s10796-019-09975-1}.
\bibitem[{Lee et~al.(2021a)Lee, Sheehan, Lee and Chang}]{lee2021continuation}
\bibinfo{author}{Lee, K.Y.}, \bibinfo{author}{Sheehan, L.}, \bibinfo{author}{Lee, K.}, \bibinfo{author}{Chang, Y.}, \bibinfo{year}{2021}a.
\newblock \bibinfo{title}{The continuation and recommendation intention of artificial intelligence-based voice assistant systems (aivas): the influence of personal traits}.
\newblock \bibinfo{journal}{Internet Research} \bibinfo{volume}{31}, \bibinfo{pages}{1899--1939}.
\bibitem[{Lee et~al.(2021b)Lee, Frank and IJsselsteijn}]{leeBrokerbotCryptocurrencyChatbot2021}
\bibinfo{author}{Lee, M.}, \bibinfo{author}{Frank, L.}, \bibinfo{author}{IJsselsteijn, W.}, \bibinfo{year}{2021}b.
\newblock \bibinfo{title}{Brokerbot: A cryptocurrency chatbot in the social-technical gap of trust}.
\newblock \bibinfo{journal}{Computer Supported Cooperative Work (CSCW)} \bibinfo{volume}{30}, \bibinfo{pages}{79--117}.
\newblock \DOIprefix\doi{10.1007/s10606-021-09392-6}.
\bibitem[{Lee et~al.(2021c)Lee, Ayyagari, Nasirian and Ahmadian}]{lee2021role}
\bibinfo{author}{Lee, O.K.D.}, \bibinfo{author}{Ayyagari, R.}, \bibinfo{author}{Nasirian, F.}, \bibinfo{author}{Ahmadian, M.}, \bibinfo{year}{2021}c.
\newblock \bibinfo{title}{Role of interaction quality and trust in use of ai-based voice-assistant systems}.
\newblock \bibinfo{journal}{Journal of Systems and Information Technology} \bibinfo{volume}{23}, \bibinfo{pages}{154--170}.
\bibitem[{Lee and Lee(2014)}]{lee2014early}
\bibinfo{author}{Lee, S.}, \bibinfo{author}{Lee, S.}, \bibinfo{year}{2014}.
\newblock \bibinfo{title}{Early diffusion of smartphones in oecd and brics countries: An examination of the effects of platform competition and indirect network effects}.
\newblock \bibinfo{journal}{Telematics and Informatics} \bibinfo{volume}{31}, \bibinfo{pages}{345--355}.
\bibitem[{Lee and Sun(2022)}]{lee2022testing}
\bibinfo{author}{Lee, S.K.}, \bibinfo{author}{Sun, J.}, \bibinfo{year}{2022}.
\newblock \bibinfo{title}{Testing a theoretical model of trust in human-machine communication: emotional experience and social presence}.
\newblock \bibinfo{journal}{Behaviour \& Information Technology} , \bibinfo{pages}{1--14}.
\bibitem[{Lewis and Hardzinski(2015)}]{lewis2015investigating}
\bibinfo{author}{Lewis, J.R.}, \bibinfo{author}{Hardzinski, M.L.}, \bibinfo{year}{2015}.
\newblock \bibinfo{title}{Investigating the psychometric properties of the speech user interface service quality questionnaire}.
\newblock \bibinfo{journal}{International Journal of Speech Technology} \bibinfo{volume}{18}, \bibinfo{pages}{479--487}.
\bibitem[{Li et~al.(2023a)Li, Erickson, Cross and Lee}]{li2023s}
\bibinfo{author}{Li, M.}, \bibinfo{author}{Erickson, I.M.}, \bibinfo{author}{Cross, E.V.}, \bibinfo{author}{Lee, J.D.}, \bibinfo{year}{2023}a.
\newblock \bibinfo{title}{It’s not only what you say, but also how you say it: Machine learning approach to estimate trust from conversation}.
\newblock \bibinfo{journal}{Human Factors} , \bibinfo{pages}{00187208231166624}.
\bibitem[{Li et~al.(2023b)Li, Kamaraj and Lee}]{li2023modeling}
\bibinfo{author}{Li, M.}, \bibinfo{author}{Kamaraj, A.V.}, \bibinfo{author}{Lee, J.D.}, \bibinfo{year}{2023}b.
\newblock \bibinfo{title}{Modeling trust dimensions and dynamics in human-agent conversation: A trajectory epistemic network analysis approach}.
\newblock \bibinfo{journal}{International Journal of Human--Computer Interaction} , \bibinfo{pages}{1--12}.
\bibitem[{Li et~al.(2019)Li, Chan and Kim}]{li2019service}
\bibinfo{author}{Li, X.}, \bibinfo{author}{Chan, K.W.}, \bibinfo{author}{Kim, S.}, \bibinfo{year}{2019}.
\newblock \bibinfo{title}{Service with emoticons: How customers interpret employee use of emoticons in online service encounters}.
\newblock \bibinfo{journal}{Journal of Consumer Research} \bibinfo{volume}{45}, \bibinfo{pages}{973--987}.
\bibitem[{Li et~al.(2008)Li, Hess and Valacich}]{liWhyWeTrust2008a}
\bibinfo{author}{Li, X.}, \bibinfo{author}{Hess, T.J.}, \bibinfo{author}{Valacich, J.S.}, \bibinfo{year}{2008}.
\newblock \bibinfo{title}{Why do we trust new technology? {{A}} study of initial trust formation with organizational information systems}.
\newblock \bibinfo{journal}{The Journal of Strategic Information Systems} \bibinfo{volume}{17}, \bibinfo{pages}{39--71}.
\newblock \DOIprefix\doi{10.1016/j.jsis.2008.01.001}.
\bibitem[{Li and Sung(2021)}]{li2021anthropomorphism}
\bibinfo{author}{Li, X.}, \bibinfo{author}{Sung, Y.}, \bibinfo{year}{2021}.
\newblock \bibinfo{title}{Anthropomorphism brings us closer: The mediating role of psychological distance in user--ai assistant interactions}.
\newblock \bibinfo{journal}{Computers in Human Behavior} \bibinfo{volume}{118}, \bibinfo{pages}{106680}.
\bibitem[{Liao et~al.(2019)Liao, Vitak, Kumar, Zimmer and Kritikos}]{liaoUnderstandingRolePrivacy2019b}
\bibinfo{author}{Liao, Y.}, \bibinfo{author}{Vitak, J.}, \bibinfo{author}{Kumar, P.}, \bibinfo{author}{Zimmer, M.}, \bibinfo{author}{Kritikos, K.}, \bibinfo{year}{2019}.
\newblock \bibinfo{title}{Understanding the {{Role}} of {{Privacy}} and {{Trust}} in {{Intelligent Personal Assistant Adoption}}}, in: \bibinfo{editor}{Taylor, N.G.}, \bibinfo{editor}{{Christian-Lamb}, C.}, \bibinfo{editor}{Martin, M.H.}, \bibinfo{editor}{Nardi, B.} (Eds.), \bibinfo{booktitle}{Information in {{Contemporary Society}}}. \bibinfo{publisher}{{Springer International Publishing}}, \bibinfo{address}{{Cham}}. volume \bibinfo{volume}{11420}, pp. \bibinfo{pages}{102--113}.
\newblock \DOIprefix\doi{10.1007/978-3-030-15742-5_9}.
\bibitem[{Lim(2003)}]{Lim2003}
\bibinfo{author}{Lim, N.}, \bibinfo{year}{2003}.
\newblock \bibinfo{title}{Consumers’ perceived risk: sources versus consequences}.
\newblock \bibinfo{journal}{Electronic Commerce Research and Applications} \bibinfo{volume}{2}, \bibinfo{pages}{216--228}.
\newblock \URLprefix \url{https://www.sciencedirect.com/science/article/pii/S1567422303000255}, \DOIprefix\doi{https://doi.org/10.1016/S1567-4223(03)00025-5}. \bibinfo{note}{selected Papers from the Pacific Asia Conference on Information Systems}.
\bibitem[{Lin et~al.(2023)Lin, Cronj{\'e}, K{\"a}thner, Pauli and Latoschik}]{lin2023measuringtrustgame}
\bibinfo{author}{Lin, J.}, \bibinfo{author}{Cronj{\'e}, J.}, \bibinfo{author}{K{\"a}thner, I.}, \bibinfo{author}{Pauli, P.}, \bibinfo{author}{Latoschik, M.E.}, \bibinfo{year}{2023}.
\newblock \bibinfo{title}{Measuring interpersonal trust towards virtual humans with a virtual maze paradigm}.
\newblock \bibinfo{journal}{IEEE Transactions on Visualization and Computer Graphics} \bibinfo{volume}{29}, \bibinfo{pages}{2401--2411}.
\bibitem[{Lin and Parkin(2020)}]{linTransferabilityPrivacyrelatedBehaviours2020a}
\bibinfo{author}{Lin, V.Z.}, \bibinfo{author}{Parkin, S.}, \bibinfo{year}{2020}.
\newblock \bibinfo{title}{Transferability of {{Privacy-related Behaviours}} to {{Shared Smart Home Assistant Devices}}}, in: \bibinfo{booktitle}{2020 7th {{International Conference}} on {{Internet}} of {{Things}}: {{Systems}}, {{Management}} and {{Security}} ({{IOTSMS}})}, pp. \bibinfo{pages}{1--8}.
\newblock \DOIprefix\doi{10.1109/IOTSMS52051.2020.9340199}.
\bibitem[{Liu et~al.(2005)Liu, Marchewka, Lu and Yu}]{liuConcernPrivacytrustbehavioralIntention2005}
\bibinfo{author}{Liu, C.}, \bibinfo{author}{Marchewka, J.T.}, \bibinfo{author}{Lu, J.}, \bibinfo{author}{Yu, C.S.}, \bibinfo{year}{2005}.
\newblock \bibinfo{title}{Beyond concern{\textemdash}a privacy-trust-behavioral intention model of electronic commerce}.
\newblock \bibinfo{journal}{Information \& Management} \bibinfo{volume}{42}, \bibinfo{pages}{289--304}.
\newblock \DOIprefix\doi{10.1016/j.im.2004.01.003}.
\bibitem[{Liu and Gal(2011)}]{liu2011bringing}
\bibinfo{author}{Liu, W.}, \bibinfo{author}{Gal, D.}, \bibinfo{year}{2011}.
\newblock \bibinfo{title}{Bringing us together or driving us apart: The effect of soliciting consumer input on consumers’ propensity to transact with an organization}.
\newblock \bibinfo{journal}{Journal of Consumer Research} \bibinfo{volume}{38}, \bibinfo{pages}{242--259}.
\bibitem[{Liu et~al.(2021)Liu, Gan, Song and Liu}]{liu2021influences}
\bibinfo{author}{Liu, Y.}, \bibinfo{author}{Gan, Y.}, \bibinfo{author}{Song, Y.}, \bibinfo{author}{Liu, J.}, \bibinfo{year}{2021}.
\newblock \bibinfo{title}{What influences the perceived trust of a voice-enabled smart home system: an empirical study}.
\newblock \bibinfo{journal}{Sensors} \bibinfo{volume}{21}, \bibinfo{pages}{2037}.
\bibitem[{Lopez et~al.(2023)Lopez, Watkins and Pak}]{lopez2023enhancing}
\bibinfo{author}{Lopez, J.}, \bibinfo{author}{Watkins, H.}, \bibinfo{author}{Pak, R.}, \bibinfo{year}{2023}.
\newblock \bibinfo{title}{Enhancing component-specific trust with consumer automated systems through humanness design}.
\newblock \bibinfo{journal}{Ergonomics} \bibinfo{volume}{66}, \bibinfo{pages}{291--302}.
\bibitem[{Lu et~al.(2011)Lu, Yang, Chau and Cao}]{lu2011dynamics}
\bibinfo{author}{Lu, Y.}, \bibinfo{author}{Yang, S.}, \bibinfo{author}{Chau, P.Y.}, \bibinfo{author}{Cao, Y.}, \bibinfo{year}{2011}.
\newblock \bibinfo{title}{Dynamics between the trust transfer process and intention to use mobile payment services: A cross-environment perspective}.
\newblock \bibinfo{journal}{Information \& management} \bibinfo{volume}{48}, \bibinfo{pages}{393--403}.
\bibitem[{Lucia-Palacios and P{\'e}rez-L{\'o}pez(2021)}]{lucia2021effects}
\bibinfo{author}{Lucia-Palacios, L.}, \bibinfo{author}{P{\'e}rez-L{\'o}pez, R.}, \bibinfo{year}{2021}.
\newblock \bibinfo{title}{Effects of home voice assistants’ autonomy on instrusiveness and usefulness: direct, indirect, and moderating effects of interactivity}.
\newblock \bibinfo{journal}{Journal of Interactive Marketing} \bibinfo{volume}{56}, \bibinfo{pages}{41--54}.
\bibitem[{Lutz and Newlands(2021)}]{lutzPrivacySmartSpeakers2021}
\bibinfo{author}{Lutz, C.}, \bibinfo{author}{Newlands, G.}, \bibinfo{year}{2021}.
\newblock \bibinfo{title}{Privacy and smart speakers: {{A}} multi-dimensional approach}.
\newblock \bibinfo{journal}{The Information Society} \bibinfo{volume}{37}, \bibinfo{pages}{147--162}.
\newblock \DOIprefix\doi{10.1080/01972243.2021.1897914}.
\bibitem[{Lv et~al.(2022)Lv, Hu, Liu and Qi}]{lv2022research}
\bibinfo{author}{Lv, Y.}, \bibinfo{author}{Hu, S.}, \bibinfo{author}{Liu, F.}, \bibinfo{author}{Qi, J.}, \bibinfo{year}{2022}.
\newblock \bibinfo{title}{Research on users’ trust in customer service chatbots based on human-computer interaction}, in: \bibinfo{booktitle}{China National Conference on Big Data and Social Computing}, \bibinfo{organization}{Springer}. pp. \bibinfo{pages}{291--306}.
\bibitem[{Maccario and Naldi(2023)}]{maccarioPrivacySmartSpeakers2023}
\bibinfo{author}{Maccario, G.}, \bibinfo{author}{Naldi, M.}, \bibinfo{year}{2023}.
\newblock \bibinfo{title}{Privacy in smart speakers: A systematic literature review}.
\newblock \bibinfo{journal}{Security and Privacy} \bibinfo{volume}{6}, \bibinfo{pages}{e274}.
\newblock \DOIprefix\doi{10.1002/spy2.274}.
\bibitem[{Madsen and Gregor(2000)}]{madsen2000measuring}
\bibinfo{author}{Madsen, M.}, \bibinfo{author}{Gregor, S.}, \bibinfo{year}{2000}.
\newblock \bibinfo{title}{Measuring human-computer trust}, in: \bibinfo{booktitle}{11th australasian conference on information systems}, \bibinfo{organization}{Citeseer}. pp. \bibinfo{pages}{6--8}.
\bibitem[{Malhotra et~al.(2004)Malhotra, Kim and Agarwal}]{malhotra2004internet}
\bibinfo{author}{Malhotra, N.}, \bibinfo{author}{Kim, S.}, \bibinfo{author}{Agarwal, J.}, \bibinfo{year}{2004}.
\newblock \bibinfo{title}{Internet users' information privacy concerns (iuipc): The construct, the scale, and a causal model}.
\newblock \bibinfo{journal}{Information Systems Research} \bibinfo{volume}{15}, \bibinfo{pages}{336--355}.
\newblock \DOIprefix\doi{10.1287/isre.1040.0032}.
\bibitem[{Malkin et~al.(2019)Malkin, Deatrick, Tong, Wijesekera, Egelman and Wagner}]{malkinPrivacyAttitudesSmart2019}
\bibinfo{author}{Malkin, N.}, \bibinfo{author}{Deatrick, J.}, \bibinfo{author}{Tong, A.}, \bibinfo{author}{Wijesekera, P.}, \bibinfo{author}{Egelman, S.}, \bibinfo{author}{Wagner, D.}, \bibinfo{year}{2019}.
\newblock \bibinfo{title}{Privacy {{Attitudes}} of {{Smart Speaker Users}}}.
\newblock \bibinfo{journal}{Proceedings on Privacy Enhancing Technologies} \bibinfo{volume}{2019}, \bibinfo{pages}{250--271}.
\newblock \DOIprefix\doi{10.2478/popets-2019-0068}.
\bibitem[{Mamonov and Benbunan-Fich(2018)}]{mamonov2018impact}
\bibinfo{author}{Mamonov, S.}, \bibinfo{author}{Benbunan-Fich, R.}, \bibinfo{year}{2018}.
\newblock \bibinfo{title}{The impact of information security threat awareness on privacy-protective behaviors}.
\newblock \bibinfo{journal}{Computers in Human Behavior} \bibinfo{volume}{83}, \bibinfo{pages}{32--44}.
\bibitem[{Manikonda et~al.(2018)Manikonda, Deotale and Kambhampati}]{manikondaWhatPrivacyUser2018a}
\bibinfo{author}{Manikonda, L.}, \bibinfo{author}{Deotale, A.}, \bibinfo{author}{Kambhampati, S.}, \bibinfo{year}{2018}.
\newblock \bibinfo{title}{What's up with {{Privacy}}?: {{User Preferences}} and {{Privacy Concerns}} in {{Intelligent Personal Assistants}}}, in: \bibinfo{booktitle}{Proceedings of the 2018 {{AAAI}}/{{ACM Conference}} on {{AI}}, {{Ethics}}, and {{Society}}}, \bibinfo{publisher}{{ACM}}, \bibinfo{address}{{New Orleans LA USA}}. pp. \bibinfo{pages}{229--235}.
\newblock \DOIprefix\doi{10.1145/3278721.3278773}.
\bibitem[{Maqableh et~al.(2021)Maqableh, Hmoud, Jaradat et~al.}]{maqableh2021integrating}
\bibinfo{author}{Maqableh, M.}, \bibinfo{author}{Hmoud, H.Y.}, \bibinfo{author}{Jaradat, M.}, et~al., \bibinfo{year}{2021}.
\newblock \bibinfo{title}{Integrating an information systems success model with perceived privacy, perceived security, and trust: the moderating role of facebook addiction}.
\newblock \bibinfo{journal}{Heliyon} \bibinfo{volume}{7}.
\bibitem[{Maranguni{\'c} and Grani{\'c}(2015)}]{marangunic2015technologyTAM}
\bibinfo{author}{Maranguni{\'c}, N.}, \bibinfo{author}{Grani{\'c}, A.}, \bibinfo{year}{2015}.
\newblock \bibinfo{title}{Technology acceptance model: a literature review from 1986 to 2013}.
\newblock \bibinfo{journal}{Universal access in the information society} \bibinfo{volume}{14}, \bibinfo{pages}{81--95}.
\bibitem[{Mari and Algesheimer(2021)}]{mari2021role}
\bibinfo{author}{Mari, A.}, \bibinfo{author}{Algesheimer, R.}, \bibinfo{year}{2021}.
\newblock \bibinfo{title}{The role of trusting beliefs in voice assistants during voice shopping}.
\newblock \bibinfo{journal}{Hawaii International Conference on System Sciences 2021 (HICSS-54)} .
\bibitem[{Maroufkhani et~al.(2022)Maroufkhani, Asadi, Ghobakhloo, Jannesari and Ismail}]{maroufkhani2022interactive}
\bibinfo{author}{Maroufkhani, P.}, \bibinfo{author}{Asadi, S.}, \bibinfo{author}{Ghobakhloo, M.}, \bibinfo{author}{Jannesari, M.T.}, \bibinfo{author}{Ismail, W.K.W.}, \bibinfo{year}{2022}.
\newblock \bibinfo{title}{How do interactive voice assistants build brands' loyalty?}
\newblock \bibinfo{journal}{Technological Forecasting and Social Change} \bibinfo{volume}{183}, \bibinfo{pages}{121870}.
\bibitem[{Masur(2019)}]{Masur2019}
\bibinfo{author}{Masur, P.K.}, \bibinfo{year}{2019}.
\newblock \bibinfo{title}{Situational {Privacy} and {Self}-{Disclosure}}.
\newblock \bibinfo{publisher}{Springer International Publishing}, \bibinfo{address}{Cham}.
\newblock \DOIprefix\doi{10.1007/978-3-319-78884-5}.
\bibitem[{Mayer and Davis(1999)}]{mayerEffectPerformanceAppraisal1999}
\bibinfo{author}{Mayer, R.C.}, \bibinfo{author}{Davis, J.H.}, \bibinfo{year}{1999}.
\newblock \bibinfo{title}{The effect of the performance appraisal system on trust for management: {{A}} field quasi-experiment.}
\newblock \bibinfo{journal}{Journal of Applied Psychology} \bibinfo{volume}{84}, \bibinfo{pages}{123--136}.
\newblock \DOIprefix\doi{10.1037/0021-9010.84.1.123}.
\bibitem[{Mayer et~al.(1995)Mayer, Davis and Schoorman}]{mayer1995integrative}
\bibinfo{author}{Mayer, R.C.}, \bibinfo{author}{Davis, J.H.}, \bibinfo{author}{Schoorman, F.D.}, \bibinfo{year}{1995}.
\newblock \bibinfo{title}{An integrative model of organizational trust}.
\newblock \bibinfo{journal}{Academy of management review} \bibinfo{volume}{20}, \bibinfo{pages}{709--734}.
\bibitem[{McCarthy et~al.(2020)McCarthy, Gaster and Legg}]{mccarthyShoutingLetterboxesStudy2020b}
\bibinfo{author}{McCarthy, A.}, \bibinfo{author}{Gaster, B.R.}, \bibinfo{author}{Legg, P.}, \bibinfo{year}{2020}.
\newblock \bibinfo{title}{Shouting {{Through Letterboxes}}: {{A}} study on attack susceptibility of voice assistants}, in: \bibinfo{booktitle}{2020 {{International Conference}} on {{Cyber Security}} and {{Protection}} of {{Digital Services}} ({{Cyber Security}})}, \bibinfo{publisher}{{IEEE}}, \bibinfo{address}{{Dublin, Ireland}}. pp. \bibinfo{pages}{1--8}.
\newblock \DOIprefix\doi{10.1109/CyberSecurity49315.2020.9138860}.
\bibitem[{Mcknight et~al.(2011)Mcknight, Carter, Thatcher and Clay}]{mcknight2011trust}
\bibinfo{author}{Mcknight, D.H.}, \bibinfo{author}{Carter, M.}, \bibinfo{author}{Thatcher, J.B.}, \bibinfo{author}{Clay, P.F.}, \bibinfo{year}{2011}.
\newblock \bibinfo{title}{Trust in a specific technology: An investigation of its components and measures}.
\newblock \bibinfo{journal}{ACM Transactions on management information systems (TMIS)} \bibinfo{volume}{2}, \bibinfo{pages}{1--25}.
\bibitem[{McKnight et~al.(2002)McKnight, Choudhury and Kacmar}]{mcknight2002impact}
\bibinfo{author}{McKnight, D.H.}, \bibinfo{author}{Choudhury, V.}, \bibinfo{author}{Kacmar, C.}, \bibinfo{year}{2002}.
\newblock \bibinfo{title}{The impact of initial consumer trust on intentions to transact with a web site: a trust building model}.
\newblock \bibinfo{journal}{The journal of strategic information systems} \bibinfo{volume}{11}, \bibinfo{pages}{297--323}.
\bibitem[{McLean and {Osei-Frimpong}(2019)}]{mcleanHeyAlexaExamine2019a}
\bibinfo{author}{McLean, G.}, \bibinfo{author}{{Osei-Frimpong}, K.}, \bibinfo{year}{2019}.
\newblock \bibinfo{title}{Hey {{Alexa}} {\ldots} examine the variables influencing the use of artificial intelligent in-home voice assistants}.
\newblock \bibinfo{journal}{Computers in Human Behavior} \bibinfo{volume}{99}, \bibinfo{pages}{28--37}.
\newblock \DOIprefix\doi{10.1016/j.chb.2019.05.009}.
\bibitem[{Mehrabian and Stefl(1995)}]{mehrabian1995basic}
\bibinfo{author}{Mehrabian, A.}, \bibinfo{author}{Stefl, C.A.}, \bibinfo{year}{1995}.
\newblock \bibinfo{title}{Basic temperament components of loneliness, shyness, and conformity}.
\newblock \bibinfo{journal}{Social Behavior and Personality: an international journal} \bibinfo{volume}{23}, \bibinfo{pages}{253--263}.
\bibitem[{Mende et~al.(2019)Mende, Scott, {van Doorn}, Grewal and Shanks}]{mendeServiceRobotsRising2019}
\bibinfo{author}{Mende, M.}, \bibinfo{author}{Scott, M.L.}, \bibinfo{author}{{van Doorn}, J.}, \bibinfo{author}{Grewal, D.}, \bibinfo{author}{Shanks, I.}, \bibinfo{year}{2019}.
\newblock \bibinfo{title}{Service {{Robots Rising}}: {{How Humanoid Robots Influence Service Experiences}} and {{Elicit Compensatory Consumer Responses}}}.
\newblock \bibinfo{journal}{Journal of Marketing Research} \bibinfo{volume}{56}, \bibinfo{pages}{535--556}.
\newblock \DOIprefix\doi{10.1177/0022243718822827}.
\bibitem[{Merritt et~al.(2013)Merritt, Heimbaugh, LaChapell and Lee}]{merritt2013trust}
\bibinfo{author}{Merritt, S.M.}, \bibinfo{author}{Heimbaugh, H.}, \bibinfo{author}{LaChapell, J.}, \bibinfo{author}{Lee, D.}, \bibinfo{year}{2013}.
\newblock \bibinfo{title}{I trust it, but i don’t know why: Effects of implicit attitudes toward automation on trust in an automated system}.
\newblock \bibinfo{journal}{Human factors} \bibinfo{volume}{55}, \bibinfo{pages}{520--534}.
\bibitem[{Metzger(2004)}]{Metzger2004ecommerce}
\bibinfo{author}{Metzger, M.J.}, \bibinfo{year}{2004}.
\newblock \bibinfo{title}{{Privacy, Trust, and Disclosure: Exploring Barriers to Electronic Commerce}}.
\newblock \bibinfo{journal}{Journal of Computer-Mediated Communication} \bibinfo{volume}{9}, \bibinfo{pages}{JCMC942}.
\newblock \DOIprefix\doi{10.1111/j.1083-6101.2004.tb00292.x}.
\bibitem[{Metzger(2007)}]{metzger2007making}
\bibinfo{author}{Metzger, M.J.}, \bibinfo{year}{2007}.
\newblock \bibinfo{title}{Making sense of credibility on the web: Models for evaluating online information and recommendations for future research}.
\newblock \bibinfo{journal}{Journal of the American society for information science and technology} \bibinfo{volume}{58}, \bibinfo{pages}{2078--2091}.
\bibitem[{Mhaidli et~al.(2020)Mhaidli, Venkatesh, Zou and Schaub}]{mhaidliListenOnlyWhen2020a}
\bibinfo{author}{Mhaidli, A.}, \bibinfo{author}{Venkatesh, M.K.}, \bibinfo{author}{Zou, Y.}, \bibinfo{author}{Schaub, F.}, \bibinfo{year}{2020}.
\newblock \bibinfo{title}{Listen {{Only When Spoken To}}: {{Interpersonal Communication Cues}} as {{Smart Speaker Privacy Controls}}}.
\newblock \bibinfo{journal}{Proceedings on Privacy Enhancing Technologies} \bibinfo{volume}{2020}, \bibinfo{pages}{251--270}.
\newblock \DOIprefix\doi{10.2478/popets-2020-0026}.
\bibitem[{Misiolek et~al.(2002)Misiolek, Zakaria and Zhang}]{misiolek2002trust}
\bibinfo{author}{Misiolek, N.I.}, \bibinfo{author}{Zakaria, N.}, \bibinfo{author}{Zhang, P.}, \bibinfo{year}{2002}.
\newblock \bibinfo{title}{Trust in organizational acceptance of information technology: A conceptual model and preliminary evidence}, in: \bibinfo{booktitle}{33rd Annual Meeting of the Decision Sciences Institute}.
\bibitem[{Moher et~al.(2009)Moher, Liberati, Tetzlaff, Altman and Group}]{moherPreferredReportingItems2009}
\bibinfo{author}{Moher, D.}, \bibinfo{author}{Liberati, A.}, \bibinfo{author}{Tetzlaff, J.}, \bibinfo{author}{Altman, D.G.}, \bibinfo{author}{Group, T.P.}, \bibinfo{year}{2009}.
\newblock \bibinfo{title}{Preferred {{Reporting Items}} for {{Systematic Reviews}} and {{Meta-Analyses}}: {{The PRISMA Statement}}}.
\newblock \bibinfo{journal}{PLOS Medicine} \bibinfo{volume}{6}, \bibinfo{pages}{e1000097}.
\newblock \DOIprefix\doi{10.1371/journal.pmed.1000097}.
\bibitem[{Mols et~al.(2022)Mols, Wang and Pridmore}]{Mols2021}
\bibinfo{author}{Mols, A.}, \bibinfo{author}{Wang, Y.}, \bibinfo{author}{Pridmore, J.}, \bibinfo{year}{2022}.
\newblock \bibinfo{title}{Household intelligent personal assistants in the netherlands: Exploring privacy concerns around surveillance, security, and platforms}.
\newblock \bibinfo{journal}{Convergence} \bibinfo{volume}{28}, \bibinfo{pages}{1841--1860}.
\newblock \DOIprefix\doi{10.1177/13548565211042234}.
\bibitem[{Moradinezhad and Solovey(2021)}]{moradinezhad2021investigating}
\bibinfo{author}{Moradinezhad, R.}, \bibinfo{author}{Solovey, E.T.}, \bibinfo{year}{2021}.
\newblock \bibinfo{title}{Investigating trust in interaction with inconsistent embodied virtual agents}.
\newblock \bibinfo{journal}{International Journal of Social Robotics} \bibinfo{volume}{13}, \bibinfo{pages}{2103--2118}.
\bibitem[{M{\"u}ller et~al.(2019)M{\"u}ller, Mattke, Maier, Weitzel and Graser}]{muller2019chatbot}
\bibinfo{author}{M{\"u}ller, L.}, \bibinfo{author}{Mattke, J.}, \bibinfo{author}{Maier, C.}, \bibinfo{author}{Weitzel, T.}, \bibinfo{author}{Graser, H.}, \bibinfo{year}{2019}.
\newblock \bibinfo{title}{Chatbot acceptance: A latent profile analysis on individuals' trust in conversational agents}, in: \bibinfo{booktitle}{Proceedings of the 2019 on Computers and People Research Conference}, pp. \bibinfo{pages}{35--42}.
\bibitem[{Mu{\~n}oz and Kremer(2023)}]{munoz2023voice}
\bibinfo{author}{Mu{\~n}oz, N.}, \bibinfo{author}{Kremer, B.A.}, \bibinfo{year}{2023}.
\newblock \bibinfo{title}{The voice era: Future acceptance of digital voice assistants and how they will transform consumers' online purchasing behaviour}.
\newblock \bibinfo{journal}{Applied Marketing Analytics} \bibinfo{volume}{8}, \bibinfo{pages}{255--270}.
\bibitem[{Ng et~al.(2020)Ng, Coopamootoo, Toreini, Aitken, Elliot and {van Moorsel}}]{ngSimulatingEffectsSocial2020a}
\bibinfo{author}{Ng, M.}, \bibinfo{author}{Coopamootoo, K.P.}, \bibinfo{author}{Toreini, E.}, \bibinfo{author}{Aitken, M.}, \bibinfo{author}{Elliot, K.}, \bibinfo{author}{{van Moorsel}, A.}, \bibinfo{year}{2020}.
\newblock \bibinfo{title}{Simulating the {{Effects}} of {{Social Presence}} on {{Trust}}, {{Privacy Concerns}} \& {{Usage Intentions}} in {{Automated Bots}} for {{Finance}}}, in: \bibinfo{booktitle}{2020 {{IEEE European Symposium}} on {{Security}} and {{Privacy Workshops}} ({{EuroS}}\&{{PW}})}, pp. \bibinfo{pages}{190--199}.
\newblock \DOIprefix\doi{10.1109/EuroSPW51379.2020.00034}.
\bibitem[{Nissenbaum(2009)}]{nissenbaum2009privacy}
\bibinfo{author}{Nissenbaum, H.}, \bibinfo{year}{2009}.
\newblock \bibinfo{title}{Privacy in Context: Technology, Policy, and the Integrity of Social Life}.
\newblock Stanford Law Books, \bibinfo{publisher}{Stanford University Press}.
\newblock \URLprefix \url{https://books.google.de/books?id=_NN1uGn1Jd8C}.
\bibitem[{Nordheim(2018)}]{nordheimTrustChatbotsCustomer}
\bibinfo{author}{Nordheim, C.B.}, \bibinfo{year}{2018}.
\newblock \bibinfo{title}{Trust in chatbots for customer service--findings from a questionnaire study}.
\newblock Master's thesis. University of Oslo.
\bibitem[{Nordheim et~al.(2019)Nordheim, F{\o}lstad and Bj{\o}rkli}]{nordheim2019initialtrust}
\bibinfo{author}{Nordheim, C.B.}, \bibinfo{author}{F{\o}lstad, A.}, \bibinfo{author}{Bj{\o}rkli, C.A.}, \bibinfo{year}{2019}.
\newblock \bibinfo{title}{An initial model of trust in chatbots for customer service—findings from a questionnaire study}.
\newblock \bibinfo{journal}{Interacting with Computers} \bibinfo{volume}{31}, \bibinfo{pages}{317--335}.
\bibitem[{Norton(2018)}]{Norton2020}
\bibinfo{author}{Norton}, \bibinfo{year}{2018}.
\newblock \bibinfo{title}{Can smart speakers be hacked? 10 tips to help stay secure}.
\newblock \bibinfo{howpublished}{[Accessed Online 13-12-2023] \url{https://us.norton.com/blog/iot/can-smart-speakers-be-hacked}}.
\bibitem[{Nowak and Rauh(2005)}]{nowak2005influence}
\bibinfo{author}{Nowak, K.L.}, \bibinfo{author}{Rauh, C.}, \bibinfo{year}{2005}.
\newblock \bibinfo{title}{The influence of the avatar on online perceptions of anthropomorphism, androgyny, credibility, homophily, and attraction}.
\newblock \bibinfo{journal}{Journal of Computer-Mediated Communication} \bibinfo{volume}{11}, \bibinfo{pages}{153--178}.
\bibitem[{Oliveira et~al.(2016)Oliveira, Thomas, Baptista and Campos}]{oliveiraMobilePaymentUnderstanding2016}
\bibinfo{author}{Oliveira, T.}, \bibinfo{author}{Thomas, M.}, \bibinfo{author}{Baptista, G.}, \bibinfo{author}{Campos, F.}, \bibinfo{year}{2016}.
\newblock \bibinfo{title}{Mobile payment: {{Understanding}} the determinants of customer adoption and intention to recommend the technology}.
\newblock \bibinfo{journal}{Computers in Human Behavior} \bibinfo{volume}{61}, \bibinfo{pages}{404--414}.
\newblock \DOIprefix\doi{10.1016/j.chb.2016.03.030}.
\bibitem[{Pal et~al.(2020)Pal, Arpnikanondt and Razzaque}]{pal2020personal}
\bibinfo{author}{Pal, D.}, \bibinfo{author}{Arpnikanondt, C.}, \bibinfo{author}{Razzaque, M.A.}, \bibinfo{year}{2020}.
\newblock \bibinfo{title}{Personal information disclosure via voice assistants: the personalization--privacy paradox}.
\newblock \bibinfo{journal}{SN Computer Science} \bibinfo{volume}{1}, \bibinfo{pages}{1--17}.
\bibitem[{Pal et~al.(2022a)Pal, Babakerkhell and Roy}]{pal2022perceptions}
\bibinfo{author}{Pal, D.}, \bibinfo{author}{Babakerkhell, M.D.}, \bibinfo{author}{Roy, P.}, \bibinfo{year}{2022}a.
\newblock \bibinfo{title}{How perceptions of trust and intrusiveness affect the adoption of voice activated personal assistants}.
\newblock \bibinfo{journal}{IEEE Access} \bibinfo{volume}{10}, \bibinfo{pages}{123094--123113}.
\bibitem[{Pal et~al.(2022b)Pal, Roy, Arpnikanondt and Thapliyal}]{palEffectTrustIts2022a}
\bibinfo{author}{Pal, D.}, \bibinfo{author}{Roy, P.}, \bibinfo{author}{Arpnikanondt, C.}, \bibinfo{author}{Thapliyal, H.}, \bibinfo{year}{2022}b.
\newblock \bibinfo{title}{The effect of trust and its antecedents towards determining users' behavioral intention with voice-based consumer electronic devices}.
\newblock \bibinfo{journal}{Heliyon} \bibinfo{volume}{8}, \bibinfo{pages}{e09271}.
\newblock \DOIprefix\doi{10.1016/j.heliyon.2022.e09271}.
\bibitem[{Panjaitan and Utomo(2023)}]{panjaitan2023influence}
\bibinfo{author}{Panjaitan, A.G.}, \bibinfo{author}{Utomo, R.G.}, \bibinfo{year}{2023}.
\newblock \bibinfo{title}{The influence of users' perception of security, privacy, and trust in using online dating applications}, in: \bibinfo{booktitle}{2023 11th International Conference on Information and Communication Technology (ICoICT)}, \bibinfo{organization}{IEEE}. pp. \bibinfo{pages}{551--556}.
\bibitem[{Park et~al.(2021)Park, Choi and Jung}]{park2021users}
\bibinfo{author}{Park, J.}, \bibinfo{author}{Choi, H.}, \bibinfo{author}{Jung, Y.}, \bibinfo{year}{2021}.
\newblock \bibinfo{title}{Users’ cognitive and affective response to the risk to privacy from a smart speaker}.
\newblock \bibinfo{journal}{International Journal of Human--Computer Interaction} \bibinfo{volume}{37}, \bibinfo{pages}{759--771}.
\bibitem[{Patil and Kulkarni(2022)}]{patilCanWeTrust2022a}
\bibinfo{author}{Patil, K.}, \bibinfo{author}{Kulkarni, M.}, \bibinfo{year}{2022}.
\newblock \bibinfo{title}{Can we trust {{Health}} and {{Wellness Chatbot}} going mobile? {{Empirical}} research using {{TAM}} and {{HBM}}}, in: \bibinfo{booktitle}{2022 {{IEEE Region}} 10 {{Symposium}} ({{TENSYMP}})}, pp. \bibinfo{pages}{1--6}.
\newblock \DOIprefix\doi{10.1109/TENSYMP54529.2022.9864368}.
\bibitem[{Patrizi et~al.(2021)Patrizi, Vernuccio and Pastore}]{patrizi2021talking}
\bibinfo{author}{Patrizi, M.}, \bibinfo{author}{Vernuccio, M.}, \bibinfo{author}{Pastore, A.}, \bibinfo{year}{2021}.
\newblock \bibinfo{title}{Talking to voice assistants: Exploring negative and positive users’ perceptions}, in: \bibinfo{booktitle}{Digital Marketing \& eCommerce Conference}, \bibinfo{organization}{Springer}. pp. \bibinfo{pages}{24--34}.
\bibitem[{Pattnaik et~al.(2023)Pattnaik, Li and Nurse}]{pattnaikSurveyUserPerspectives2023c}
\bibinfo{author}{Pattnaik, N.}, \bibinfo{author}{Li, S.}, \bibinfo{author}{Nurse, J.R.C.}, \bibinfo{year}{2023}.
\newblock \bibinfo{title}{A {{Survey}} of {{User Perspectives}} on {{Security}} and {{Privacy}} in a {{Home Networking Environment}}}.
\newblock \bibinfo{journal}{ACM Computing Surveys} \bibinfo{volume}{55}, \bibinfo{pages}{1--38}.
\newblock \DOIprefix\doi{10.1145/3558095}.
\bibitem[{Pavlou(2003)}]{pavlou2003consumer}
\bibinfo{author}{Pavlou, P.A.}, \bibinfo{year}{2003}.
\newblock \bibinfo{title}{Consumer acceptance of electronic commerce: Integrating trust and risk with the technology acceptance model}.
\newblock \bibinfo{journal}{International journal of electronic commerce} \bibinfo{volume}{7}, \bibinfo{pages}{101--134}.
\bibitem[{Pavlou and Gefen(2004)}]{pavlou2004building}
\bibinfo{author}{Pavlou, P.A.}, \bibinfo{author}{Gefen, D.}, \bibinfo{year}{2004}.
\newblock \bibinfo{title}{Building effective online marketplaces with institution-based trust}.
\newblock \bibinfo{journal}{Information systems research} \bibinfo{volume}{15}, \bibinfo{pages}{37--59}.
\bibitem[{Perez~Garcia and Saffon~Lopez(2018)}]{perez2018buildingQualitative}
\bibinfo{author}{Perez~Garcia, M.}, \bibinfo{author}{Saffon~Lopez, S.}, \bibinfo{year}{2018}.
\newblock \bibinfo{title}{Building trust between users and telecommunications data driven virtual assistants}, in: \bibinfo{booktitle}{IFIP International Conference on Artificial Intelligence Applications and Innovations}, \bibinfo{organization}{Springer}. pp. \bibinfo{pages}{628--637}.
\bibitem[{Pesonen(2021)}]{pesonen2021you}
\bibinfo{author}{Pesonen, J.A.}, \bibinfo{year}{2021}.
\newblock \bibinfo{title}{‘are you ok?’students’ trust in a chatbot providing support opportunities}, in: \bibinfo{booktitle}{International Conference on Human-Computer Interaction}, \bibinfo{organization}{Springer}. pp. \bibinfo{pages}{199--215}.
\bibitem[{Pettersson et~al.(2018)Pettersson, Lachner, Frison, Riener and Butz}]{pettersson2018}
\bibinfo{author}{Pettersson, I.}, \bibinfo{author}{Lachner, F.}, \bibinfo{author}{Frison, A.K.}, \bibinfo{author}{Riener, A.}, \bibinfo{author}{Butz, A.}, \bibinfo{year}{2018}.
\newblock \bibinfo{title}{A bermuda triangle? a review of method application and triangulation in user experience evaluation}, in: \bibinfo{booktitle}{Proceedings of the 2018 CHI Conference on Human Factors in Computing Systems}, \bibinfo{publisher}{Association for Computing Machinery}, \bibinfo{address}{New York, NY, USA}. p. \bibinfo{pages}{1–16}.
\newblock \URLprefix \url{https://doi.org/10.1145/3173574.3174035}, \DOIprefix\doi{10.1145/3173574.3174035}.
\bibitem[{Pham and Ho(2015)}]{pham2015effects}
\bibinfo{author}{Pham, T.T.T.}, \bibinfo{author}{Ho, J.C.}, \bibinfo{year}{2015}.
\newblock \bibinfo{title}{The effects of product-related, personal-related factors and attractiveness of alternatives on consumer adoption of nfc-based mobile payments}.
\newblock \bibinfo{journal}{Technology in society} \bibinfo{volume}{43}, \bibinfo{pages}{159--172}.
\bibitem[{Pitardi and Marriott(2021)}]{pitardi2021alexa}
\bibinfo{author}{Pitardi, V.}, \bibinfo{author}{Marriott, H.R.}, \bibinfo{year}{2021}.
\newblock \bibinfo{title}{Alexa, she's not human but… unveiling the drivers of consumers' trust in voice-based artificial intelligence}.
\newblock \bibinfo{journal}{Psychology \& Marketing} \bibinfo{volume}{38}, \bibinfo{pages}{626--642}.
\bibitem[{Prakash et~al.(2023)Prakash, Joshi, Nim and Das}]{prakash2023determinants}
\bibinfo{author}{Prakash, A.V.}, \bibinfo{author}{Joshi, A.}, \bibinfo{author}{Nim, S.}, \bibinfo{author}{Das, S.}, \bibinfo{year}{2023}.
\newblock \bibinfo{title}{Determinants and consequences of trust in ai-based customer service chatbots}.
\newblock \bibinfo{journal}{The Service Industries Journal} \bibinfo{volume}{43}, \bibinfo{pages}{642--675}.
\bibitem[{Preibusch(2013)}]{preibuschGuideMeasuringPrivacy2013}
\bibinfo{author}{Preibusch, S.}, \bibinfo{year}{2013}.
\newblock \bibinfo{title}{Guide to measuring privacy concern: {{Review}} of survey and observational instruments}.
\newblock \bibinfo{journal}{International Journal of Human-Computer Studies} \bibinfo{volume}{71}, \bibinfo{pages}{1133--1143}.
\newblock \DOIprefix\doi{10.1016/j.ijhcs.2013.09.002}.
\bibitem[{Purwanto et~al.(2020)Purwanto, Kuswandi and Fatmah}]{purwanto2020interactive}
\bibinfo{author}{Purwanto, P.}, \bibinfo{author}{Kuswandi, K.}, \bibinfo{author}{Fatmah, F.}, \bibinfo{year}{2020}.
\newblock \bibinfo{title}{Interactive applications with artificial intelligence: The role of trust among digital assistant users}.
\newblock \bibinfo{journal}{Foresight and STI Governance} \bibinfo{volume}{14}, \bibinfo{pages}{64--75}.
\newblock \URLprefix \url{https://foresight-journal.hse.ru/en/2020-14-2/370913009.html}.
\bibitem[{Quinn(2016)}]{quinn2016we}
\bibinfo{author}{Quinn, K.}, \bibinfo{year}{2016}.
\newblock \bibinfo{title}{Why we share: A uses and gratifications approach to privacy regulation in social media use}.
\newblock \bibinfo{journal}{Journal of Broadcasting \& Electronic Media} \bibinfo{volume}{60}, \bibinfo{pages}{61--86}.
\bibitem[{Rahman et~al.(2021)Rahman, Rohan, Pal and Kanthamanon}]{rahmanHumanFactorsCybersecurity2021}
\bibinfo{author}{Rahman, T.}, \bibinfo{author}{Rohan, R.}, \bibinfo{author}{Pal, D.}, \bibinfo{author}{Kanthamanon, P.}, \bibinfo{year}{2021}.
\newblock \bibinfo{title}{Human {{Factors}} in {{Cybersecurity}}: {{A Scoping Review}}}, in: \bibinfo{booktitle}{The 12th {{International Conference}} on {{Advances}} in {{Information Technology}}}, \bibinfo{publisher}{{ACM}}, \bibinfo{address}{{Bangkok Thailand}}. pp. \bibinfo{pages}{1--11}.
\newblock \DOIprefix\doi{10.1145/3468784.3468789}.
\bibitem[{Rajapaksha et~al.(2021)Rajapaksha, Thakrar, Kinzler, Sun, Smith and Perouli}]{rajapakshaFieldStudyUsability2021b}
\bibinfo{author}{Rajapaksha, S.}, \bibinfo{author}{Thakrar, S.}, \bibinfo{author}{Kinzler, M.}, \bibinfo{author}{Sun, H.}, \bibinfo{author}{Smith, J.}, \bibinfo{author}{Perouli, D.}, \bibinfo{year}{2021}.
\newblock \bibinfo{title}{Field {{Study}} on {{Usability}} and {{Security Perceptions Surrounding Social Robots}}}, in: \bibinfo{booktitle}{2021 {{IEEE}} 45th {{Annual Computers}}, {{Software}}, and {{Applications Conference}} ({{COMPSAC}})}, pp. \bibinfo{pages}{1593--1598}.
\newblock \DOIprefix\doi{10.1109/COMPSAC51774.2021.00237}.
\bibitem[{Rauschnabel et~al.(2017)Rauschnabel, Rossmann and tom Dieck}]{rauschnabel2017adoption}
\bibinfo{author}{Rauschnabel, P.A.}, \bibinfo{author}{Rossmann, A.}, \bibinfo{author}{tom Dieck, M.C.}, \bibinfo{year}{2017}.
\newblock \bibinfo{title}{An adoption framework for mobile augmented reality games: The case of pok{\'e}mon go}.
\newblock \bibinfo{journal}{Computers in human behavior} \bibinfo{volume}{76}, \bibinfo{pages}{276--286}.
\bibitem[{Renz et~al.(2023)Renz, Neff, Baldauf and Maier}]{renzAuthenticationMethodsVoice2023}
\bibinfo{author}{Renz, A.}, \bibinfo{author}{Neff, T.}, \bibinfo{author}{Baldauf, M.}, \bibinfo{author}{Maier, E.}, \bibinfo{year}{2023}.
\newblock \bibinfo{title}{Authentication methods for voice services on smart speakers {\textendash} a multi-method study on perceived security and ease of use}.
\newblock \bibinfo{journal}{i-com} \bibinfo{volume}{22}, \bibinfo{pages}{67--81}.
\newblock \DOIprefix\doi{10.1515/icom-2022-0039}.
\bibitem[{Rese et~al.(2020)Rese, Ganster and Baier}]{reseChatbotsRetailersCustomer2020a}
\bibinfo{author}{Rese, A.}, \bibinfo{author}{Ganster, L.}, \bibinfo{author}{Baier, D.}, \bibinfo{year}{2020}.
\newblock \bibinfo{title}{Chatbots in retailers' customer communication: {{How}} to measure their acceptance?}
\newblock \bibinfo{journal}{Journal of Retailing and Consumer Services} \bibinfo{volume}{56}, \bibinfo{pages}{102176}.
\newblock \DOIprefix\doi{10.1016/j.jretconser.2020.102176}.
\bibitem[{Rohan et~al.(2023)Rohan, Pal, Hautam{\"a}ki, Funilkul, Chutimaskul and Thapliyal}]{rohanSystematicLiteratureReview2023}
\bibinfo{author}{Rohan, R.}, \bibinfo{author}{Pal, D.}, \bibinfo{author}{Hautam{\"a}ki, J.}, \bibinfo{author}{Funilkul, S.}, \bibinfo{author}{Chutimaskul, W.}, \bibinfo{author}{Thapliyal, H.}, \bibinfo{year}{2023}.
\newblock \bibinfo{title}{A systematic literature review of cybersecurity scales assessing information security awareness}.
\newblock \bibinfo{journal}{Heliyon} \bibinfo{volume}{9}, \bibinfo{pages}{e14234}.
\newblock \DOIprefix\doi{10.1016/j.heliyon.2023.e14234}.
\bibitem[{Rotter(1967)}]{rotter1967new}
\bibinfo{author}{Rotter, J.B.}, \bibinfo{year}{1967}.
\newblock \bibinfo{title}{A new scale for the measurement of interpersonal trust.}
\newblock \bibinfo{journal}{Journal of personality} .
\bibitem[{Saffarizadeh et~al.(2017)Saffarizadeh, Boodraj, Alashoor et~al.}]{saffarizadeh2017conversational}
\bibinfo{author}{Saffarizadeh, K.}, \bibinfo{author}{Boodraj, M.}, \bibinfo{author}{Alashoor, T.M.}, et~al., \bibinfo{year}{2017}.
\newblock \bibinfo{title}{Conversational assistants: Investigating privacy concerns, trust, and self-disclosure.}, in: \bibinfo{booktitle}{ICIS}.
\bibitem[{Salah et~al.(2023)Salah, Alhalbusi, Ismail and Abdelfattah}]{salah2023chatting}
\bibinfo{author}{Salah, M.}, \bibinfo{author}{Alhalbusi, H.}, \bibinfo{author}{Ismail, M.M.}, \bibinfo{author}{Abdelfattah, F.}, \bibinfo{year}{2023}.
\newblock \bibinfo{title}{Chatting with chatgpt: Decoding the mind of chatbot users and unveiling the intricate connections between user perception, trust and stereotype perception on self-esteem and psychological well-being}.
\newblock \bibinfo{journal}{Research Square} , \bibinfo{pages}{1--26}.
\bibitem[{Salisbury et~al.(2001)Salisbury, Pearson, Pearson and Miller}]{salisburyPerceivedSecurityWorld2001}
\bibinfo{author}{Salisbury, W.D.}, \bibinfo{author}{Pearson, R.A.}, \bibinfo{author}{Pearson, A.W.}, \bibinfo{author}{Miller, D.W.}, \bibinfo{year}{2001}.
\newblock \bibinfo{title}{Perceived security and {{World Wide Web}} purchase intention}.
\newblock \bibinfo{journal}{Industrial Management \& Data Systems} \bibinfo{volume}{101}, \bibinfo{pages}{165--177}.
\newblock \DOIprefix\doi{10.1108/02635570110390071}.
\bibitem[{Schanke et~al.(2021)Schanke, Burtch and Ray}]{schanke2021estimating}
\bibinfo{author}{Schanke, S.}, \bibinfo{author}{Burtch, G.}, \bibinfo{author}{Ray, G.}, \bibinfo{year}{2021}.
\newblock \bibinfo{title}{Estimating the impact of “humanizing” customer service chatbots}.
\newblock \bibinfo{journal}{Information Systems Research} \bibinfo{volume}{32}, \bibinfo{pages}{736--751}.
\bibitem[{Schepman and Rodway(2023)}]{schepman2023general}
\bibinfo{author}{Schepman, A.}, \bibinfo{author}{Rodway, P.}, \bibinfo{year}{2023}.
\newblock \bibinfo{title}{The general attitudes towards artificial intelligence scale (gaais): Confirmatory validation and associations with personality, corporate distrust, and general trust}.
\newblock \bibinfo{journal}{International Journal of Human--Computer Interaction} \bibinfo{volume}{39}, \bibinfo{pages}{2724--2741}.
\bibitem[{Schmidt et~al.(2020)Schmidt, Biessmann and Teubner}]{schmidt2020transparency}
\bibinfo{author}{Schmidt, P.}, \bibinfo{author}{Biessmann, F.}, \bibinfo{author}{Teubner, T.}, \bibinfo{year}{2020}.
\newblock \bibinfo{title}{Transparency and trust in artificial intelligence systems}.
\newblock \bibinfo{journal}{Journal of Decision Systems} \bibinfo{volume}{29}, \bibinfo{pages}{260--278}.
\bibitem[{Schomakers et~al.(2021)Schomakers, Biermann and Ziefle}]{schomakersUsersPreferencesSmart2021}
\bibinfo{author}{Schomakers, E.M.}, \bibinfo{author}{Biermann, H.}, \bibinfo{author}{Ziefle, M.}, \bibinfo{year}{2021}.
\newblock \bibinfo{title}{Users' {{Preferences}} for {{Smart Home Automation}} {\textendash} {{Investigating Aspects}} of {{Privacy}} and {{Trust}}}.
\newblock \bibinfo{journal}{Telematics and Informatics} \bibinfo{volume}{64}, \bibinfo{pages}{101689}.
\newblock \DOIprefix\doi{10.1016/j.tele.2021.101689}.
\bibitem[{Schomakers et~al.(2022)Schomakers, Lidynia and Ziefle}]{schomakers2022role}
\bibinfo{author}{Schomakers, E.M.}, \bibinfo{author}{Lidynia, C.}, \bibinfo{author}{Ziefle, M.}, \bibinfo{year}{2022}.
\newblock \bibinfo{title}{The role of privacy in the acceptance of smart technologies: Applying the privacy calculus to technology acceptance}.
\newblock \bibinfo{journal}{International Journal of Human--Computer Interaction} \bibinfo{volume}{38}, \bibinfo{pages}{1276--1289}.
\bibitem[{Schrepp and Thomaschewski(2019)}]{SchreppUEQ}
\bibinfo{author}{Schrepp, M.}, \bibinfo{author}{Thomaschewski, J.}, \bibinfo{year}{2019}.
\newblock \bibinfo{title}{Design and validation of a framework for the creation of user experience questionnaires}.
\newblock \bibinfo{journal}{International Journal of Interactive Multimedia and Artificial Intelligence} \bibinfo{volume}{5}, \bibinfo{pages}{88--95}.
\newblock \URLprefix \url{https://www.ijimai.org/journal/sites/default/files/files/2019/06/ijimai20195_7_9_pdf_19082.pdf}, \DOIprefix\doi{10.9781/ijimai.2019.06.006}.
\bibitem[{Schreuter et~al.(2021)Schreuter, van~der Putten and Lamers}]{schreuter2021trust}
\bibinfo{author}{Schreuter, D.}, \bibinfo{author}{van~der Putten, P.}, \bibinfo{author}{Lamers, M.H.}, \bibinfo{year}{2021}.
\newblock \bibinfo{title}{Trust me on this one: conforming to conversational assistants}.
\newblock \bibinfo{journal}{Minds and Machines} \bibinfo{volume}{31}, \bibinfo{pages}{535--562}.
\bibitem[{Schuetzler et~al.(2020)Schuetzler, Grimes and Scott~Giboney}]{schuetzler2020impact}
\bibinfo{author}{Schuetzler, R.M.}, \bibinfo{author}{Grimes, G.M.}, \bibinfo{author}{Scott~Giboney, J.}, \bibinfo{year}{2020}.
\newblock \bibinfo{title}{The impact of chatbot conversational skill on engagement and perceived humanness}.
\newblock \bibinfo{journal}{Journal of Management Information Systems} \bibinfo{volume}{37}, \bibinfo{pages}{875--900}.
\bibitem[{Seymour(2023)}]{seymourIgnoranceBlissEffect}
\bibinfo{author}{Seymour, W.}, \bibinfo{year}{2023}.
\newblock \bibinfo{title}{Ignorance is {{Bliss}}? {{The Effect}} of {{Explanations}} on {{Perceptions}} of {{Voice Assistants}}}.
\newblock \bibinfo{journal}{Proceedings of the ACM on Human-Computer Interaction} \bibinfo{volume}{7}, \bibinfo{pages}{1--24}.
\bibitem[{Shlega et~al.(2022)Shlega, Maqsood and Chiasson}]{shlega2022users}
\bibinfo{author}{Shlega, M.}, \bibinfo{author}{Maqsood, S.}, \bibinfo{author}{Chiasson, S.}, \bibinfo{year}{2022}.
\newblock \bibinfo{title}{Users, smart homes, and digital assistants: impact of technology experience and adoption}, in: \bibinfo{booktitle}{International Conference on Human-Computer Interaction}, \bibinfo{organization}{Springer}. pp. \bibinfo{pages}{422--443}.
\bibitem[{Shofolahan and Kang(2018)}]{Shofolahan2018}
\bibinfo{author}{Shofolahan, T.O.}, \bibinfo{author}{Kang, J.}, \bibinfo{year}{2018}.
\newblock \bibinfo{title}{An {{Integrated Framework}} for {{Modeling}} the {{Influential Factors Affecting}} the {{Use}} of {{Voice-Enabled IoT Devices}} : {{A Case Study}} of {{Amazon Echo}}}.
\newblock \bibinfo{journal}{Asia Pacific Journal of Information Systems} \bibinfo{volume}{28}, \bibinfo{pages}{320--349}.
\newblock \DOIprefix\doi{10.14329/apjis.2018.28.4.320}.
\bibitem[{Shuhaiber and Mashal(2019)}]{shuhaiber2019understanding}
\bibinfo{author}{Shuhaiber, A.}, \bibinfo{author}{Mashal, I.}, \bibinfo{year}{2019}.
\newblock \bibinfo{title}{Understanding users’ acceptance of smart homes}.
\newblock \bibinfo{journal}{Technology in Society} \bibinfo{volume}{58}, \bibinfo{pages}{101110}.
\bibitem[{Skjuve and Brandzaeg(2019)}]{skjuve2019measuring}
\bibinfo{author}{Skjuve, M.}, \bibinfo{author}{Brandzaeg, P.B.}, \bibinfo{year}{2019}.
\newblock \bibinfo{title}{Measuring user experience in chatbots: An approach to interpersonal communication competence}, in: \bibinfo{booktitle}{Internet Science: INSCI 2018 International Workshops, St. Petersburg, Russia, October 24--26, 2018, Revised Selected Papers 5}, \bibinfo{organization}{Springer}. pp. \bibinfo{pages}{113--120}.
\bibitem[{{Smith} et~al.(2011){Smith}, {Dinev} and {Xu}}]{smithInformationPrivacyResearch2011}
\bibinfo{author}{{Smith}}, \bibinfo{author}{{Dinev}}, \bibinfo{author}{{Xu}}, \bibinfo{year}{2011}.
\newblock \bibinfo{title}{Information {{Privacy Research}}: {{An Interdisciplinary Review}}}.
\newblock \bibinfo{journal}{MIS Quarterly} \bibinfo{volume}{35}, \bibinfo{pages}{989}.
\newblock \DOIprefix\doi{10.2307/41409970}, \href{http://arxiv.org/abs/10.2307/41409970}{\tt arXiv:10.2307/41409970}.
\bibitem[{Smith et~al.(1996)Smith, Milberg and Burke}]{smithInformationPrivacyMeasuring1996b}
\bibinfo{author}{Smith, H.J.}, \bibinfo{author}{Milberg, S.J.}, \bibinfo{author}{Burke, S.J.}, \bibinfo{year}{1996}.
\newblock \bibinfo{title}{Information privacy: {{Measuring}} individuals' concerns about organizational practices}.
\newblock \bibinfo{journal}{MIS Quarterly} \bibinfo{volume}{20}, \bibinfo{pages}{167--196}.
\newblock \DOIprefix\doi{10.2307/249477}.
\bibitem[{Son and Kim(2008)}]{son2008internet}
\bibinfo{author}{Son, J.Y.}, \bibinfo{author}{Kim, S.S.}, \bibinfo{year}{2008}.
\newblock \bibinfo{title}{Internet users' information privacy-protective responses: A taxonomy and a nomological model}.
\newblock \bibinfo{journal}{MIS quarterly} , \bibinfo{pages}{503--529}.
\bibitem[{Song et~al.(2022a)Song, Du, Xing and Mou}]{song2022should}
\bibinfo{author}{Song, M.}, \bibinfo{author}{Du, J.}, \bibinfo{author}{Xing, X.}, \bibinfo{author}{Mou, J.}, \bibinfo{year}{2022}a.
\newblock \bibinfo{title}{Should the chatbot “save itself” or “be helped by others”? the influence of service recovery types on consumer perceptions of recovery satisfaction}.
\newblock \bibinfo{journal}{Electronic Commerce Research and Applications} \bibinfo{volume}{55}, \bibinfo{pages}{101199}.
\bibitem[{Song et~al.(2022b)Song, Xing, Duan, Cohen and Mou}]{song2022will}
\bibinfo{author}{Song, M.}, \bibinfo{author}{Xing, X.}, \bibinfo{author}{Duan, Y.}, \bibinfo{author}{Cohen, J.}, \bibinfo{author}{Mou, J.}, \bibinfo{year}{2022}b.
\newblock \bibinfo{title}{Will artificial intelligence replace human customer service? the impact of communication quality and privacy risks on adoption intention}.
\newblock \bibinfo{journal}{Journal of Retailing and Consumer Services} \bibinfo{volume}{66}, \bibinfo{pages}{102900}.
\bibitem[{Suplet et~al.(2009)Suplet, G{\'o}mez~Su{\'a}rez and {D{\'i}az-Mart{\'i}n}}]{supletCustomerPerceptionsPerceived2009}
\bibinfo{author}{Suplet, M.}, \bibinfo{author}{G{\'o}mez~Su{\'a}rez, M.}, \bibinfo{author}{{D{\'i}az-Mart{\'i}n}, A.}, \bibinfo{year}{2009}.
\newblock \bibinfo{title}{Customer perceptions of perceived risk in generic drugs: {{The Spanish}} market}.
\newblock \bibinfo{journal}{Innovar} \bibinfo{volume}{19}, \bibinfo{pages}{53--64}.
\bibitem[{Tabassum et~al.(2019)Tabassum, Kosi{\'n}ski, Frik, Malkin, Wijesekera, Egelman and Lipford}]{tabassumInvestigatingUsersPreferences2019a}
\bibinfo{author}{Tabassum, M.}, \bibinfo{author}{Kosi{\'n}ski, T.}, \bibinfo{author}{Frik, A.}, \bibinfo{author}{Malkin, N.}, \bibinfo{author}{Wijesekera, P.}, \bibinfo{author}{Egelman, S.}, \bibinfo{author}{Lipford, H.R.}, \bibinfo{year}{2019}.
\newblock \bibinfo{title}{Investigating {{Users}}' {{Preferences}} and {{Expectations}} for {{Always-Listening Voice Assistants}}}.
\newblock \bibinfo{journal}{Proceedings of the ACM on Interactive, Mobile, Wearable and Ubiquitous Technologies} \bibinfo{volume}{3}, \bibinfo{pages}{1--23}.
\newblock \DOIprefix\doi{10.1145/3369807}.
\bibitem[{Taddei and Contena(2013)}]{taddei2013privacy}
\bibinfo{author}{Taddei, S.}, \bibinfo{author}{Contena, B.}, \bibinfo{year}{2013}.
\newblock \bibinfo{title}{Privacy, trust and control: Which relationships with online self-disclosure?}
\newblock \bibinfo{journal}{Computers in human behavior} \bibinfo{volume}{29}, \bibinfo{pages}{821--826}.
\bibitem[{Tan and Sutherland(2005)}]{tan2005consumer}
\bibinfo{author}{Tan, F.B.}, \bibinfo{author}{Sutherland, P.}, \bibinfo{year}{2005}.
\newblock \bibinfo{title}{Consumer trust: A multi-dimensional model}.
\newblock \bibinfo{journal}{Advanced Topics in Electronic Commerce, Volume 1} \bibinfo{volume}{1}, \bibinfo{pages}{188}.
\bibitem[{Tastemirova et~al.(2022)Tastemirova, Schneider, Kruse, Heinzle and Brocke}]{tastemirova2022microexpressions}
\bibinfo{author}{Tastemirova, A.}, \bibinfo{author}{Schneider, J.}, \bibinfo{author}{Kruse, L.C.}, \bibinfo{author}{Heinzle, S.}, \bibinfo{author}{Brocke, J.v.}, \bibinfo{year}{2022}.
\newblock \bibinfo{title}{Microexpressions in digital humans: perceived affect, sincerity, and trustworthiness}.
\newblock \bibinfo{journal}{Electronic Markets} \bibinfo{volume}{32}, \bibinfo{pages}{1603--1620}.
\bibitem[{Tenhundfeld et~al.(2021)Tenhundfeld, Barr, Emily and Weger}]{tenhundfeld2021myISMYSIRI}
\bibinfo{author}{Tenhundfeld, N.L.}, \bibinfo{author}{Barr, H.M.}, \bibinfo{author}{Emily, H.}, \bibinfo{author}{Weger, K.}, \bibinfo{year}{2021}.
\newblock \bibinfo{title}{Is my siri the same as your siri? an exploration of users’ mental model of virtual personal assistants, implications for trust}.
\newblock \bibinfo{journal}{IEEE Transactions on Human-Machine Systems} \bibinfo{volume}{52}, \bibinfo{pages}{512--521}.
\bibitem[{Tennant et~al.(2022)Tennant, Allana, Mercer and Burns}]{Tennant2022Caregiver}
\bibinfo{author}{Tennant, R.}, \bibinfo{author}{Allana, S.}, \bibinfo{author}{Mercer, K.}, \bibinfo{author}{Burns, C.M.}, \bibinfo{year}{2022}.
\newblock \bibinfo{title}{Caregiver expectations of interfacing with voice assistants to support complex home care: Mixed methods study}.
\newblock \bibinfo{journal}{JMIR Hum Factors} \bibinfo{volume}{9}, \bibinfo{pages}{e37688}.
\newblock \DOIprefix\doi{10.2196/37688}.
\bibitem[{Tlili et~al.(2023)Tlili, Shehata, Adarkwah, Bozkurt, Hickey, Huang and Agyemang}]{tliliWhatIfDevil2023a}
\bibinfo{author}{Tlili, A.}, \bibinfo{author}{Shehata, B.}, \bibinfo{author}{Adarkwah, M.A.}, \bibinfo{author}{Bozkurt, A.}, \bibinfo{author}{Hickey, D.T.}, \bibinfo{author}{Huang, R.}, \bibinfo{author}{Agyemang, B.}, \bibinfo{year}{2023}.
\newblock \bibinfo{title}{What if the devil is my guardian angel: {{ChatGPT}} as a case study of using chatbots in education}.
\newblock \bibinfo{journal}{Smart Learning Environments} \bibinfo{volume}{10}, \bibinfo{pages}{15}.
\newblock \DOIprefix\doi{10.1186/s40561-023-00237-x}.
\bibitem[{Toader et~al.(2019)Toader, Boca, Toader, M{\u{a}}celaru, Toader, Ighian and R{\u{a}}dulescu}]{toader2019effect}
\bibinfo{author}{Toader, D.C.}, \bibinfo{author}{Boca, G.}, \bibinfo{author}{Toader, R.}, \bibinfo{author}{M{\u{a}}celaru, M.}, \bibinfo{author}{Toader, C.}, \bibinfo{author}{Ighian, D.}, \bibinfo{author}{R{\u{a}}dulescu, A.T.}, \bibinfo{year}{2019}.
\newblock \bibinfo{title}{The effect of social presence and chatbot errors on trust}.
\newblock \bibinfo{journal}{Sustainability} \bibinfo{volume}{12}, \bibinfo{pages}{256}.
\bibitem[{Trepte et~al.(2020)Trepte, Scharkow and Dienlin}]{treptePrivacyCalculusContextualized2020}
\bibinfo{author}{Trepte, S.}, \bibinfo{author}{Scharkow, M.}, \bibinfo{author}{Dienlin, T.}, \bibinfo{year}{2020}.
\newblock \bibinfo{title}{The privacy calculus contextualized: {{The}} influence of affordances}.
\newblock \bibinfo{journal}{Computers in Human Behavior} \bibinfo{volume}{104}, \bibinfo{pages}{106115}.
\newblock \DOIprefix\doi{10.1016/j.chb.2019.08.022}.
\bibitem[{Trivedi(2019)}]{trivedi2019examining}
\bibinfo{author}{Trivedi, J.}, \bibinfo{year}{2019}.
\newblock \bibinfo{title}{Examining the customer experience of using banking chatbots and its impact on brand love: The moderating role of perceived risk}.
\newblock \bibinfo{journal}{Journal of internet Commerce} \bibinfo{volume}{18}, \bibinfo{pages}{91--111}.
\bibitem[{Tsu~Wei et~al.(2009)Tsu~Wei, Marthandan, Yee-Loong~Chong, Ooi and Arumugam}]{tsuweiWhatDrivesMalaysian2009}
\bibinfo{author}{Tsu~Wei, T.}, \bibinfo{author}{Marthandan, G.}, \bibinfo{author}{Yee-Loong~Chong, A.}, \bibinfo{author}{Ooi, K.B.}, \bibinfo{author}{Arumugam, S.}, \bibinfo{year}{2009}.
\newblock \bibinfo{title}{What drives {{Malaysian}} m-commerce adoption? {{An}} empirical analysis}.
\newblock \bibinfo{journal}{Industrial Management \& Data Systems} \bibinfo{volume}{109}, \bibinfo{pages}{370--388}.
\newblock \DOIprefix\doi{10.1108/02635570910939399}.
\bibitem[{Uysal et~al.(2022)Uysal, Alavi and Bezen{\c c}on}]{uysalTrojanHorseUseful2022a}
\bibinfo{author}{Uysal, E.}, \bibinfo{author}{Alavi, S.}, \bibinfo{author}{Bezen{\c c}on, V.}, \bibinfo{year}{2022}.
\newblock \bibinfo{title}{Trojan horse or useful helper? {{A}} relationship perspective on artificial intelligence assistants with humanlike features}.
\newblock \bibinfo{journal}{Journal of the Academy of Marketing Science} \bibinfo{volume}{50}, \bibinfo{pages}{1153--1175}.
\newblock \DOIprefix\doi{10.1007/s11747-022-00856-9}.
\bibitem[{{van Bussel} et~al.(2022){van Bussel}, {Odekerken{\textendash}Schr{\"o}der}, Ou, Swart and Jacobs}]{vanbusselAnalyzingDeterminantsAccept2022b}
\bibinfo{author}{{van Bussel}, M.J.P.}, \bibinfo{author}{{Odekerken{\textendash}Schr{\"o}der}, G.J.}, \bibinfo{author}{Ou, C.}, \bibinfo{author}{Swart, R.R.}, \bibinfo{author}{Jacobs, M.J.G.}, \bibinfo{year}{2022}.
\newblock \bibinfo{title}{Analyzing the determinants to accept a virtual assistant and use cases among cancer patients: A mixed methods study}.
\newblock \bibinfo{journal}{BMC Health Services Research} \bibinfo{volume}{22}, \bibinfo{pages}{890}.
\newblock \DOIprefix\doi{10.1186/s12913-022-08189-7}.
\bibitem[{Van Der~Goot and Pilgrim(2020)}]{vandergootExploringAgeDifferences2020}
\bibinfo{author}{Van Der~Goot, M.J.}, \bibinfo{author}{Pilgrim, T.}, \bibinfo{year}{2020}.
\newblock \bibinfo{title}{Exploring {{Age Differences}} in {{Motivations}} for and {{Acceptance}} of {{Chatbot Communication}} in a {{Customer Service Context}}}, in: \bibinfo{editor}{F{\o}lstad, A.}, \bibinfo{editor}{Araujo, T.}, \bibinfo{editor}{Papadopoulos, S.}, \bibinfo{editor}{Law, E.L.C.}, \bibinfo{editor}{Granmo, O.C.}, \bibinfo{editor}{Luger, E.}, \bibinfo{editor}{Brandtzaeg, P.B.} (Eds.), \bibinfo{booktitle}{Chatbot {{Research}} and {{Design}}}. \bibinfo{publisher}{{Springer International Publishing}}, \bibinfo{address}{{Cham}}. volume \bibinfo{volume}{11970}, pp. \bibinfo{pages}{173--186}.
\newblock \DOIprefix\doi{10.1007/978-3-030-39540-7_12}.
\bibitem[{Van~Deursen et~al.(2016)Van~Deursen, Helsper and Eynon}]{van2016development}
\bibinfo{author}{Van~Deursen, A.J.}, \bibinfo{author}{Helsper, E.J.}, \bibinfo{author}{Eynon, R.}, \bibinfo{year}{2016}.
\newblock \bibinfo{title}{Development and validation of the internet skills scale (iss)}.
\newblock \bibinfo{journal}{Information, communication \& society} \bibinfo{volume}{19}, \bibinfo{pages}{804--823}.
\bibitem[{Venkatesh and Davis(2000)}]{venkatesh2000theoretical}
\bibinfo{author}{Venkatesh, V.}, \bibinfo{author}{Davis, F.D.}, \bibinfo{year}{2000}.
\newblock \bibinfo{title}{A theoretical extension of the technology acceptance model: Four longitudinal field studies}.
\newblock \bibinfo{journal}{Management science} \bibinfo{volume}{46}, \bibinfo{pages}{186--204}.
\bibitem[{Venkatesh et~al.(2012)Venkatesh, Thong and Xu}]{venkatesh2012consumer}
\bibinfo{author}{Venkatesh, V.}, \bibinfo{author}{Thong, J.Y.}, \bibinfo{author}{Xu, X.}, \bibinfo{year}{2012}.
\newblock \bibinfo{title}{Consumer acceptance and use of information technology: extending the unified theory of acceptance and use of technology}.
\newblock \bibinfo{journal}{MIS quarterly} , \bibinfo{pages}{157--178}.
\bibitem[{Vimalkumar et~al.(2021)Vimalkumar, Sharma, Singh and Dwivedi}]{vimalkumarOkayGoogleWhat2021c}
\bibinfo{author}{Vimalkumar, M.}, \bibinfo{author}{Sharma, S.K.}, \bibinfo{author}{Singh, J.B.}, \bibinfo{author}{Dwivedi, Y.K.}, \bibinfo{year}{2021}.
\newblock \bibinfo{title}{`{{Okay}} google, what about my privacy?': {{User}}'s privacy perceptions and acceptance of voice based digital assistants}.
\newblock \bibinfo{journal}{Computers in Human Behavior} \bibinfo{volume}{120}, \bibinfo{pages}{106763}.
\newblock \DOIprefix\doi{10.1016/j.chb.2021.106763}.
\bibitem[{Vitak(2015)}]{vitakBalancingPrivacyConcerns}
\bibinfo{author}{Vitak, J.}, \bibinfo{year}{2015}.
\newblock \bibinfo{title}{Balancing {{Privacy Concerns}} and {{Impression Management Strategies}} on {{Facebook}}}, in: \bibinfo{booktitle}{Symposium on usable privacy and security (SOUPS)}, pp. \bibinfo{pages}{22--24}.
\bibitem[{Vitak(2016)}]{vitakDigitalPathHappiness2016}
\bibinfo{author}{Vitak, J.}, \bibinfo{year}{2016}.
\newblock \bibinfo{title}{A {{Digital Path}} to {{Happiness}}?: {{Applying Communication Privacy Management Theory}} to {{Mediated Interactions}}}, in: \bibinfo{booktitle}{The {{Routledge Handbook}} of {{Media Use}} and {{Well-Being}}}, \bibinfo{publisher}{{Routledge}}. pp. \bibinfo{pages}{274--288}.
\bibitem[{{Vixen Labs}(2023)}]{VixenLabs2023}
\bibinfo{author}{{Vixen Labs}}, \bibinfo{year}{2023}.
\newblock \bibinfo{title}{{AI} {Consumer} {Index}}.
\newblock \bibinfo{howpublished}{[Accessed Online 13-12-2023] \url{https://vixenlabs.co/research/ai-consumer-index-2023}}.
\bibitem[{Wald et~al.(2021)Wald, Heijselaar and Bosse}]{wald2021make}
\bibinfo{author}{Wald, R.}, \bibinfo{author}{Heijselaar, E.}, \bibinfo{author}{Bosse, T.}, \bibinfo{year}{2021}.
\newblock \bibinfo{title}{Make your own: The potential of chatbot customization for the development of user trust}, in: \bibinfo{booktitle}{Adjunct Proceedings of the 29th ACM Conference on User Modeling, Adaptation and Personalization}, pp. \bibinfo{pages}{382--387}.
\bibitem[{Wang and Benbasat(2016)}]{wang2016empirical}
\bibinfo{author}{Wang, W.}, \bibinfo{author}{Benbasat, I.}, \bibinfo{year}{2016}.
\newblock \bibinfo{title}{Empirical assessment of alternative designs for enhancing different types of trusting beliefs in online recommendation agents}.
\newblock \bibinfo{journal}{Journal of management information systems} \bibinfo{volume}{33}, \bibinfo{pages}{744--775}.
\bibitem[{Wang and Wang(2022)}]{wang2022development}
\bibinfo{author}{Wang, Y.Y.}, \bibinfo{author}{Wang, Y.S.}, \bibinfo{year}{2022}.
\newblock \bibinfo{title}{Development and validation of an artificial intelligence anxiety scale: An initial application in predicting motivated learning behavior}.
\newblock \bibinfo{journal}{Interactive Learning Environments} \bibinfo{volume}{30}, \bibinfo{pages}{619--634}.
\bibitem[{Wang et~al.(2017)Wang, Mao, Li and Liu}]{wang2017smile}
\bibinfo{author}{Wang, Z.}, \bibinfo{author}{Mao, H.}, \bibinfo{author}{Li, Y.J.}, \bibinfo{author}{Liu, F.}, \bibinfo{year}{2017}.
\newblock \bibinfo{title}{Smile big or not? effects of smile intensity on perceptions of warmth and competence}.
\newblock \bibinfo{journal}{Journal of Consumer Research} \bibinfo{volume}{43}, \bibinfo{pages}{787--805}.
\bibitem[{Waytz et~al.(2010)Waytz, Cacioppo and Epley}]{waytz2010sees}
\bibinfo{author}{Waytz, A.}, \bibinfo{author}{Cacioppo, J.}, \bibinfo{author}{Epley, N.}, \bibinfo{year}{2010}.
\newblock \bibinfo{title}{Who sees human? the stability and importance of individual differences in anthropomorphism}.
\newblock \bibinfo{journal}{Perspectives on Psychological Science} \bibinfo{volume}{5}, \bibinfo{pages}{219--232}.
\bibitem[{Weidm{\"u}ller(2022)}]{weidmuller2022human}
\bibinfo{author}{Weidm{\"u}ller, L.}, \bibinfo{year}{2022}.
\newblock \bibinfo{title}{Human, hybrid, or machine?: Exploring the trustworthiness of voice-based assistants}.
\newblock \bibinfo{journal}{Human-Machine Communication} \bibinfo{volume}{4}, \bibinfo{pages}{85--110}.
\bibitem[{Weitz et~al.(2019)Weitz, Schiller, Schlagowski, Huber and Andr{\'e}}]{weitz2019you}
\bibinfo{author}{Weitz, K.}, \bibinfo{author}{Schiller, D.}, \bibinfo{author}{Schlagowski, R.}, \bibinfo{author}{Huber, T.}, \bibinfo{author}{Andr{\'e}, E.}, \bibinfo{year}{2019}.
\newblock \bibinfo{title}{"do you trust me?" increasing user-trust by integrating virtual agents in explainable ai interaction design}, in: \bibinfo{booktitle}{Proceedings of the 19th ACM International Conference on Intelligent Virtual Agents}, pp. \bibinfo{pages}{7--9}.
\bibitem[{Xiao and Benbasat(2002)}]{xiao2002impact}
\bibinfo{author}{Xiao, S.}, \bibinfo{author}{Benbasat, I.}, \bibinfo{year}{2002}.
\newblock \bibinfo{title}{The impact of internalization and familiarity on trust and adoption of recommendation agents}.
\newblock \bibinfo{journal}{Unpublished Working Paper} .
\bibitem[{Xu et~al.(2008)Xu, Dinev, Smith and Hart}]{xuExaminingFormationIndividual2008}
\bibinfo{author}{Xu, H.}, \bibinfo{author}{Dinev, T.}, \bibinfo{author}{Smith, H.}, \bibinfo{author}{Hart, P.}, \bibinfo{year}{2008}.
\newblock \bibinfo{title}{Examining the {{Formation}} of {{Individual}}'s {{Privacy Concerns}}: {{Toward}} an {{Integrative View}}}.
\newblock \bibinfo{journal}{ICIS 2008 Proceedings} .
\bibitem[{Xu et~al.(2011a)Xu, Dinev, Smith and Hart}]{xu2011information}
\bibinfo{author}{Xu, H.}, \bibinfo{author}{Dinev, T.}, \bibinfo{author}{Smith, J.}, \bibinfo{author}{Hart, P.}, \bibinfo{year}{2011}a.
\newblock \bibinfo{title}{Information privacy concerns: Linking individual perceptions with institutional privacy assurances}.
\newblock \bibinfo{journal}{Journal of the Association for Information Systems} \bibinfo{volume}{12}, \bibinfo{pages}{1}.
\bibitem[{Xu and Gupta(2009)}]{xu2009effects}
\bibinfo{author}{Xu, H.}, \bibinfo{author}{Gupta, S.}, \bibinfo{year}{2009}.
\newblock \bibinfo{title}{The effects of privacy concerns and personal innovativeness on potential and experienced customers’ adoption of location-based services}.
\newblock \bibinfo{journal}{Electronic Markets} \bibinfo{volume}{19}, \bibinfo{pages}{137--149}.
\bibitem[{Xu et~al.(2012)Xu, Gupta, Rosson and Carroll}]{xuMeasuringMobileUsers2012}
\bibinfo{author}{Xu, H.}, \bibinfo{author}{Gupta, S.}, \bibinfo{author}{Rosson, M.}, \bibinfo{author}{Carroll, J.}, \bibinfo{year}{2012}.
\newblock \bibinfo{title}{Measuring {{Mobile Users}}' {{Concerns}} for {{Information Privacy}}}.
\newblock \bibinfo{journal}{ICIS 2012 Proceedings} .
\bibitem[{Xu et~al.(2011b)Xu, Luo, Carroll and Rosson}]{xu2011personalization}
\bibinfo{author}{Xu, H.}, \bibinfo{author}{Luo, X.R.}, \bibinfo{author}{Carroll, J.M.}, \bibinfo{author}{Rosson, M.B.}, \bibinfo{year}{2011}b.
\newblock \bibinfo{title}{The personalization privacy paradox: An exploratory study of decision making process for location-aware marketing}.
\newblock \bibinfo{journal}{Decision support systems} \bibinfo{volume}{51}, \bibinfo{pages}{42--52}.
\bibitem[{Xu et~al.(2022)Xu, Chan-Olmsted and Liu}]{xuSmartSpeakersRequire2022}
\bibinfo{author}{Xu, K.}, \bibinfo{author}{Chan-Olmsted, S.}, \bibinfo{author}{Liu, F.}, \bibinfo{year}{2022}.
\newblock \bibinfo{title}{Smart {{Speakers Require Smart Management}}: {{Two Routes From User Gratifications}} to {{Privacy Settings}}}.
\newblock \bibinfo{journal}{International Journal of Communication} \bibinfo{volume}{16}, \bibinfo{pages}{23}.
\bibitem[{Yamagishi and Yamagishi(1994)}]{yamagishi1994trust}
\bibinfo{author}{Yamagishi, T.}, \bibinfo{author}{Yamagishi, M.}, \bibinfo{year}{1994}.
\newblock \bibinfo{title}{Trust and commitment in the united states and japan}.
\newblock \bibinfo{journal}{Motivation and emotion} \bibinfo{volume}{18}, \bibinfo{pages}{129--166}.
\bibitem[{Yang et~al.(2017)Yang, Lee and Zo}]{yangUserAcceptanceSmart2017}
\bibinfo{author}{Yang, H.}, \bibinfo{author}{Lee, H.}, \bibinfo{author}{Zo, H.}, \bibinfo{year}{2017}.
\newblock \bibinfo{title}{User acceptance of smart home services: An extension of the theory of planned behavior}.
\newblock \bibinfo{journal}{Industrial Management \& Data Systems} \bibinfo{volume}{117}, \bibinfo{pages}{68--89}.
\newblock \DOIprefix\doi{10.1108/IMDS-01-2016-0017}.
\bibitem[{Yang et~al.(2015)Yang, Liu, Li and Yu}]{yang2015understanding}
\bibinfo{author}{Yang, Y.}, \bibinfo{author}{Liu, Y.}, \bibinfo{author}{Li, H.}, \bibinfo{author}{Yu, B.}, \bibinfo{year}{2015}.
\newblock \bibinfo{title}{Understanding perceived risks in mobile payment acceptance}.
\newblock \bibinfo{journal}{Industrial Management \& Data Systems} \bibinfo{volume}{115}, \bibinfo{pages}{253--269}.
\bibitem[{Yao et~al.(2019)Yao, Basdeo, Mcdonough and Wang}]{yaoPrivacyPerceptionsDesigns2019d}
\bibinfo{author}{Yao, Y.}, \bibinfo{author}{Basdeo, J.R.}, \bibinfo{author}{Mcdonough, O.R.}, \bibinfo{author}{Wang, Y.}, \bibinfo{year}{2019}.
\newblock \bibinfo{title}{Privacy {{Perceptions}} and {{Designs}} of {{Bystanders}} in {{Smart Homes}}}.
\newblock \bibinfo{journal}{Proceedings of the ACM on Human-Computer Interaction} \bibinfo{volume}{3}, \bibinfo{pages}{1--24}.
\newblock \DOIprefix\doi{10.1145/3359161}.
\bibitem[{Ye et~al.(2019)Ye, Ying, Zhou and Wang}]{yeEnhancingCustomerTrust2019}
\bibinfo{author}{Ye, S.}, \bibinfo{author}{Ying, T.}, \bibinfo{author}{Zhou, L.}, \bibinfo{author}{Wang, T.}, \bibinfo{year}{2019}.
\newblock \bibinfo{title}{Enhancing customer trust in peer-to-peer accommodation: {{A}} ``soft'' strategy via social presence}.
\newblock \bibinfo{journal}{International Journal of Hospitality Management} \bibinfo{volume}{79}, \bibinfo{pages}{1--10}.
\newblock \DOIprefix\doi{10.1016/j.ijhm.2018.11.017}.
\bibitem[{Yoo et~al.(2010)Yoo, Lee and Park}]{yoo2010role}
\bibinfo{author}{Yoo, W.S.}, \bibinfo{author}{Lee, Y.}, \bibinfo{author}{Park, J.}, \bibinfo{year}{2010}.
\newblock \bibinfo{title}{The role of interactivity in e-tailing: Creating value and increasing satisfaction}.
\newblock \bibinfo{journal}{Journal of retailing and consumer services} \bibinfo{volume}{17}, \bibinfo{pages}{89--96}.
\bibitem[{Zeissig et~al.(2017)Zeissig, Lidynia, Vervier, Gadeib and Ziefle}]{zeissig2017online}
\bibinfo{author}{Zeissig, E.M.}, \bibinfo{author}{Lidynia, C.}, \bibinfo{author}{Vervier, L.}, \bibinfo{author}{Gadeib, A.}, \bibinfo{author}{Ziefle, M.}, \bibinfo{year}{2017}.
\newblock \bibinfo{title}{Online privacy perceptions of older adults}, in: \bibinfo{booktitle}{Human Aspects of IT for the Aged Population. Applications, Services and Contexts: Third International Conference, ITAP 2017, Held as Part of HCI International 2017, Vancouver, BC, Canada, July 9-14, 2017, Proceedings, Part II 3}, \bibinfo{organization}{Springer}. pp. \bibinfo{pages}{181--200}.
\bibitem[{Zeng et~al.(2017)Zeng, Mare and Roesner}]{zeng2017end}
\bibinfo{author}{Zeng, E.}, \bibinfo{author}{Mare, S.}, \bibinfo{author}{Roesner, F.}, \bibinfo{year}{2017}.
\newblock \bibinfo{title}{End user security and privacy concerns with smart homes}, in: \bibinfo{booktitle}{thirteenth symposium on usable privacy and security (SOUPS 2017)}, pp. \bibinfo{pages}{65--80}.
\bibitem[{Zhang et~al.(2021)Zhang, Meng, Chen, Yang and Zhao}]{zhang2021motivation}
\bibinfo{author}{Zhang, S.}, \bibinfo{author}{Meng, Z.}, \bibinfo{author}{Chen, B.}, \bibinfo{author}{Yang, X.}, \bibinfo{author}{Zhao, X.}, \bibinfo{year}{2021}.
\newblock \bibinfo{title}{Motivation, social emotion, and the acceptance of artificial intelligence virtual assistants—trust-based mediating effects}.
\newblock \bibinfo{journal}{Frontiers in Psychology} \bibinfo{volume}{12}, \bibinfo{pages}{728495}.
\bibitem[{Zhou(2011)}]{zhouImpactPrivacyConcern2011}
\bibinfo{author}{Zhou, T.}, \bibinfo{year}{2011}.
\newblock \bibinfo{title}{The impact of privacy concern on user adoption of location-based services}.
\newblock \bibinfo{journal}{Industrial Management \& Data Systems} \bibinfo{volume}{111}, \bibinfo{pages}{212--226}.
\newblock \DOIprefix\doi{10.1108/02635571111115146}.
\bibitem[{Zlatolas et~al.(2019)Zlatolas, Welzer, Hölbl, Hericko and Kamisalic}]{Zlatolas2019AMO}
\bibinfo{author}{Zlatolas, L.N.}, \bibinfo{author}{Welzer, T.}, \bibinfo{author}{Hölbl, M.}, \bibinfo{author}{Hericko, M.}, \bibinfo{author}{Kamisalic, A.}, \bibinfo{year}{2019}.
\newblock \bibinfo{title}{A model of perception of privacy, trust, and self-disclosure on online social networks}.
\newblock \bibinfo{journal}{Entropy} \bibinfo{volume}{21}.
\newblock \URLprefix \url{https://api.semanticscholar.org/CorpusID:201264486}.
\bibitem[{Zwakman et~al.(2021)Zwakman, Pal and Arpnikanondt}]{zwakmanUsabilityEvaluationArtificial2021c}
\bibinfo{author}{Zwakman, D.S.}, \bibinfo{author}{Pal, D.}, \bibinfo{author}{Arpnikanondt, C.}, \bibinfo{year}{2021}.
\newblock \bibinfo{title}{Usability {{Evaluation}} of {{Artificial Intelligence-Based Voice Assistants}}: {{The Case}} of {{Amazon Alexa}}}.
\newblock \bibinfo{journal}{SN Computer Science} \bibinfo{volume}{2}, \bibinfo{pages}{28}.
\newblock \DOIprefix\doi{10.1007/s42979-020-00424-4}.

\end{thebibliography}

\end{document}